\documentclass[a4paper,11pt]{article}
\pdfoutput=1 

\usepackage{jheppub} 

\usepackage{amsmath, mathtools}
\usepackage{amsfonts}
\usepackage{amssymb}
\usepackage{mathtools}
\usepackage{accents}
\usepackage{comment}
\usepackage{braket}
\usepackage{enumitem}
\usepackage{xcolor}
\usepackage[force]{feynmp-auto}
 \newcommand{\marrow}[5]{%
    \fmfcmd{style_def marrow#1
    expr p = drawarrow subpath (1/4, 3/4) of p shifted 6 #2 withpen pencircle scaled 0.4;
    label.#3(btex #4 etex, point 0.5 of p shifted 6 #2);
    enddef;}
    \fmf{marrow#1,tension=0}{#5}}

\newcommand{\Marrow}[6]{%
    \fmfcmd{style_def marrow#1
    expr p = drawarrow subpath (1/4, 3/4) of p shifted #6 #2 withpen pencircle scaled 0.4;
    label.#3(btex #4 etex, point 0.5 of p shifted #6 #2);
    enddef;}
    \fmf{marrow#1,tension=0}{#5}}

\numberwithin{equation}{section}

\newcommand{\pri}[1]{\accentset{\prime}{#1}}

\newcommand{\Rads}{R_{\text{AdS}}}
\newcommand{\AdS}[1]{\text{AdS}_{#1}}
\newcommand{\CFT}{\text{CFT}}
\newcommand{\AdSSS}{\text{AdS}_{3}\times\text{S}^3\times\text{S}^3\times\text{S}^1}
\renewcommand{\S}{\text{S}}
\newcommand{\cX}{\mathcal{X}}
\newcommand{\T}{\text{T}}
\newcommand{\psu}{\mathfrak{psu}}
\newcommand{\dalg}{\mathfrak{d}(2,1;\alpha)}
\newcommand{\su}{\mathfrak{su}}
\renewcommand{\sl}{\mathfrak{sl}}
\newcommand{\Li}[1]{\text{Li}_2\Big[{#1}\Big]}

\title{Integrable S~matrix, mirror TBA and spectrum for the stringy $\AdSSS$ WZW model}

\author{A.~Dei,}
\author{A.~Sfondrini}

\affiliation{Institut f\"ur theoretische Physik, ETH Z\"urich\\ Wolfgang-Pauli-Strasse 27, 8093 Z\"urich, Switzerland}

\emailAdd{adei@itp.phys.ethz.ch}
\emailAdd{sfondria@itp.phys.ethz.ch}

\abstract{%
We compute the tree-level bosonic S~matrix in light-cone gauge for superstrings on pure-NSNS $\AdSSS$. We show that it is proportional to the identity and that it takes the same form as for $\AdS{3}\times\S^3\times\T^4$ and for flat space. Based on this, we make a conjecture for the exact worldsheet S~matrix and derive the mirror thermodynamic Bethe ansatz (TBA) equations describing the spectrum. Despite a non-trivial vacuum~energy, they can be solved in closed form and coincide with a simple set of Bethe ansatz equations---again much like $\AdS{3}\times\S^3\times\T^4$ and flat space. This suggests that the model may have an integrable spin-chain interpretation. 
Finally, as a check of our proposal, we compute the spectrum from the worldsheet~CFT in the case of highest-weight representations of the underlying Ka\v{c}-Moody algebras, and show that the mirror-TBA prediction matches it on the~nose.
}

\begin{document}
\maketitle
\flushbottom

\section{Introduction}

A particularly interesting instance of the correspondence between strings on $d$-dimensional anti-de~Sitter spaces ($\AdS{d}$) and conformal field theories in $(d-1)$ dimensions ($\CFT_{d-1}$)~\cite{Maldacena:1997re, Gubser:1998bc, Witten:1998qj} is $\AdS{3}/\CFT_2$, see \textit{e.g.}\ ref.~\cite{David:2002wn} for a review. Restricting to the most supersymmetric string backgrounds, which preserve 16 Killing spinors, we encounter two classes of geometries: $\AdS{3}\times\S^3\times\cX$, with $\cX=\T^4$ or K3, and $\AdSSS$. In the former case, supersymmetry requires the radius of $\AdS{3}$ and that of $S^3$ to be equal, $\Rads=R_{\S}$. The supersymmetries arrange themselves into two copies of the $\psu(1,1|2)$ superalgebra, which is enhanced to the ``small'' $\mathcal{N}=(4,4)$ superconformal algebra in the dual $\CFT_2$~\cite{Giveon:1998ns}.
In the case of $\AdSSS$, the radius $\Rads$ of $\text{AdS}_3$ and the radii of the two three-spheres $R_1$ and $R_2$ must obey the curvature constraint
\begin{equation}
\dfrac{1}{\Rads^2} = \dfrac{1}{R^2_1} + \dfrac{1}{R^2_2} \,,
\end{equation} 
so that we have a one-parameter of backgrounds; the relative radii of the spheres are encoded in $\alpha$, $0<\alpha<1$,
\begin{equation}
\label{eq:alphadef}
\dfrac{\Rads^2}{R^2_1} = \alpha,  \qquad \dfrac{\Rads^2}{R^2_2} = 1 - \alpha\,. 
\end{equation}
The super-isometries of the background then close into two copies of the exceptional superalgebra $\dalg$, and the dual $\CFT_2$ enjoys ``large'' $\mathcal{N}=(4,4)$ superconformal symmetry~\cite{Elitzur:1998mm} first described in ref.~\cite{Sevrin:1988ew}.

All these backgrounds can be realised from a system of D1, D5, NS5 branes and fundamental strings, yielding a mixture of Ramond-Ramond (RR) and Neveu-Schwarz-Neveu-Schwarz (NSNS) fluxes---see ref.~\cite{OhlssonSax:2018hgc} for a detailed discussion of the moduli in the $\AdS{3}\times\S^3\times\T^4$ case. The case of NS5 branes and fundamental strings is particularly appealing in that the resulting string theory, in absence of other moduli~\cite{OhlssonSax:2018hgc}, can be studied relatively easily by CFT techniques \textit{on the string worldsheet}. In fact, it can be described by a stringy supersymmetric Wess-Zumino-Witten model involving $\sl(2)$ and $\su(2)$ Ka\v{c}-Moody algebras~\cite{Giveon:1998ns,Elitzur:1998mm}. This can be analysed in detail following the seminal work of Maldacena and Ooguri~\cite{Maldacena:2000hw,Maldacena:2000kv,Maldacena:2001km}, see also refs.~\cite{Pakman:2003cu,Israel:2003ry,Raju:2007uj,Giribet:2007wp, Ferreira:2017pgt}.
On the other hand, it is much harder to study the worldsheet CFT when RR fluxes are present~\cite{Berkovits:1999im}, see also ref.~\cite{Eberhardt:2018exh} and references therein for recent work in this direction. More generally, perturbative computations~\cite{Hoare:2013pma, Babichenko:2014yaa,Sax:2014mea, Hernandez:2014eta,Nieto:2018jzi} seem to suggest that non-protected observables take a substantially more involved form for backgrounds with at least some RR~flux.

It is possible to study $\AdS{3}$ string backgrounds without relying on conformal symmetry on the worldsheet. As it turns out, the Green-Schwarz action for geometries  supported by RR fluxes is \textit{classically integrable}~\cite{Babichenko:2009dk,Sundin:2012gc}, in close analogy with what happens for superstrings on $\AdS{5}\times\S^5$. Moreover, backgrounds supported by a mixture of RR and NSNS fluxes are classically integrable too~\cite{Cagnazzo:2012se}. This raised hopes of understanding the (planar) spectrum of strings following the strategy that proved extremely successful in $\AdS{5}/\CFT_4$ see \textit{e.g.} refs.~\cite{Arutyunov:2009ga,Beisert:2010jr} (see also ref.~\cite{Sfondrini:2014via} for a review of $\AdS{3}/\CFT_2$ integrability).
A key ingredient is the existence of an integrable worldsheet S~matrix for the light-cone gauge-fixed non-linear sigma model (NLSM). Symmetry arguments~\cite{Borsato:2012ud, Borsato:2013qpa,Hoare:2013ida, Borsato:2014exa, Borsato:2014hja, Lloyd:2014bsa, Borsato:2015mma} and explicit computations~\cite{Rughoonauth:2012qd, Sundin:2013ypa, Hoare:2013pma, Hoare:2013lja, Sundin:2014ema, Sundin:2015uva, Sundin:2016gqe} resulted in a proposal for such an S~matrix up to some overall scalar factors, the so-called dressing factors%
\footnote{%
Only in the case of the pure-RR $\AdS{3}\times\S^3\times\T^4$ background a proposal exists for the dressing factors too~\cite{Borsato:2013hoa,Borsato:2016xns,Borsato:2016kbm}, based on the requirement of crossing symmetry~\cite{Janik:2006dc}.
}%
, valid for almost any value of the background fluxes. Somewhat unexpectedly, the symmetry arguments fail precisely for pure-NSNS fluxes, \textit{i.e.}\ when the string theory can be described as a WZW model (see in particular ref.~\cite{Lloyd:2014bsa} for a discussion on this point). This is a little counterintuive, as we might expect the worldsheet description to be simplest precisely at this point. Moreover, perturbative computations take a qualitatively different form at this point~\cite{Hoare:2013pma}, as worldsheet excitations become \textit{chiral} even in light-cone gauge---as it could be expected from the pp-wave analysis of ref.~\cite{Berenstein:2002jq}.

Recently, an exact integrable S~matrix was proposed for the pure-NSNS $\AdS{3}\times\S^3\times\T^4$ theory~\cite{Baggio:2018gct, Dei:2018mfl} in uniform light-cone gauge~\cite{Arutyunov:2004yx, Arutyunov:2005hd, Arutyunov:2006gs}%
\footnote{See also ref.~\cite{Arutyunov:2009ga} for a pedagogical introduction to S-matrix integrability for strings in light-cone gauge (in $\AdS{5}\times\S^5$), and ref.~\cite{Sfondrini:2014via} for a review of some of the peculiarities of the $\AdS{3}$ set-up.}. It takes a remarkably simple form, being proportional to the identity up to a simple dressing factor.
For the two-particle S~matrix,
\begin{equation}
\label{eq:Smatschematically}
\mathbf{S}(p_1,p_2) = e^{i\,\Phi(p_1,p_2)}\,\mathbf{1}\,,
\end{equation}
which trivially satisfies the Yang-Baxter equation.
Moreover, the dressing phase~$\Phi(p_1,p_2)$ takes the Dray--'t~Hooft shockwave form~\cite{Dray:1984ha} and is identical to the one found in ref.~\cite{Dubovsky:2012wk} for strings in flat space. This suggests that the worldsheet theory in light-cone gauge is an integrable deformation (of the ``$T\bar{T}$'' type~\cite{Smirnov:2016lqw,Cavaglia:2016oda}) of a free theory. The resulting mirror Thermodynamic Bethe Ansatz (TBA) equations~\cite{Yang:1968rm,Zamolodchikov:1989cf} are so simple that they trivialize and yield back the ``asymptotic'' Bethe equations---showing that finite-size ``wrapping'' effects~\cite{Ambjorn:2005wa} cancel exactly here. Moreover, the equations can be solved in closed form~\cite{Baggio:2018gct} yielding the known stringy WZW result~\cite{Dei:2018mfl}, including the sectors corresponding to ``spectrally flowed'' Ka\v{c}-Moody representations~\cite{Maldacena:2000hw}. The fact that the mirror TBA equations trivialize strongly suggests that the underlying integrable model can be interpreted as a quantum-mechanical model---a spin~chain.
It is worth stressing that it is quite unusual that a set of Bethe equations---let alone TBA equations---can be solved in closed form, even for the simplest integrable models; one notable example which shares many similarities with our case arises in $\AdS{5}/\CFT_4$ when restricting to a $\su(1|1)$ sector~\cite{Alday:2005jm,Arutyunov:2005hd}.

It is a little surprising that the worldsheet S~matrix takes the  form~\eqref{eq:Smatschematically} with \textit{the same dressing phase}~$\Phi(p_1,p_2)$ both for $\AdS{3}\times\S^3\times\T^4$ and flat space. It is natural to wonder whether this universality holds for more general NSNS backgrounds. The main aim of this paper is to investigate this question for the pure-NSNS $\AdSSS$ background. Despite its relative simplicity, this background is qualitatively different from $\AdS{3}\times\S^3\times\T^4$: the Killing spinors close into a different superalgebra, light-cone gauge-fixing preserves at most one quarter (rather than one half) of the original supersymmetries, and the underlying integrable structure is believed to be related to an \textit{alternating}, rather then homogeneous, spin-chain~\cite{OhlssonSax:2011ms, Borsato:2012ud, Borsato:2012ss}. Additionally, it is worth noting that there are ways of fixing light-cone gauge that break \textit{all} manifest supersymmetry, which nonetheless need to be considered to reproduce the whole spectrum of the model, as it can be seen \textit{e.g.}\  from a pp-wave analysis~\cite{Dei:2018yth}.

In this paper we will compute the tree-level bosonic S~matrix of light-cone gauge-fixed strings on the pure-NSNS $\AdSSS$ background. This will end up depending on three parameters parameters: $k$, the level of the (supersymmetric) WZW model, which is the square-radius of $\AdS{3}$ in appropriate units, $\alpha$ of eq.~\eqref{eq:alphadef}, which encodes the relative curvature of the two spheres, and an additional parameter~$\vartheta$ which fixes our choice of (non-necessarily supersymmetric) uniform light-cone gauge, \textit{cf.} ref.~\cite{Dei:2018yth}. We introduce these quantities, as well as fix our notations for the background and string action, in section~\ref{sec:background}.
The computation of the tree-level S-matrix follows a standard route~\cite{Klose:2006zd,Arutyunov:2009ga}, and is presented in section~\ref{sec:smatrix}. We find that indeed the (perturbative) S~matrix is proportional to the identity---\textit{i.e.}, compatible with the expansion of~\eqref{eq:Smatschematically}. Moreover, the tree-level phase is the same as the one of flat space~\cite{Dubovsky:2012wk} and $\AdS{3}\times\S^3\times\T^4$~\cite{Baggio:2018gct}. As a result, we propose that the all-loop S~matrix is simply given by the exponentiation of the tree-level result, like in the previous cases. To substantiate our claim, in section~\ref{sec:TBA} we write down the equations describing the specturm. We find again that a simple set of asymptotic Bethe ansatz equations reproduce the WZW spectrum---pointing again to a spin-chain interpretation.
A novel subtlety arises when the choice of light-cone gauge does not preserve manifest supersymmetry. In that case, the energy spectrum is shifted by a (negative) ground-state energy, which in the spin~chain can be understood as a redefinition of a Fermi sea. We derive this energy shift by working out the Thermodynamic Bethe Ansatz (TBA) equations~\cite{Yang:1968rm,Zamolodchikov:1989cf} for the mirror model~\cite{Arutyunov:2007tc}. We find that indeed the  mirror TBA is not entirely trivial when no supersymmetry is manifestly preserved, but all finite-size corrections boil down to an overall shift of \textit{all} energy levels, indeed compatibly with a redefinition of the Fermi sea.
We comment on our results in section~\ref{sec:conclusions} and relegate some technical details to five appendices.

\section{String theory on \texorpdfstring{$\AdSSS$}{AdS3xS3xS3xS1}}
\label{sec:background}
We start by briefly recalling some facts about the $\AdSSS$ background supported by pure-NSNS three-form flux.

\subsection{The supergravity background}

Let us write the metric of $\text{AdS}_3 \times \text{S}^3 \times \text{S}^3 \times \text{S}^1$ as 
\begin{equation}
ds^2 = \Rads^2 \, G_{\mu \nu} \, dX^\mu dX^\nu = \Rads^2 \left( \ ds^2_{\text{AdS}_3} + \frac{ds^2_{\text{S}^3_1}}{\alpha} +  \frac{ds^2_{\text{S}^3_2}}{1- \alpha} \  + ds^2_{\text{S}^1} \right) \,. 
\end{equation}
Following the notation of \cite{Borsato:2015mma}, we write the metric of $\text{AdS}_3$ and $\text{S}^3$ as 
\begin{equation}
\begin{aligned}
ds^2_{\text{AdS}_3} & = - \biggl( \dfrac{1 + \frac{z_1^2 + z_2^2}{4}}{1 - \frac{z_1^2 + z_2^2}{4}} \bigg)^2 dt^2 + \biggl( \dfrac{1}{1 - \frac{z_1^2 + z_2^2}{4}} \bigg)^2 (dz_1^2 + dz_2^2) \,,  \\ 
ds^2_{\text{S}^3_1} & = \biggl( \dfrac{1 - \frac{y_3^2 + y_4^2}{4}}{1 + \frac{y_3^2 + y_4^2}{4}} \bigg)^2 d\phi_5^2 + \biggl( \dfrac{1}{1 + \frac{y_3^2 + y_4^2}{4}} \bigg)^2 (dy_3^2 + dy_4^2) \,, \\
ds^2_{\text{S}^3_2} & = \biggl( \dfrac{1 - \frac{x_6^2 + x_7^2}{4}}{1 + \frac{x_6^2 + x_7^2}{4}} \bigg)^2 d\phi_8^2 + \biggl( \dfrac{1}{1 + \frac{x_6^2 + x_7^2}{4}} \bigg)^2 (dx_6^2 + dx_7^2) \,, \\ 
ds^2_{\text{S}^1} & = dw^2 \,, 
\end{aligned}
\end{equation}
where $t, \phi_5$ and $\phi_8$ are isometric coordinates. We consider a background supported by pure NSNS three-form flux: 
\begin{equation}
dB = \Rads^2 \, db = 2 \Rads^2 \biggl( \text{Vol}(\text{AdS}_3) + \dfrac{ \text{Vol}(\text{S}^3_1)}{\alpha} + \dfrac{\text{Vol}(\text{S}^3_2)}{1- \alpha}  \biggr) \,,  
\end{equation}
where we have introduced the rescaled Kalb-Ramond field $b_{\mu \nu} = B_{\mu \nu}/ \Rads^2$. This background is maximally supersymmetric: the 16 Killing spinors close on two copies of the exceptional super Lie algebra $\mathfrak{d}(2,1; \alpha)$ \cite{Sevrin:1988ew}. We will refer to these two copies as \emph{left} and \emph{right} and denote them as $\mathfrak{d}(2,1; \alpha)_L$ and $\mathfrak{d}(2,1; \alpha)_R$ respectively. Each copy of $\mathfrak{d}(2,1; \alpha)$ contains $\mathfrak{su}(1,1) \oplus \mathfrak{su}(2)_{(1)} \oplus \mathfrak{su}(2)_{(2)}$ as a bosonic subalgebra, so that $\mathfrak{so}(2,2) \oplus \mathfrak{so}(4)_{(1)} \oplus \mathfrak{so}(4)_{(2)}$ is the global isometry algebra. In the following we will denote with $\ell$ the lowest weight of $\mathfrak{su}(1,1)_L$ and with $j_1$ and $j_2$ the spins of $\mathfrak{su}(2)_{(1)L} \oplus \mathfrak{su}(2)_{(2)L}$. Similarly $(\tilde \ell, \tilde \jmath_1, \tilde \jmath_2)$ will label the corresponding quantum numbers of $\mathfrak{d}(2,1; \alpha)_R$. 

\subsection{String action}

The bosonic string action is
\begin{equation}
S = -\dfrac{k}{4 \pi} \int d^2 \sigma \left( \gamma^{\alpha \beta} G_{\mu \nu} \partial_\alpha X^\mu \partial_\beta X^\nu + \epsilon^{\alpha \beta} b_{\mu \nu}\partial_\alpha X^\mu \partial_\beta X^\nu \right), 
\label{eq:polyakov-action}
\end{equation}
where the determinant of the worldsheet metric $\gamma^{\alpha \beta}$ has been set to $-1$ and the $1/ \Rads^2$ dependence of the string tension canceled against the $\Rads^2$ dependence of metric and Kalb-Ramond field. Working in first order formalism, define conjugate momenta 
\begin{equation} \label{eq:conjugate-momenta}
p_\mu = \dfrac{\delta S}{\delta \dot X^\mu} = - \dfrac{k}{2 \pi} \left( \gamma^{0 \beta} G_{\mu \nu} \partial_\beta X^\nu + b_{\mu \nu} \pri{X}^\nu \right) 
\end{equation}
and rewrite \eqref{eq:polyakov-action} as \cite{Lloyd:2014bsa, Borsato:2015mma}
\begin{equation}
S = \int d \tau d \sigma \left( p_\mu \dot X^\mu + \dfrac{\gamma^{10}}{\gamma^{00}}C_1 + \dfrac{\pi}{k \gamma^{00}}C_2 \right) 
\end{equation}
where 
\begin{equation}
\begin{gathered}
    C_1 = p_\mu \pri{X}^\mu \ , \\[4pt]
    C_2 = G^{\mu \nu} p_\mu p_\nu + \dfrac{k^2}{4 \pi^2}G_{\mu \nu} \pri{X}^\mu \pri{X}^\nu + 2 \dfrac{k}{2 \pi} G^{\mu \nu} B_{\nu \rho} p_\mu \pri{X}^\rho + \dfrac{k^2}{4 \pi^2} G^{\mu \nu} b_{\mu \rho} b_{\nu \sigma} \pri{X}^\rho \pri{X}^\sigma \  .
\end{gathered}
\label{eq:virasoro-constraints}
\end{equation}
Above we have denoted with dot and prime, respectively, derivatives with respect to $\tau$ and $\sigma$. The Virasoro constraints amount to the equations of motion for the auxiliary field $\gamma^{\alpha \beta}$. For the bosonic action they are equivalent to setting $C_1 = 0$ and $C_2 = 0$. The Virasoro constraints impose~\cite{Borsato:2015mma}
\begin{equation}
\begin{aligned}
\gamma^{11} G_{\mu \nu} \dot{X}^\mu \pri{X}^\nu + \gamma^{01} G_{\mu \nu} \dot{X}^\mu \dot{X}^\nu &= 0 \,, \\
\gamma^{00} G_{\mu \nu} \dot{X}^\mu \dot{X}^\nu -  \gamma^{11} G_{\mu \nu} \pri{X}^\mu \pri{X}^\nu &= 0\,. 
\end{aligned}
\label{eq:virasoro-constraints-bis}
\end{equation}

\subsection{Gauge fixing and decompactification limit}
\label{sec:gaugefix}
In order to streamline the notation let us rewrite equation~\eqref{eq:alphadef} in terms of an angle $\varphi$
\begin{equation}
\dfrac{\Rads^2}{R^2_1} = \alpha \equiv \cos^2 \varphi,  \qquad \dfrac{\Rads^2}{R^2_2} = 1 - \alpha \equiv \sin^2 \varphi\,. 
\end{equation}
Geodesics in $\text{AdS}_3 \times \text{S}^3 \times \text{S}^3 \times \text{S}^1$ can be parametrized by a constant motion along the isometric coordinates, 
\begin{equation}
(t, \phi_5, \phi_8) = 2 \pi \alpha' \left( \frac{\ell + \tilde \ell}{\Rads^2}, \frac{j_1 + \tilde \jmath_1}{R_1^2}, \frac{j_2 + \tilde \jmath_2}{R_2^2} \right) \ \tau \,, 
\label{eq:geodesics}
\end{equation}
where $\tau$ is the affine parameter. Requiring the geodesic to be null amounts to the Pythagorean constraint
\begin{equation}
(\ell + \tilde \ell )^2  = \cos^2 \varphi \ (j_1 + \tilde \jmath_1)^2 + \sin^2 \varphi \ (j_2 + \tilde \jmath_2)^2 \,, 
\end{equation}
that can be solved in terms of the angle $\vartheta$ as
\begin{equation}
\label{eq:thetadef}
(j_1 + \tilde \jmath_1) \cos \varphi = (\ell + \tilde \ell) \cos \vartheta  \,, \qquad (j_2 + \tilde \jmath_2) \sin \varphi = (\ell + \tilde \ell) \sin \vartheta \,.  
\end{equation}
For $\vartheta = \varphi$, the geodesic is $\frac{1}{4}$-BPS and~\cite{Babichenko:2009dk,Baggio:2017kza,Eberhardt:2017fsi} 
\begin{equation}
\ell + \tilde \ell = j_1 + \tilde \jmath_1 = j_2 + \tilde \jmath_2 \,. 
\end{equation} 
We introduce light-cone coordinates $x^\pm$ and a transverse angle $\psi$ adapted to the geodesic in~\eqref{eq:geodesics},  
\begin{equation}
\begin{aligned}
x^+ &= (1-a) \ t + a \frac{\cos \vartheta}{\cos \varphi} \ \phi_5 + a \frac{\sin \vartheta}{\sin \varphi} \ \phi_8 \,,  \\
x^- &= \frac{1}{2} \left(  -t + \frac{\cos \vartheta}{\cos \varphi} \ \phi_5 +  \frac{\sin \vartheta}{\sin \varphi} \ \phi_8 \right) \,, \\
\psi &= - \frac{\sin \vartheta}{\cos \varphi} \ \phi_5 + \frac{\cos \vartheta}{\sin \varphi} \ \phi_8 \,,
\end{aligned}
\label{eq:lightcone-coord}
\end{equation}
where $0 \leq a \leq 1$ is a
parameter which will be useful to fix a rather general uniform light-cone gauge~\cite{Arutyunov:2004yx, Arutyunov:2005hd, Arutyunov:2006gs}
\begin{equation}
x^+ = \tau \,, \qquad p_- = \dfrac{\delta S}{\delta \dot x^-} = 2 \,, \qquad \dfrac{\delta S}{\delta \pri x^-} = 0 \,. 
\label{eq:lightcone-gauge}
\end{equation}
This completely fixes the dynamics of the lightcone coordinates $x^\pm$ and allows to study the theory in terms of eight transverse worldsheet bosons. The worldsheet Hamiltonian density is related to the lightcone momentum,  $\mathcal{H} = - p_+$ with $p_+ = \delta S / \delta \dot x^+ $\cite{Arutyunov:2009ga},  so that we find 
\begin{equation}
H = \int d^2 \sigma \ \mathcal H = \ell + \tilde \ell - \mathcal J \,, 
\end{equation}
where we have defined 
\begin{equation}
\label{eq:defJ}
\mathcal J = \cos \varphi \cos \vartheta \ (j_1 + \tilde \jmath_1) + \sin \varphi \sin \vartheta \ (j_2 + \tilde \jmath_2) \,.  
\end{equation} 
The second gauge-fixing condition in eq.~\eqref{eq:lightcone-gauge} relates the lightcone momentum charge $P_-$ to the length of the worldsheet~$R$
\begin{equation}
P_- = \int_0^R d \sigma \ p_- = 2 R \,. 
\end{equation}
In the following we will mainly work in the so-called decompactification limit where ${P_- \to \infty}$ and the worldsheet effectively decompactifies into a plane. The change of coordinates in eq.~\eqref{eq:lightcone-coord} implies 
\begin{equation}
\label{eq:Rdef}
R = \mathcal J + a H \,.
\end{equation} 
Gauge dependence (namely, $a$-dependence) of the worldsheet length is a well known feature of uniform light-cone gauge, which gives a direct link to $T\bar{T}$ deformations see refs.~\cite{Baggio:2018gct,Baggio:2018rpv} for a discussion of this point). We will discuss in section~\ref{sec:ABA} the gauge \textit{invariance} of the spectrum, see also ref.~\cite{Staudacher:2004tk}.

\subsection{Near-BMN expansion}
\label{sec:near-BMN-expansion}

On top of the decompactification limit described in the previous section, one can perform perturbative computations in the $k \to \infty$ limit~\cite{Klose:2006zd}. This is nicely explained in \cite{Arutyunov:2009ga} and here we briefly recall the procedure. As the effective string tension $\frac{k}{2 \pi}$ always appears together with the $\sigma$ derivative of a field we rescale $\sigma$ as $\sigma \to \frac{k}{2 \pi} \sigma$, so that the gauge-fixed action can be written as 
\begin{equation}
S = \dfrac{k}{2 \pi} \int d^2 \sigma \, \mathcal{L} \,,
\end{equation}
where the integrand $\mathcal{L}$ does no longer depend on $k$. Rescaling the transverse fields by $\sqrt{2 \pi / k}$ the gauge-fixed action becomes 
\begin{equation}
S = \int d^2 \sigma \left( \mathcal{L}_2 + \sqrt{\dfrac{2 \pi}{k}} \mathcal{L}_3 + \dfrac{2 \pi}{k} \mathcal{L}_4 + \cdots\right) \,, 
\end{equation}
where the subscripts indicate the number of fields in each term. This amounts to a near-BMN expansion~\cite{Berenstein:2002jq}, where the strict BMN limit corresponds to the quadratic part of the action and higher order terms provide corrections to it. In the following we will frequently omit factors of $\tfrac{2 \pi}{k}$, which can be reinstated by counting the interaction order. Moreover, in order to lighten the notation, we will adopt the conventions of \cite{Borsato:2015mma}, where the coordinates of the three-spheres have been rescaled by the radius of the respective sphere and $\Rads = 1$. In the near-BMN limit, one can solve for the worldsheet metric eq.~\eqref{eq:lightcone-gauge} together with the condition ${\text{det} \gamma = -1}$.  Notice that the worldsheet metric is not constant and is in general a function of the transverse fields. The reader can find the explicit result up to cubic order in appendix~\ref{app:details-BMN-expansion}.  The Virasoro constraints \eqref{eq:virasoro-constraints-bis} give the derivatives with respect to $\tau$ and $\sigma$ of the lightcone coordinate $x^-$. Again we relegate to appendix~\ref{app:details-BMN-expansion} the explicit result. 

Once lightcone gauge has been fixed, one can expand~\eqref{eq:polyakov-action} in the number of transverse fields and compute the bosonic Lagrangian perturbatively. We will express the result in terms of the four complex bosons
\begin{equation} \label{eq:complex-bosons}
\begin{aligned}
Z = & -z_2 + i z_1 \,, \quad Y =-y_3 -i y_4 \,, \quad X =  -x_6 - i x_7 \,, \quad W = w - i \psi \,, \\
\bar Z = & -z_2 - i z_1 \,, \quad \bar Y =-y_3 +i y_4 \,, \quad \bar X = -x_6 + i x_7 \,, \quad \bar W = w + i \psi \,.
\end{aligned}
\end{equation}
In terms of $\AdSSS$ coordinates introduced above, we see that $Z,\bar{Z}$ are related to the transverse directions of $\AdS{3}$, $Y,\bar{Y}$ and $X,\bar{X}$ to the transverse modes on the two spheres, and $W,\bar{W}$ are combinations of the  flat coordinate $w$ along $S^1$ and the transverse angle $\psi$. The complex bosons in eq.~\eqref{eq:complex-bosons} have a well-defined angular momentum on $\text{AdS}_3$ and on the two spheres, see table~\ref{tab:S-matrix-basis}. It is convenient to introduce the quantum number $\mu$ defined as 
\begin{equation}
\mu  = (\ell-\tilde{\ell})-\cos\varphi\cos\vartheta(j_1-\tilde{\jmath}_1) -\sin\varphi\sin\vartheta(j_2-\tilde{\jmath}_2) \,.
\end{equation}
In fact, $\mu$ is a central element of the $\mathfrak{su}(1|1)_{\text{c.e.}}$ algebra which is preserved after gauge fixing and it identifies (short) irreducible representations of that algebra \cite{Borsato:2012ud}. In computing the Lagrangian we discard total derivative terms proportional to $\dot x^-$. Then at quadratic order we find: 
\begin{equation}
\begin{aligned}
\mathcal{L}_2 = & \ \dfrac{1}{2} \left( \dot W \dot{\bar W} - \pri W \pri{\bar W} \right) + \dfrac{1}{2} \left( \dot Z \dot{\bar Z} - \pri Z \pri{\bar Z} - i \pri Z \bar Z + i Z \pri{\bar Z} - Z \bar Z \right) \\
& + \dfrac{1}{2} \left(\dot X \dot{\bar X} - \pri X \pri{\bar X} - i \sin \varphi \sin \vartheta \ \pri X \bar X + i \sin \varphi \sin \vartheta \ X \pri{\bar X} - \sin^2 \varphi \sin^2 \vartheta \ X \bar X \right) \\
& + \dfrac{1}{2} \left(\dot Y \dot{\bar Y} - \pri Y \pri{\bar Y} - i \cos \varphi \cos \vartheta \ \pri Y \bar Y + i \cos \varphi \cos \vartheta \ Y \pri{\bar Y} - \cos^2 \varphi \cos^2 \vartheta \ Y \bar Y \right) \,.  
\end{aligned}
\label{eq:L2}
\end{equation}
Differently from the $\text{AdS}_3 \times \text{S}^3 \times \text{T}^4$ case, we find a non-vanishing cubic term 
\begin{equation}
\begin{aligned}
\mathcal{L}_3 = & \  \dfrac{1}{4} \cos \vartheta \sin \varphi \ (\pri W - \pri{\bar W})(X \dot{\bar X} - \dot X \bar X) \\
& \ + \dfrac{1}{4} \cos \vartheta \sin \varphi \ (\dot{\bar W} - \dot W)(X \pri{\bar X} - \bar X \pri X + 2i \sin \varphi \sin \vartheta \ X \bar X)   \\
 & \ - \dfrac{1}{4} \cos \varphi \sin \vartheta \ (\pri W - \pri{\bar W})(Y \dot{\bar Y} - \dot Y \bar Y ) \\
 & \ - \dfrac{1}{4} \cos \varphi \sin \vartheta \ (\dot{\bar W} - \dot W)(Y \pri{\bar Y} - \bar Y \pri Y + 2i \cos \varphi \cos \vartheta \ Y \bar Y )    \,. 
\end{aligned}
\label{eq:L3}
\end{equation}
The quartic Lagrangian is a sum of terms each containing at most two different fields:
\begin{equation}
\mathcal{L}_4 = \mathcal{L}_4^{\scriptscriptstyle{WW}} + \mathcal{L}_4^{\scriptscriptstyle{XX}} + \mathcal{L}_4^{\scriptscriptstyle{YY}}+ \mathcal{L}_4^{\scriptscriptstyle{ZZ}} + \mathcal{L}_4^{\scriptscriptstyle{WX}} + \mathcal{L}_4^{\scriptscriptstyle{WY}} + \mathcal{L}_4^{\scriptscriptstyle{WZ}} + \mathcal{L}_4^{\scriptscriptstyle{XY}} + \mathcal{L}_4^{\scriptscriptstyle{XZ}} + \mathcal{L}_4^{\scriptscriptstyle{YZ}} \,, 
\label{eq:L4}
\end{equation}
where the superscripts indicate the different fields contained in each term and $\mathcal{L}_4^{\scriptscriptstyle{WW}}, \mathcal{L}_4^{\scriptscriptstyle{XX}}, \dots$ etc.~are explicitly defined in appendix~\ref{app:quartic-L}.

\section{The perturbative S~matrix}
\label{sec:smatrix}

\begin{table}[t]
\centering
\begin{tabular}{|c|c|c|c|c|c|}
\hline
Field$\phantom{\Big|}$   & $\ell - \tilde \ell$ & $\ (j_1 - \tilde \jmath_1)$ & $\ (j_2 - \tilde \jmath_2)$ & $\mu$ & $\mathcal J$  \\
\hline
$W$  &  &  &  &  &    \\
$\bar W$  & & & &  &    \\
\hline
$X$ &  &  & $+1$ & $+s$ &  $-s$ \\
$\bar X$  &  &  & $-1$ & $-s$ &  $+s$ \\
\hline
$Y$ &  & $+1$ &  &  $+c$ & $-c$  \\
$\bar Y$  &  & $-1$ &  & $-c$ & $+c$  \\
\hline
$Z$  & $+1$ &  &  & $+1$ &  \\
$\bar Z$  & $-1$ &  &  & $-1$ &   \\
\hline
\hline
$\Psi_{\pm \pm \pm}\phantom{\Bigg|}$ & $\pm \dfrac{1}{2}$ & $\pm \dfrac{1}{2}$ & $\pm \dfrac{1}{2}$ & $\dfrac{\pm 1 \pm c \pm s}{2}$ & $\mp \dfrac{c}{2} \mp \dfrac{s}{2} $ \\
\hline
\end{tabular}
\caption{We list the spins of the bosonic excitations $W,X,Y,Z$ and of their conjugate, as well as of the eight physical fermions of the Green-Schwarz string, see ref.~\cite{Dei:2018yth}. Spins are quantised in integers and half-integers for bosons and fermions, respectively.
The combination $\mu$, which plays the role of the excitations' mass in the dispersion relation~\eqref{eq:omega} is an appropriately rescaled combination of the spins, which we write in terms of
 $s = \sin \varphi \sin \vartheta$ and $c =\cos \varphi \cos \vartheta$.
The total angular momentum of $\S^3\times\S^3$ is $\mathcal{J}$, see eq.~\eqref{eq:defJ}; its sign depends on the worldsheet chirality of the particles, as we discuss below. Here we list the values for positive-chirality particles.
}
\label{tab:S-matrix-basis}
\end{table}

The 2-to-2 perturbative S~matrix can be computed using standard QFT techniques, see \textit{e.g.}~\cite{Arutyunov:2009ga} for a review. From eqs.~(\ref{eq:L2}--\ref{eq:L4}) one can derive Feynman rules, that we list in appendix~\ref{app:feynman-rules}.  The  2-to-2 perturbative S~matrix can then be defined in terms of scattering amplitudes. We take two particles worldsheet momentum $p_1, p_2$ as incoming and obtain particles of momenta $p_3, p_4$ as outgoing\footnote{Both the worldsheet momenta and the conjugate momenta, cf eq.~\eqref{eq:conjugate-momenta}, are denoted by $p_i$. We hope that this will not lead to any ambiguity as the context should clarify which quantity is being discussed.}.  Let us define a basis for the S~matrix 
\begin{equation}
\mathbf{S} = \mathbf{1} + i \mathbf{T} \,, 
\end{equation}
\begin{equation}
i\, \mathbf{T} \ket{p_1, \mu_1 ; p_2, \mu_2} = i\, {T}_{12}^{34} \ket{p_3, \mu_3; p_4, \mu_4 } \,,
\end{equation}
where the set of quantum numbers $p_i$ and $\mu_i$ uniquely identifies a particle, according to Table \ref{tab:S-matrix-basis}. The perturbative S-matrix can be computed in terms of scattering amplitudes, as
\begin{equation}
\begin{aligned}
i\,{T}_{12}^{34} = &  \braket{p_3, \mu_3; p_4, \mu_4|i \mathbf{T} |p_1, \mu_1; p_2, \mu_2} \\
= &  \int dp_3 dp_4 \dfrac{\delta(p_1 + p_2 - p_3 - p_4) \delta(\omega_1  + \omega_2 - \omega_3 - \omega_4)}{\sqrt{\omega_1 \omega_2 \omega_3 \omega_4}} i \mathcal{M}_{12 \to 34} \,,  
\end{aligned}
\label{eq:full-S-expression}
\end{equation}
where $i \mathcal{M}_{12 \to 34}$ is the scattering amplitude and the tree-level dispersion relation $\omega(p)$ can be read off the quadratic Lagrangian~\eqref{eq:L2},
\begin{equation}
\label{eq:omega}
\omega_i = \omega(p_i) = |p_i + \mu_i| \,. 
\end{equation} 
Notice that this dispersion is \textit{chiral}, \textit{i.e.}\ it distinguishes from left- and right-moving particles on the worldsheet. Given that the dispersion reproduces the energy $\ell+\tilde{\ell}-\mathcal{J}$ of a single excitation, we see that the (quantised) contribution of $\mathcal{J}$, which can be read off at $p=0$, comes with a sign depending on the particle's chirality.
We are now going to illustrate some examples to clarify the main steps of the computation.

\subsection{Example: a (vanishing) off-diagonal term}

As a first example, we are going to present the computation of the off-diagonal S-matrix element
\begin{equation}
\braket{\bar{X}(p_3) \bar{W}(p_4) |\,i\, \mathbf{T}\, |\bar{X}(p_1) W(p_2) } \,, 
\label{eq:off-diagonal-S}
\end{equation}
representing a ``{left}'' $W$ mode and a ``{right}'' $\bar{X}$ mode scattering into a ``{right}'' $\bar{W}$ mode and a ``{right}'' $X$ mode. Notice that here ``{left}'' and ``{right}'' refer to target space chirality, not to chirality \emph{on the worldsheet}. That is to say, here {left} and {right} are not synonyms of {chiral} and {anti-chiral}, but label the two creation/annihilation operators entering the mode expansion of a complex boson.
Let us emphasise that this process should vanish by conservation of the $\mathfrak{u}(1)$ charge relative to particle~$W$. It is still non-trivial to see how this comes about, since the charge of $W$-particles does not affect the dispersion relation. The amplitude associated to the process in \eqref{eq:off-diagonal-S} is
\begin{equation}
\begin{fmffile}{off-diagonal-amplitude}
\begin{gathered}
\begin{fmfgraph*}(75,75)
  \fmfleft{i1,i2}
  \fmfright{o1,o2}
  \fmfblob{.5w}{v1}
  \fmf{fermion,foreground=(1,,0.1,,0.1)}{v1,i2}
  \fmf{scalar}{i1,v1}
  \fmf{scalar}{o1,v1}
  \fmf{fermion,foreground=(1,,0.1,,0.1)}{o2,v1}
  	\Marrow{a}{left}{lft}{$p_1$}{i2,v1}{7}    
  	\Marrow{b}{left}{lft}{$p_2$}{i1,v1}{8}    
  	\Marrow{c}{right}{rt}{$ \ p_3$}{v1,o2}{8}   
  	\Marrow{d}{right}{rt}{$\ p_4$}{v1,o1}{8}    
	        \end{fmfgraph*}
\end{gathered} = i \mathcal{M}_{W \bar{X} \to \bar{W} \bar{X}} = \sum_{j=1}^5 i \mathcal{M}_j
\end{fmffile}
\label{eq:off-diagonal-amplitude}
\end{equation}
where time flows from left to right and the Feynman diagrams contributing to the amplitude \eqref{eq:off-diagonal-amplitude} can be  represented by the following diagrams

\begin{fmffile}{off-diagonal-scattering}
\begin{equation}
\begin{aligned}
i \mathcal{M}_1 = & \quad 
\begin{gathered}
\begin{fmfgraph*}(100,60)
  \fmfleft{i1,i2}
  \fmfright{o1,o2}
  \fmf{fermion,foreground=(1,,0.1,,0.1)}{v1,i2}
  \fmf{scalar}{i1,v1}
  \fmf{scalar}{o1,v2}
  \fmf{fermion,foreground=(1,,0.1,,0.1)}{o2,v2}
  \fmf{fermion,foreground=(1,,0.1,,0.1)}{v2,v1}
  	\Marrow{a}{left}{lft}{$p_1$}{i2,v1}{7}    
  	\Marrow{b}{left}{lft}{$p_2$}{i1,v1}{8}    
  	\Marrow{c}{right}{rt}{$ \ p_3$}{v2,o2}{8}   
  	\Marrow{d}{right}{rt}{$\ p_4$}{v2,o1}{8}  
  	\Marrow{e}{up}{top}{$p$}{v2,v1}{8}    
	        \end{fmfgraph*}
\end{gathered} \quad \,, & \quad \quad i \mathcal{M}_2 = & \quad \begin{gathered}
\begin{fmfgraph*}(100,60)
  \fmfleft{i1,i2}
  \fmfright{o1,o2}
  \fmf{fermion,foreground=(1,,0.1,,0.1)}{v1,v2}
  \fmf{scalar}{i1,v1}
  \fmf{scalar}{o1,v2}
   \fmf{phantom}{i2,v1}
  \fmf{phantom}{o2,v2}
  \fmf{fermion,tension=0,foreground=(1,,0.1,,0.1)}{o2,v1}
  \fmf{fermion,tension=0,foreground=(1,,0.1,,0.1)}{v2,i2}
  	\Marrow{a}{left}{bot}{$p_1 \ \ \ $}{i2,v2}{20} 
  	\Marrow{b}{left}{lft}{$p_2$}{i1,v1}{8} 
  	\Marrow{c}{right}{bot}{$ \ \ \  p_3$}{v1,o2}{20}   
  	\Marrow{d}{right}{rt}{$\ p_4$}{v2,o1}{8}   
  	\Marrow{e}{down}{bot}{$p$}{v1,v2}{8}     
  	        \end{fmfgraph*} 
\end{gathered} \quad \,, \\
& & & \\
i \mathcal{M}_3 = & \quad \begin{gathered}
\begin{fmfgraph*}(100,60)
  \fmfleft{i1,i2}
  \fmfright{o1,o2}
  \fmf{fermion,foreground=(1,,0.1,,0.1)}{v2,v1}
  \fmf{phantom}{i1,v1}
  \fmf{phantom}{o1,v2}
  \fmf{scalar,tension=0}{i1,v2}
  \fmf{scalar,tension=0}{o1,v1}
  \fmf{fermion,foreground=(1,,0.1,,0.1)}{v1,i2}
  \fmf{fermion,foreground=(1,,0.1,,0.1)}{o2,v2}   
  	\Marrow{a}{left}{lft}{$p_1$}{i2,v1}{7}
  	\Marrow{b}{left}{top}{$p_2 \ \ \ $}{i1,v2}{20}    
  	\Marrow{c}{right}{rt}{$ \ p_3$}{v2,o2}{8}  
  	\Marrow{d}{right}{top}{$\ \ \ p_4$}{v1,o1}{20}     
  	\Marrow{e}{up}{top}{$p$}{v2,v1}{8}    
  	        \end{fmfgraph*} 
\end{gathered} \quad \,, & \quad \quad i \mathcal{M}_4 = & \quad \begin{gathered}
\begin{fmfgraph*}(100,60)
  \fmfleft{i1,i2}
  \fmfright{o1,o2}
  \fmf{fermion,foreground=(1,,0.1,,0.1)}{v1,v2}
   \fmf{phantom}{i2,v1}
  \fmf{phantom}{o2,v2}
  \fmf{phantom}{i1,v1}
  \fmf{phantom}{o1,v2}
  \fmf{fermion,tension=0,foreground=(1,,0.1,,0.1)}{o2,v1}
  \fmf{fermion,tension=0,foreground=(1,,0.1,,0.1)}{v2,i2}
  \fmf{scalar,tension=0}{i1,v2}
  \fmf{scalar,tension=0}{o1,v1}
  	\Marrow{a}{left}{bot}{$p_1 \ \ \ $}{i2,v2}{20} 
  	\Marrow{b}{left}{top}{$p_2 \ \ \ $}{i1,v2}{20}    
  	\Marrow{c}{right}{bot}{$ \ \ \  p_3$}{v1,o2}{20}  
  	\Marrow{d}{right}{top}{$\ \ \ p_4$}{v1,o1}{20}  
  	 	        \end{fmfgraph*} 
\end{gathered} \quad \,, \\
& & & \\
i \mathcal{M}_5 = & \quad 
\begin{gathered}
\begin{fmfgraph*}(100,60)
  \fmfleft{i1,i2}
  \fmfright{o1,o2}
  \fmf{fermion,foreground=(1,,0.1,,0.1)}{v1,i2}
  \fmf{scalar}{i1,v1}
  \fmf{scalar}{o1,v1}
  \fmf{fermion,foreground=(1,,0.1,,0.1)}{o2,v1}
  	\Marrow{a}{left}{bot}{$p_1 \ \ \ $}{i2,v1}{15}    
  	\Marrow{b}{left}{top}{$p_2 \ \ \ $}{i1,v1}{15}    
  	\Marrow{c}{right}{bot}{$ \ \ \ p_3$}{v1,o2}{15}   
  	\Marrow{d}{right}{top}{$\ \ \  p_4$}{v1,o1}{15}  
  		        \end{fmfgraph*}
\end{gathered} \quad \,.  & &
\end{aligned}
\end{equation}
\end{fmffile}
Exploiting the Feynman rules in appendix~\ref{app:feynman-rules} we find 
\begin{equation}
\begin{aligned}
i \mathcal{M}_1 = & \ i \mathcal{M}_4 = - \dfrac{i \sin^2 \varphi \cos^2 \vartheta}{16} \dfrac{[p_2(\omega - \omega_1) - \omega_2 (p - p_1 + 2 \sin \varphi \sin \vartheta)]}{\omega^2 - (p + \sin \varphi \sin \vartheta)^2 + i \varepsilon} \times \\
 & \qquad \qquad \qquad \times [- p_4(\omega - \omega_3) + \omega_4(p - p_3 + 2 \sin \varphi \sin \vartheta)]\,, \\
i \mathcal{M}_2 = & \ i \mathcal{M}_3 = - \dfrac{i \sin^2 \varphi \cos^2 \vartheta}{16} \dfrac{[p_2(\omega - \omega_3) - \omega_2 (p - p_3 + 2 \sin^2 \varphi)]}{\omega^2 - (p + \sin \varphi \sin \vartheta)^2 + i \varepsilon} \times \\
& \qquad \qquad \qquad \times [- p_4(\omega - \omega_1) + \omega_4(p - p_1 + 2\sin^2 \varphi)] \,, \\
i \mathcal{M}_5 = & - \dfrac{i}{4} \sin^2 \varphi \cos^2 \vartheta \  (p_2 p_4 - \omega_2 \omega_4) \,,
\end{aligned}
\label{eq:off-diagonal-diagrams}
\end{equation}
where the dispersion relation for $p_1 , \dots, p_4$ is  
\begin{equation}
\omega_1 = |p_1 - \sin \varphi \sin \vartheta| \,,  \qquad \omega_2 = |p_2| \,,  \qquad \omega_3 = |p_3 - \sin \varphi \sin \vartheta| \,,  \qquad \omega_4 = |p_4| \,. 
\label{eq:off-diagonal-dispersions}
\end{equation}
The energy $\omega$ and the momentum $p$ of the virtual particle can be found for each diagram by imposing energy and momentum conservation at each vertex. So, for example, for $i\mathcal{M}_2$ we have
\begin{equation}
\omega = \omega_2 - \omega_3 = \omega_4 - \omega_1 \,, \qquad p = p_2 -p_3 = p_4 - p_1  \,. 
\label{eq:off-diagonal-w-and-p}
\end{equation} 
In order to further simplify the expressions in \eqref{eq:off-diagonal-diagrams}, one needs to resolve the absolute values in \eqref{eq:off-diagonal-dispersions}. In order for a scattering event to take place, the two incoming particles (and similarly the two outgoing ones) must have opposite chirality. In particular we will assume $p_1$ to be anti-chiral and $p_2$ to be chiral. We hence have the following possibilities: 
\begin{enumerate}[label=(\roman*)]
\item $\qquad \omega_1 = -p_1 + \sin \varphi \sin \vartheta \,,  \qquad \omega_2 =  p_2 \,,  \qquad \omega_3 = -p_3 + \sin \varphi \sin \vartheta \,, \qquad \omega_4 = p_4 \,,  $
\item $\qquad \omega_1 = -p_1 + \sin \varphi \sin \vartheta \,,  \qquad \omega_2 =  p_2 \,,  \qquad \omega_3 = p_3 - \sin \varphi \sin \vartheta \,,  \qquad \omega_4 = -p_4 \,, $
\end{enumerate} 
For (i) the delta functions in \eqref{eq:full-S-expression} reduce to 
\begin{equation}
\delta(p_1 + p_2 - p_3 - p_4) \ \delta(\omega_1  + \omega_2 - \omega_3 - \omega_4) = \dfrac{1}{2} \delta (p_1 - p_3)  \delta (p_2 - p_4) \,, 
\end{equation}
while for (ii) we find
\begin{equation}
\delta(p_1 + p_2 - p_3 - p_4) \ \delta(\omega_1  + \omega_2 - \omega_3 - \omega_4) = \dfrac{1}{2} \delta (p_1 - p_4 - \sin^2 \varphi)  \delta (p_2 - p_3 + \sin^2 \varphi) \,.  
\end{equation}
Enforcing energy and momentum conservation, we obtain
\begin{equation}
\begin{aligned} 
i \mathcal{M}_1  & = \ i \mathcal{M}_4 = - \dfrac{i \sin^2 \varphi \cos^2 \vartheta}{4}\  p_2(p_1 - \sin \varphi \sin \vartheta) \,, \\
i \mathcal{M}_2  & = \ i \mathcal{M}_3 = \dfrac{i \sin^2 \varphi \cos^2 \vartheta}{4}\  p_2(p_1 - \sin \varphi \sin \vartheta) \,, \\ 
i \mathcal{M}_5 & =0\,,
\end{aligned}
\label{eq:off-diagonal-diagrams-values-1}
\end{equation}
for (i) and 
\begin{equation}
\begin{aligned} 
i \mathcal{M}_1  & = \ i \mathcal{M}_4 = \dfrac{i \sin^2 \varphi \cos^2 \vartheta}{4}\  p_2(p_1 - \sin \varphi \sin \vartheta) \,, \\
i \mathcal{M}_2  & = \ i \mathcal{M}_3 = 0 \,, \\ 
i \mathcal{M}_5 & = -\dfrac{i \sin^2 \varphi \cos^2 \vartheta}{2}\  p_2(p_1 - \sin \varphi \sin \vartheta) \,,
\end{aligned}
\label{eq:off-diagonal-diagrams-values-2}
\end{equation}
for (ii). Hence, we find 
\begin{equation}
i \mathcal{M}_{W \bar{X} \to \bar{W} \bar{X}}=0\,,
\end{equation}
and therefore
\begin{equation}
\braket{\bar{X}(p_3) \bar{W}(p_4) |i \mathbf{T} |\bar{X}(p_1) W(p_2) } = 0 \,.  
\end{equation}
Notice that the vanishing of $i \mathcal{M}_{W \bar{X} \to \bar{W} \bar{X}}$ for (ii) is expected from integrability, as the set of initial and final momenta do not coincide. The vanishing of $i \mathcal{M}_2$ and $i \mathcal{M}_3$ in~\eqref{eq:off-diagonal-diagrams-values-2} comes out in a slightly more subtle way. In fact, once we enforce energy and momentum conservation, the two vertices vanish and at the same time the propagator of the virtual particle is on-shell. Loosely speaking, one may write 
\begin{equation}
i \mathcal{M}_2 = i \mathcal{M}_3 = \dfrac{i \sin^2 \varphi \cos^2 \vartheta}{16} \dfrac{p_2 p_4 (\omega + p + m)(\omega - p - m)}{(\omega + p + m)(\omega - p - m) + i \varepsilon} \sim \dfrac{0}{0 + i \varepsilon} = 0 \,,
\end{equation} 
where in the next-to-the-last equality we have made use of eq.~\eqref{eq:off-diagonal-w-and-p}. In this respect, we see that the $i \varepsilon$ prescription for regularizing the Feynman propagator plays an essential role. 

\subsection{Example: a (non-vanishing) diagonal term}

 We are now going to illustrate how the computation of the non-vanishing diagonal terms works by means of an example: we are going to compute the S-matrix element 
\begin{equation}
\braket{X(p_3) Y(p_4)|\,i\, \mathbf{T}\, |X(p_1) Y(p_2)} \,. 
\label{eq:diagonal-S}
\end{equation}
In principle this may contain two channels: one where the momenta are transmitted along with particles' flavours (the diagonal process), and one where they undergo a reflection. We will see that only the former is non-vanishing.
The amplitude associated to the process in \eqref{eq:diagonal-S} is
\begin{equation}
\begin{fmffile}{diagonal-amplitude}
\begin{gathered}
\begin{fmfgraph*}(75,75)
  \fmfleft{i1,i2}
  \fmfright{o1,o2}
  \fmfblob{.5w}{v1}
  \fmf{fermion,foreground=(1,,0.1,,0.1)}{i2,v1}
  \fmf{fermion,foreground=(0.1,,0.1,,1)}{i1,v1}
  \fmf{fermion,foreground=(0.1,,0.1,,1)}{v1,o1}
  \fmf{fermion,foreground=(1,,0.1,,0.1)}{v1,o2}
  	\Marrow{a}{left}{lft}{$p_1$}{i2,v1}{7}    
  	\Marrow{b}{left}{lft}{$p_2$}{i1,v1}{8}    
  	\Marrow{c}{right}{rt}{$ \ p_3$}{v1,o2}{8}   
  	\Marrow{d}{right}{rt}{$\ p_4$}{v1,o1}{8}    
	        \end{fmfgraph*}
\end{gathered} = i \mathcal{M}_{X Y \to X Y} = \sum_{j=1}^5 i \tilde{\mathcal{M}}_j
\end{fmffile}
\label{eq:diagonal-amplitude}
\end{equation}
where time flows from left to right and the Feynman diagrams contributing to the amplitude \eqref{eq:diagonal-amplitude} are 
\begin{fmffile}{diagonal-scattering}
\begin{equation}
\begin{aligned}
i \tilde{\mathcal{M}}_1 = & \quad 
\begin{gathered}
\begin{fmfgraph*}(72,72)
  \fmfleft{i1,i2}
  \fmfright{o1,o2}
  \fmf{fermion,foreground=(1,,0.1,,0.1)}{i2,v2}
  \fmf{fermion,foreground=(0.1,,0.1,,1)}{i1,v1}
  \fmf{fermion,foreground=(0.1,,0.1,,1)}{v1,o1}
  \fmf{fermion,foreground=(1,,0.1,,0.1)}{v2,o2}
  \fmf{scalar}{v1,v2}
  	\Marrow{a}{left}{bot}{$p_1 \ \ $}{i2,v2}{15}    
  	\Marrow{b}{left}{top}{$p_2 \ \ $}{i1,v1}{15}    
  	\Marrow{c}{right}{bot}{$\ \ p_3$}{v2,o2}{15}   
  	\Marrow{d}{right}{top}{$\ \ p_4$}{v1,o1}{15}  
  	\Marrow{e}{right}{rt}{$p$}{v1,v2}{7}    
	        \end{fmfgraph*}
\end{gathered} \quad \,, & \quad \quad i \tilde{\mathcal{M}}_2 = & \begin{gathered}
\begin{fmfgraph*}(72,72)
  \fmfleft{i1,i2}
  \fmfright{o1,o2}
  \fmf{fermion,foreground=(1,,0.1,,0.1)}{i2,v2}
  \fmf{fermion,foreground=(0.1,,0.1,,1)}{i1,v1}
  \fmf{fermion,foreground=(0.1,,0.1,,1)}{v1,o1}
  \fmf{fermion,foreground=(1,,0.1,,0.1)}{v2,o2}
  \fmf{scalar}{v2,v1}
  	\Marrow{a}{left}{bot}{$p_1 \ \ $}{i2,v2}{15}    
  	\Marrow{b}{left}{top}{$p_2 \ \ $}{i1,v1}{15}    
  	\Marrow{c}{right}{bot}{$\ \ p_3$}{v2,o2}{15}   
  	\Marrow{d}{right}{top}{$\ \ p_4$}{v1,o1}{15}  
  	\Marrow{e}{right}{rt}{$p$}{v2,v1}{7}    
	        \end{fmfgraph*}
\end{gathered} \quad \,, \\
& & & \\
i \tilde{\mathcal{M}}_3 = & \quad \begin{gathered}
\begin{fmfgraph*}(72,72)
  \fmfleft{i1,i2}
  \fmfright{o1,o2}
  \fmf{phantom}{i1,v1}
  \fmf{phantom}{o1,v1}
  \fmf{phantom}{i2,v2}
  \fmf{phantom}{o2,v2}
  \fmf{fermion,tension=0,foreground=(1,,0.1,,0.1)}{i2,v1}
  \fmf{fermion,tension=0,foreground=(0.1,,0.1,,1)}{i1,v2}
  \fmf{fermion,tension=0,foreground=(0.1,,0.1,,1)}{v2,o1}
  \fmf{fermion,tension=0,foreground=(1,,0.1,,0.1)}{v1,o2}
  \fmf{scalar}{v2,v1}
  	\Marrow{a}{left}{lft}{$ \ p_1$}{i2,v1}{10}    
  	\Marrow{b}{down}{bot}{$\ \ \ \ p_2$}{i1,v2}{15}    
  	\Marrow{c}{up}{lft}{$\ p_3$}{v1,o2}{15}   
  	\Marrow{d}{right}{rt}{$\ p_4$}{v2,o1}{7}  
	        \end{fmfgraph*}
\end{gathered} \quad \,, & \quad \quad i \tilde{\mathcal{M}}_4 = & \quad \begin{gathered}
\begin{fmfgraph*}(72,72)
  \fmfleft{i1,i2}
  \fmfright{o1,o2}
  \fmf{phantom}{i1,v1}
  \fmf{phantom}{o1,v1}
  \fmf{phantom}{i2,v2}
  \fmf{phantom}{o2,v2}
  \fmf{fermion,tension=0,foreground=(1,,0.1,,0.1)}{i2,v1}
  \fmf{fermion,tension=0,foreground=(0.1,,0.1,,1)}{i1,v2}
  \fmf{fermion,tension=0,foreground=(0.1,,0.1,,1)}{v2,o1}
  \fmf{fermion,tension=0,foreground=(1,,0.1,,0.1)}{v1,o2}
  \fmf{scalar}{v1,v2}
  	\Marrow{a}{left}{lft}{$ \ p_1$}{i2,v1}{10}    
  	\Marrow{b}{down}{bot}{$\ \ \ \ p_2$}{i1,v2}{15}    
  	\Marrow{c}{up}{lft}{$\ p_3$}{v1,o2}{15}   
  	\Marrow{d}{right}{rt}{$\ p_4$}{v2,o1}{7}  
	        \end{fmfgraph*}
\end{gathered} \quad \,, \\
& & & \\
i \tilde{\mathcal{M}}_5 = & \quad 
\begin{gathered}
\begin{fmfgraph*}(72,72)
  \fmfleft{i1,i2}
  \fmfright{o1,o2}
  \fmf{fermion,foreground=(1,,0.1,,0.1)}{i2,v1}
  \fmf{fermion,foreground=(0.1,,0.1,,1)}{i1,v1}
  \fmf{fermion,foreground=(0.1,,0.1,,1)}{v1,o1}
  \fmf{fermion,foreground=(1,,0.1,,0.1)}{v1,o2}
  	\Marrow{a}{left}{lft}{$p_1$}{i2,v1}{7}    
  	\Marrow{b}{left}{lft}{$p_2$}{i1,v1}{8}    
  	\Marrow{c}{right}{rt}{$ \ p_3$}{v1,o2}{8}   
  	\Marrow{d}{right}{rt}{$\ p_4$}{v1,o1}{8}    
	        \end{fmfgraph*}
\end{gathered} \quad \,.  & &
\end{aligned}
\label{eq:diag-scattering-example}
\end{equation}
\end{fmffile}
Proceeding like in the previous section, we can use the Feynman rules in appendix~\ref{app:feynman-rules} to compute the diagrams in eq.~\eqref{eq:diag-scattering-example}. Again, we can resolve the absolute values entering the dispersion relations in two different ways. For one of these choices the energy and momentum conservation delta functions reduce to 
\begin{equation}
\delta(p_1 + p_2 - p_3 - p_4) \ \delta(\omega_1  + \omega_2 - \omega_3 - \omega_4) = \dfrac{1}{2} \delta (p_1 - p_3)  \delta (p_2 - p_4) \,, 
\end{equation}
while for the other one the set of initial momenta does not coincide with the set of final momenta. Such a process would violate the integrability of the theory, we would hope for it to vanish. Indeed exploiting energy and momentum conservation we find  
\begin{equation}
\begin{aligned} 
i \tilde{\mathcal{M}}_1  & = \ i \tilde{\mathcal{M}}_2 = i \tilde{\mathcal{M}}_3 = i \tilde{\mathcal{M}}_4 = \dfrac{i \sin \varphi \sin \vartheta  \cos \varphi \cos \vartheta}{4}(p_1 + \sin \varphi \sin \vartheta)(p_3 + \sin \varphi \sin \vartheta) \,, \\
i \tilde{\mathcal{M}}_5 & = - i \sin \varphi \sin \vartheta \cos \varphi \cos \vartheta \ (p_1 + \sin \varphi \sin \vartheta)(p_3 + \sin \varphi \sin \vartheta)
\end{aligned}
\label{eq:diagonal-diagrams-values-2}
\end{equation}
so that the amplitude vanishes. Instead, when the sets of initial and final momenta coincide we obtain
\begin{equation}
\begin{aligned} 
i \tilde{\mathcal{M}}_1  & = \ i \tilde{\mathcal{M}}_2 = i \tilde{\mathcal{M}}_3 = i \tilde{\mathcal{M}}_4 = 0 \,, \\
i \tilde{\mathcal{M}}_5 & = 2i (p_1 + \sin \varphi \sin \vartheta)(p_2 + \cos \varphi \cos \vartheta) \times \\
 & \qquad \times [(1-2a)p_1 p_2 + (1-a) \sin \varphi \sin \vartheta \ p_2 + (1-a) \cos \varphi \cos \vartheta \ p_1 ]
\end{aligned}
\label{eq:diagonal-diagrams-values-1}
\end{equation}
The vanishing of $i \tilde{\mathcal{M}}_i$ for $i = 1 \dots 4$, arises in an particular way: due to energy and momentum conservation, we find $p = \omega = 0$, meaning that the virtual particle has null energy and null momentum. We see that also in this case the $i \varepsilon$ prescription for the Feynman propagator plays an essential role. We can now exploit \eqref{eq:full-S-expression} to compute the S-matrix element in \eqref{eq:diagonal-S}. We find 
\begin{equation}
\begin{aligned}
\braket{X(p_3) Y(p_4)|i \mathbf{T} |X(p_1) Y(p_2)}  = &  -i(1-2a)p_1 p_2  -i (1-a) \sin \varphi \sin \vartheta \ p_2  \\
& - i (1-a) \cos \varphi \cos \vartheta \ p_1 \,. 
\end{aligned}
\end{equation}

\subsection{Tree-level S~matrix}

The complete computation on the lines of the two previous sections gives us the following result for the S-matrix for a anti-chiral--chiral scattering process: 
\begin{equation}
i\,{T}_{12}^{34} = \delta_1^3 \ \delta_2^4 \ f(p_1, p_2 ) \,, 
\end{equation}
with 
\begin{equation}
f(p_1, p_2 ) 
 = -i\,p_1 p_2 - i (p_1 r_2 - p_2 r_1) \,,
\label{eq:compact-S}
\end{equation}
where we defined the (gauge-dependent) combination $r_i = \mathcal{J}_i + a\, \omega_i$. Indeed this combination takes the same form as eq.~\eqref{eq:Rdef} (as well as flat-space strings~\cite{Dubovsky:2012wk} and pure-NSNS $\AdS{3}\times\S^3\times\T^4$~\cite{Baggio:2018gct}), and compensates the gauge-dependence of the worldsheet volume~\cite{Staudacher:2004tk}. It is convenient to re-express the scattering phase in terms of $p_i$, $\omega_i$, which emphasises its anti-symmetry:
\begin{equation}
f(p_1, p_2 ) 
 = \dfrac{i}{2} \Big[p_1 (\omega_2-\hat \mu_2) - p_2 (\omega_1-\hat \mu_1)\Big] - i (p_1 r_2 - p_2 r_1) \,,
\label{eq:Stree}
\end{equation}
where $\hat{\mu}$ is the non-anomaloues (\textit{i.e.}, $p$-independent) part of the energy $\ell+\tilde{\ell}-\mathcal{J}$, and can be defined as
\begin{equation}
\hat \mu_i = \mu_i \, \text{sgn}\big[p_i+\mu_i\big]\,.
\end{equation}
This also emphasises that the perturbative S-matrix is consistent with the requirements of crossing symmetry~\cite{Janik:2006dc}. Using that under the crossing transformation $p\to-p$, $\omega\to-\omega$, so that $\mathcal{J}\to-\mathcal{J}$ and $\hat\mu\to-\hat\mu$, we get as expected
\begin{equation}
f\big(p_1\big|_{\text{crossed}}, p_2 \big) +f(p_1,p_2)
=0\,.
\end{equation}

\section{All-loop S~matrix and spectrum}
\label{sec:TBA}

Building on the simple form of the tree-level bosonic S~matrix~\eqref{eq:Stree}, we can now make a proposal for the all loop integrable S~matrix by assuming that the tree-level phase-shift simply exponentiates, compatibly with what expected from unitarity. To check whether this is the case, we will compute the finite-volume spectrum of the theory and compare it with the WZW model prediction.

\subsection{All-loop S~matrix and asymptotic Bethe ansatz}
\label{sec:ABA}
We propose the two-particle S-matrix of this model takes the form
\begin{equation}
\mathbf{S}(p_i,p_j)=e^{i\Phi(p_i,p_j)}\,\mathbf{1},
\end{equation}
with, in the $a$-gauge of section~\ref{sec:gaugefix}
\begin{equation}
\Phi(p_i,p_j) = \frac{1}{2}\Big[p_i (H_j-\hat{\mu}_j)-p_j (H_i-\hat{\mu}_i)\Big] -\Big[p_i(\mathcal{J}_j+a H_j)-p_j (\mathcal{J}_i+a H_i)\Big]\,,
\end{equation}
where for a single excitation we have%
\footnote{Since we are no longer interested in the near-BMN expansion, $P_- = 2 R$ does not scale with $k$ and accordingly we reinstated the scaling of $p$ with $k/2\pi$.}
\begin{equation}
H_i=H(p_i) = \Big|\frac{k}{2\pi}p_i+\mu_i\Big|,\qquad
\hat{\mu}_i= \mu_i\,\text{sgn}\Big[\frac{k}{2\pi}p_i+\mu_i\Big]\,,
\end{equation}
while $\mathcal{J}_i$ is the contribution to the angular momentum on $\S^3\times\S^3$, \textit{i.e.}\ the charge under  $\mathcal{J}$ of the $i$-th particle, \textit{cf.}\ eq.~\eqref{eq:defJ}. Here we are assuming that $H(p)$ does not receive corrections in $1/k$ with respect to its one-loop counterpart $\omega(p)$---\textit{i.e.}, minding the factor of $k/2\pi$ in front of $p$, $H(p)=\omega(\tfrac{k}{2\pi}p)$. In the supersymmetric case $\varphi=\vartheta$, the equality between $H(p)$ and $\omega(p)$ is the consequence of a shortening condition~\cite{Borsato:2012ud, Borsato:2015mma}, while more generally this is an assumption. As for the value of the  masses, recalling that $\mu_i$ is a combination of~spins,
\begin{equation}
\big[(\ell-\tilde{\ell})-\cos\varphi\cos\vartheta(j_1-\tilde{\jmath}_1) -\sin\varphi\sin\vartheta(j_2-\tilde{\jmath}_2)\big]\,|p_i\rangle= \mu_i\,|p_i\rangle\,,
\end{equation}
which are quantised, we take the masses of bosons and fermions to be the same as in the pp-wave limit~\cite{Dei:2018yth}, see table~\ref{tab:S-matrix-basis}. Let us emphasise once more that, when $\vartheta\neq\varphi$, the masses of the bosons and those of the fermions are different.

With this input, it is immediate to write down the asymptotic Bethe ansatz (ABA) equations for an $M$-particle state, which are
\begin{equation}
\label{eq:aba}
e^{i p_i\,R}\,\prod_{j\neq i}^M e^{i\Phi(p_i,p_j)}=1\,,\qquad i=1,\dots M\,,
\end{equation} 
which we supplement by the definition of the worldsheet size $R$ in the $a$-gauge---\textit{cf.} eq.~\eqref{eq:Rdef}---and by the level matching condition,
\begin{equation}
\label{eq:abadefs}
R= \mathcal{J}_{\text{tot}}+ a H_{\text{tot}}\,, \qquad
\mathcal{J}_{\text{tot}}=\mathcal{J}_0+\sum_{i=1}^M \mathcal{J}_i\,, \quad
H_{\text{tot}}=\sum_{i=1}^M H_i\,,\quad
 P_{\text{tot}}= \sum_{i=1}^M p_i =0\,,
\end{equation}
where $\mathcal{J}_0$ is the vacuum R-charge and the last equality is the level-matching condition%
\footnote{One can generalise this equation and its solution to the case $P_{\text{tot}}=2\pi w$ with $w\in\mathbb{Z}$~\cite{Arutyunov:2009ga} following the discussion of ref.~\cite{Dei:2018mfl}.}%
.
Taking the logarithm of eq.~\eqref{eq:aba} we get%
\footnote{%
Notice that, as expected, the gauge dependence has dropped from the ABA equations~\cite{Staudacher:2004tk}.
}
\begin{equation}	
p_i\, \Big[\mathcal{J}_0 + \frac{1}{2}\sum_{j=1}^M(H_j-\hat{\mu}_j)\Big] =  2\pi n_i,\qquad
n_i\in\mathbb{Z}\,.
\end{equation}
where we used the level matching condition. These equations can be solved much  like in ref.~\cite{Dei:2018mfl}.  Let us briefly recall how this goes in the simplest case, corresponding to the highest weight representations of the WZW model; this will be sufficient to illustrate some novel features with respect to the $\AdS{3}\times\S^3\times\T^4$ case.
We have
\begin{equation}
p_i = \frac{2\pi\,n_i}{R_{\text{eff}}},\qquad R_{\text{eff}}= \mathcal{J}_0 + \frac{H_{\text{tot}}-\hat\mu_{\text{tot}}}{2}.
\end{equation}
Plugging this in the dispersion relation, and assuming that we can take the positive (resp. negative) branch of the dispersion for $p_i>0$ (resp.\ $p_i<0$)%
\footnote{This is only true for states corresponding to those coming from highest-weight representations in the WZW model. 
When $R_{\text{eff}}$ is sufficiently large with respect to~$k$, the various spectrally flowed sectors arise, see ref.~\cite{Dei:2018mfl} for a detailed discussion.
}
we find
\begin{equation}
H_{\text{tot}}=\frac{k}{R_{\text{eff}}}\sum_{i=1}^M |n_i| +\hat\mu_{\text{tot}}=\frac{2k\mathcal{N}}{\mathcal{J}_0+\tfrac{1}{2}(H_{\text{tot}}-\hat\mu_{\text{tot}})}\,,\qquad
\mathcal{N}= \sum_{i:\,n_i>0}n_i=-\sum_{i:\,n_i<0}n_i,
\end{equation}
 which gives
\begin{equation}
\label{eq:ABAsol}
H_{\text{tot}}=\ell+\tilde{\ell} - \mathcal{J}
= \sqrt{\mathcal{J}_0^2+4k\,\mathcal{N}}+\hat\mu_{\text{tot}}\,.
\end{equation}
We have arrived at a \textit{closed-form} expression for the energy, like in ref.~\cite{Dei:2018mfl}. As remarked, note that this is a rather unique occurrence even in the theory of integrable models. Interestingly, a strikingly similar result was found in the context of $\AdS{5}/\CFT_4$~\cite{Alday:2005jm,Arutyunov:2005hd}, and it would be interesting to investigate the similarities of these two set-ups further.

Following again ref.~\cite{Dei:2018mfl}, we expect this to match with the WZW result where we identify the Green-Schwarz vacuum, labelled by $\mathcal{J}$, with a specific (physical) state arising from the $\su(2)_{(1)}\oplus\su(2)_{(2)}$ Ka\v{c}-Moody modules with highest weights $(j_{0,1},j_{0,2})$. For $\AdS{3}\times\S^3\times\T^4$ we observed that the matching required taking as a reference state the lightest physical states \textit{in the Ramond-Ramond sector} of the theory. Here there is one such state, and it has R-charge
\begin{equation}
\mathcal{J}_0 = \sin\varphi\sin\vartheta\, (2j_{0,1}+1) + \cos\varphi\cos\vartheta\, (2j_{0,2}+1)\,.
\end{equation}
The expression for the vacuum R-charge can be further simplified by recalling that $\vartheta$ encodes the ratio of R-charges of the light-cone-gauge vacuum along the two spheres as in eq.~\eqref{eq:thetadef}, so that specialising to our case
\begin{equation}
\tan \vartheta = \tan\varphi\,\frac{2j_{0,1}+1}{2j_{0,2}+1}\,,
\end{equation} 
and
\begin{equation}
\mathcal{J}_0 = \sqrt{\sin^2\varphi\,(2j_{0,1}+1)^2+\cos^2\varphi\,(2j_{0,2}+1)^2}\,.
\end{equation}
Using this expression we reproduce the WZW energy formula of appendix~\ref{app:WZW}, seemingly for any value of $\varphi$ and $\vartheta$: not only we reproduce the mode-number-dependence in the square root, but we can also match the $\hat\mu_{\text{tot}}$ charge with the charges of the WZW description.

More in detail, in the supersymmetric case when $\vartheta = \varphi$ (that is, when the highest-weights are $j_{0,1}=j_{0,2}\equiv j_0$),  the vacuum (the reference state in the Ramond-Ramond sector) is BPS and it has R-charge $\mathcal{J}_0=2j_0+1$. Note that for each given $j_0=j_{0,1}=j_{0,2}$ there are \textit{four} BPS states: the one which we considered in the R-R sector, and three more in the NS-R, R-NS and NS-NS sectors; in the Green-Schwarz language, they are related to each other by acting with a zero-mode of a massless fermion~\cite{Baggio:2017kza,Eberhardt:2017fsi}. Things are more involved in the non supersymmetric case, when $j_{0,1}\neq j_{0,2}$. The energy $H_{\text{tot}}=\Delta-\mathcal{J}$ of the R-R reference state is still zero---as correctly reproduced by the asymptotic Bethe ansatz---but this reference state \textit{is no longer the lowest-energy state}. In fact, of the four ``would-be-BPS states'' described above, the NS-R, R-NS, and NS-NS states acquire a negative energy, as we schematically represent in the diamond below
\begin{equation}
\begin{gathered}
H_{\text{NS-NS}} = H_0\,,\\
H_{\text{NS-R}} = \tfrac{1}{2}H_0\,,\qquad
H_{\text{R-NS}} = \tfrac{1}{2}H_0\,,\\
H_{\text{R-R}} = 0\,.
\end{gathered}
\end{equation}
with
\begin{equation}
\label{eq:H0}
H_0=\sin\varphi\sin\vartheta + \cos\varphi\cos\vartheta-1 \leq 0\,.
\end{equation}
Such a shift is to be expected for a non-supersymmetric theory, and arises from the polarisation of the vacuum, see ref.~\cite{Dei:2018yth} for its explicit computation. On the other hand, the energy shifts are proportional to the energies of the lightest fermions. In view of this, we can treat the state with no excitations in the asymptotic Bethe ansatz (the R-R state) as a \textit{pseudo-vacuum}, and postulate that the true vacuum (the NS-NS state) is obtained by acting with on the pseudo-vacuum with two fermion zero-modes \textit{with negative light-cone energy}. This redefinition of the Fermi sea is harmless, and allows us to interpret this model as one where \textit{the asymptotic Bethe ansatz} is exact, \textit{i.e.}\ where no particle production occurs, even at the microscopic level.%
\footnote{%
It is interesting to note that, \textit{despite the lack of manifest supersymmetry}, there exists always a pseudo-vacuum state which has vanishing light-cone energy. Note that this is necessary for the asymptotic Bethe ansatz to have any chance at all to describe the spectrum. Indeed throughout our construction we have implicitly assumed the existence of such a pseudo-vacuum, see for instance eq.~\eqref{eq:abadefs}.
}

Thus, we have found a set of asymptotic Bethe ansatz equations that match the (exact) known WZW spectrum---suggesting the existence of an underlying integrable spin chain. The main novelty with respect to $\AdSSS$ is the necessity to treat the ABA vacuum as a pseudo-vacuum, and the appearance of states with negative light-cone energy in the sectors with $\vartheta \neq \varphi$. At this stage, the analysis of the non-supersymmetric sectors relied on the known WZW result and on an \textit{ad-hoc} prescription. We know however that, if our proposal for the S~matrix is correct, the exact spectrum including the true ground state should emerge from the mirror thermodynamic Bethe ansatz (TBA). We shall see below that this is indeed the case, and highlight as we progress the physical reasons for the drastic simplification of the mirror TBA equations with respect to a generic integrable model.


\subsection{The mirror theory and TBA equations}
The use of the thermodynamic Bethe ansatz~\cite{Yang:1968rm} to compute the finite-volume spectrum of two-dimensional integrable models is well-established~\cite{Zamolodchikov:1989cf}. The basic idea is that the finite-volume properties of the model of interest are encoded in the finite-temperature properties of a \textit{mirror model}, where the roles of time and space have been interchanged. When the integrable theory is not Poincar\'e invariant, as it is the case for light-cone gauge-fixed string NLSM and indeed for our model, the mirror model is a genuinely new theory~\cite{Arutyunov:2007tc}, see also ref.~\cite{vanTongeren:2016hhc} for a pedagogical introduction to the (mirror) thermodynamic Bethe ansatz. 

As explained in detail in refs.~\cite{Arutyunov:2007tc,vanTongeren:2016hhc}, we can think of the mirror model as being related to the original one by an analytic continuation which sends the momentum of the original model $p$ into the \textit{energy} of the mirror one, $\bar{H}$, and similarly energy in the mirror momentum:
\begin{equation}
p \to i\,\bar{H}\,,\qquad
H \to i\, \bar{p}\,.
\end{equation}
Here and below we denote with a ``bar'' all quantities computed in the mirror model. In this way, from the original dispersion relation we find the mirror dispersion,
\begin{equation}
H(p)=\Big|\frac{k}{2\pi}p+\mu\Big|\quad\to
\quad
\bar{H}(\bar{p}) = \frac{|\bar{p}|+ i\mu}{k/2\pi}=
\frac{
(\bar{p}+i\mu)\Theta(\bar{p})+(-\bar{p}+i\mu)\Theta(-\bar{p})
}{k/2\pi}
\,,
\end{equation}
where we have explicitly highlighted the dispersion of chiral and anti-chiral particles by means of Heaviside's Theta function. We have that $\bar{p}>0$ for chiral particles, and $\bar{p}<0$ for anti-chiral ones. In the mirror theory, the ``mass'' $\mu$ plays the role of a purely imaginary twist of the Bethe equations. These take the form
\begin{equation}
e^{i \beta\, \bar{p}_i}\, \prod_{j\neq i}^M S(\bar{p}_i,\bar{p}_j)=(-1)^{F_i}\,,\qquad i=1,\dots M\,
\end{equation}
where we made it explicit that the fermions of the mirror model are \textit{anti-periodic} (while the ones arising from the Green-Schwarz string were periodic)~\cite{Arutyunov:2007tc}, and the ``volume'' $\beta$ is of course the inverse temperature of the original model. 
The mirror Bethe equations are written in terms of the ``mirror-mirror'' scattering matrix~$S(\bar{p}_i,\bar{p}_j)$, which is the analytic continuation of the original S~matrix. We have
\begin{equation}
\label{eq:mirrorABA}
\bar{p}_i\beta +\sum_{j=1}^M \Phi(\bar{p}_i, \bar{p}_{j}) =2\pi m_i + \pi F_i\,,\qquad i=1,\dots M\,,
\end{equation}
where\footnote{%
We define the mirror model by incorporating a shift of $\sum_i(\mathcal{J}_j+a H_j)$ due to the linear terms in the definition of $\beta$. This will simplify the form of the mirror TBA equations.
}
\begin{equation}
\Phi(\bar{p}_i, \bar{p}_{j}) =
\begin{cases}
+\dfrac{k}{2\pi} \bar{H}(\bar{p}_i)\,\bar{H}(\bar{p}_j)\,\Theta(-\bar{p}_i)\Theta(+\bar{p}_j),\phantom{\Bigg|}
\\
-\dfrac{k}{2\pi} \bar{H}(\bar{p}_i)\,\bar{H}(\bar{p}_j)\, \Theta(+\bar{p}_i)\Theta(-\bar{p}_j),\phantom{\Bigg|}
\end{cases}
\end{equation}
where we highlighted split into chiral--anti-chiral and anti-chiral--chiral scattering.

 The only subtle step remaining before obtaining the mirror TBA equations is to formulate the \textit{string hypothesis}, \textit{i.e.}\ describing all solutions of the equations~\eqref{eq:mirrorABA} in the thermodynamic limit. In general these include particles, bound states and ``Bethe strings'', see \textit{e.g.}\ refs.~\cite{Arutyunov:2007tc,Arutyunov:2009zu}. Here, like for pure-NSNS $\AdS{3}\times\S^3\times\T^4$, due to the linearity of the equations~\eqref{eq:mirrorABA}, we assume that only the $(8|8)$ fundamental particles appear in the thermodynamic limit. The mirror TBA equations for the ground state then are written in terms of $(8|8)$ pseudo-energies $\epsilon_a(u)$, where $u$ is some appropriate parametrisation, and take the familiar form%
\footnote{See appendix~C in ref.~\cite{Dei:2018mfl} for a brief derivation of the mirror TBA equations for a non-relativistic theory with diagonal scattering, and \textit{e.g.}\ ref.~\cite{vanTongeren:2015uha} for a more comprehensive derivation and discussion.}
\begin{equation}
\label{eq:mirrorTBA}
\epsilon_a(u) = \psi_a + R\,\bar{H}(u) - \sum_{b}^{(8|8)} \int\limits_{\gamma}\text{d} v\, \Lambda_b (v)\, K_{ba}(v,u),
\end{equation}
where $\gamma$ is an appropriate contour which depends on the state we consider and
\begin{equation}
\psi_a=\begin{cases}
0, &a\ \text{boson},\\
i\pi, &a\ \text{fermion },
\end{cases}
\qquad
\Lambda_b(v)=\begin{cases}
-\log(1-e^{-\epsilon_b(v)}), &b\ \text{boson},\\
+\log(1+e^{-\epsilon_b(v)}), &b\ \text{fermion},
\end{cases}
\end{equation}
and finally
\begin{equation}
K_{ba}(v,u) = \frac{1}{2\pi}\frac{\partial}{\partial v} \, \Phi\big[\bar{p}_b(v),\bar{p}_a(u)\big].
\end{equation}
The total energy and total momentum are then expressed in terms of the pseudo-energies:
\begin{equation}
\label{eq:TBAHP}
H_{\text{tot}}= -\frac{1}{2\pi}\sum_{a}^{(8|8)} \int\limits_{\gamma}\text{d} v\,\frac{\partial \bar{p}}{\partial v}  \Lambda_a (v),\qquad
P_{\text{tot}}= -\frac{1}{2\pi}\sum_{a}^{(8|8)} \int\limits_{\gamma}\text{d} v\,\frac{\partial \bar{H}}{\partial v}  \Lambda_a (v).
\end{equation}

\subsection{Ground-state energy}
The ground-state mirror TBA equations follow from eq.~\eqref{eq:mirrorTBA} by taking the contour $\gamma$ to run on the mirror-momentum line. It is convenient to take the parameter $v$ to be the mirror momentum $\bar{p}$ itself. Then
\begin{equation}
K_{ba}(\bar{p}_b,\bar{p}_a) =- \frac{1}{2\pi} \Big[\Theta(\mp \bar{p}_b) +i\mu_b\,\delta(\bar{p}_b)\Big]\,\bar{H}(\bar{p}_a)\,\Theta(\pm \bar{p}_a).
\end{equation}
Let us explicitly split the pseudo-energies between chiral and anti-chiral particles, which we denote with $\epsilon^\pm_a$. Then
\begin{equation}
\epsilon_a^\pm(\bar{p}) = \psi_a + \Big(\frac{2\pi R+ K_{\mp}+\kappa_{\mp}}{k}\Big)\,(\pm \bar{p}+i\mu_a)
,\qquad \pm \bar{p} \geq 0, 
\end{equation}
where 
\begin{equation}
\label{eq:intdef}
\begin{aligned}
K_{+}&= \sum_{b}^{(8|8)} \int\limits_0^{+\infty}\text{d}\bar{p} (-1)^{F_b+1} \log\Big(1-(-1)^{F_b}e^{-\epsilon^+_b(\bar{p})}\Big),\\
K_{-}&= \sum_{b}^{(8|8)} \int\limits_{-\infty}^0\text{d}v (-1)^{F_b+1} \log\Big(1-(-1)^{F_b}e^{-\epsilon^-_b(\bar{p})}\Big),\\
\kappa_\pm&=-i\sum_{b}^{(8|8)}\mu_b\,(-1)^{F_b} \log\Big(1-(-1)^{F_b}e^{-\epsilon_b^\pm(0)}\Big).
\end{aligned}
\end{equation}
The ground state energy and momentum~\eqref{eq:TBAHP} are given by
\begin{equation}
H_{0}= -\frac{K_+ + K_-}{2\pi},\qquad
P_{0}= \frac{K_+ - K_- + \kappa_+ - \kappa_-}{k}.
\end{equation}

We start by imposing level-matching condition for the ground state, $P_0=0$, which gives $K_+ +\kappa_+ =  K_- + \kappa_-$ so that from the expressions of~\eqref{eq:intdef} we find
\begin{equation}
\epsilon^+_a(\bar{p}) = \epsilon^-_a(-\bar{p})\,,\qquad
K_+ = K_-\,,\qquad \kappa_+ = \kappa_-\,,
\end{equation}
Using the linearity of the pseudo-energies,
\begin{equation}
\epsilon_a^\pm(\bar{p}) = (\pm\bar{p} + i \mu_a)\,c + \psi_a, \qquad
\pm\bar{p}\geq 0
\end{equation}
the integral can be computed analytically, as we do in appendix~\ref{app:integrals}. 
This computation relies on essentially two physical features of the model: first, crossing symmetry requires that the masses $\{\mu_i\}$ come in pairs of different signs. This enforces reality of the TBA integrals (despite $\bar{H}(\bar{p})$ being complex for real $\bar{p}$) and greatly simplifies their analytic form.
 Secondly, supersymmetry of the original model dictates that---even when the choice of the light-cone geodesic does not preserve supersymmetry manifestly---the sum of the square masses is the same for the bosons as it is for the fermions. We refer the reader to appendix~\ref{app:integrals} for details of how affects the computation of the integrals. 
Eventually we find that both $\kappa_\pm$ and $K_\pm$ are proportional to the differences of the bosonic and fermionic masses,
\begin{equation}
\label{eq:Ksolution}
K_\pm=-\kappa_\pm= -\pi \big(\sin\varphi\sin\vartheta+\cos\varphi\cos\vartheta-1\big)\,,
\end{equation}
so that
\begin{equation}
H_0 = \sin\varphi\sin\vartheta+\cos\varphi\cos\vartheta-1 \leq 0,
\end{equation}
perfectly matching eq.~\eqref{eq:H0}.
Moreover, in the case where the theory is manifestly supersymmetric, and $\vartheta=\varphi$, we find that the ground-state energy vanishes, as expected.
The simplicity of the solution is quite striking, though not entirely surprising in light of the simplicity of the stringy WZW spectrum. Still, it might be quite interesting to further investigate this structure by computing the finite-size corrections~\cite{Arutyunov:2006gs} to giant-magnon solutions~\cite{Hoare:2013lja}.

\subsection{Excited states from contour deformation}
Following ref.~\cite{Dorey:1996re}, we take the equation for the excited states to have the same forms of eq.~\eqref{eq:mirrorTBA} up to modifying the integration contour~$\gamma$. Different states are given by different  and inequivalent choices of~$\gamma$; for two contours to be inequivalent, they should encompass different singularities of the integral. These can arise only from singularities of the logarithm in $\Lambda(v)$, namely when the rapidity $v$ takes a value $v^*$ such that
\begin{equation}
\label{eq:TBAsing}
e^{-\epsilon_a^\pm(v^*)}=(-1)^{F_a}\,.
\end{equation}
To make the equations more transparent, we can deform the contour~$\gamma$ back to the real mirror-momentum line (like we had for the ground-state equations), picking up appropriate residues in the process, so that for a state where the contour encompassed  $M$ singularities of the form~\eqref{eq:TBAsing} we have
\begin{equation}
\label{eq:mirrorTBAexcited}
\epsilon_a(u) = \psi_a + R\,\bar{H}(u) - \sum_{b}^{(8|8)} \int\limits_{\gamma}\text{d} v\, \Lambda_b (v)\, K_{ba}(v,u) +i \sum_{j=1}^M \Phi\big[\bar{p}_{b}(v^*_{j}),\bar{p}_a(u)\big].
\end{equation}
The equations for the total energy and momentum~\eqref{eq:TBAHP} are similarly modified:
\begin{equation}
\begin{aligned}
H_{\text{tot}}&= -\frac{1}{2\pi}\sum_{a}^{(8|8)} \int\limits_{\gamma}\text{d} v\,\frac{\partial \bar{p}}{\partial v}  \Lambda_a (v)+i \sum_{j=1}^M \bar{p}(v^*_{j}),\\
P_{\text{tot}}&= -\frac{1}{2\pi}\sum_{a}^{(8|8)} \int\limits_{\gamma}\text{d} v\,\frac{\partial \bar{H}}{\partial v}  \Lambda_a (v)+i \sum_{j=1}^M \bar{H}(v^*_{j}).
\end{aligned}
\end{equation}

Using the form of the scattering phase and of the kernel we see that yet again $\epsilon_a^\pm(v) \propto \bar{H}_a(v)$ up to a fermion sign~$\psi_a$. A necessary condition for eq.~\eqref{eq:TBAsing} to have a solution is that $\epsilon_a^\pm(v^*)$ is purely imaginary, which as expected happens precisely when $v^*$ is on the real-momentum (\textit{not} mirror momentum) line and $\bar{H}_a(v^*)=i p^*$. In other words, such a singularity correspond to a particle in the original (string) theory. Using this information we can write
\begin{equation}
\begin{aligned}
\epsilon_a^\pm(\bar{p}) &= \psi_a + \Big(\frac{2\pi R+ K_{\mp}+\kappa_{\mp} \mp k P_{\mp}^*}{k}\Big)\,(\pm \bar{p}+i\mu_a)
,\qquad \pm \bar{p} \geq 0, \\
H_{\text{tot}}&= \frac{K_+ + K_-}{2\pi}+\sum_{j=1}^M H(p^*_{j}),\\
P_{\text{tot}}&= \frac{K_+ +\kappa_+ +k P_{+}^*-(K_- +\kappa_- -k P_{-}^*)}{k},
\end{aligned}
\end{equation}
where we indicated the sum of all momenta of (anti-)chiral excitations arising from the residues at $\{v_j^*\}_{j=1,\dots M}$ as $P^*_{\pm}$.
Once again, by solving the level-matching constraint we find that $K_+=K_-$ and $\kappa_+=\kappa_-$ is a solution. Moreover, the integrals give the exact same  values as in eq.~\eqref{eq:Ksolution}, see appendix~\ref{app:integrals}. Using these expressions we find simply
\begin{equation}
\epsilon_a^\pm(v) = \psi_a + \big(R+ \tfrac{k}{2\pi} P_{\mp}^*\big)\,\bar{H}_a(v),\qquad
H_{\text{tot}}= H_0+\sum_{j=1}^M H(p^*_{j}),\quad
P_{\text{tot}}= P_{+}^* + P_{-}^*.
\end{equation}
Imposing the quantisation condition~\eqref{eq:TBAsing} we have that
\begin{equation}
(R\mp \tfrac{k}{2\pi} P_{\mp}^*)\, p_j^* = \pm 2\pi \nu_j,\qquad \nu_j\in\mathbb{N},
\end{equation}
where we already picked the sign of $\nu_j$. As before, we restrict to the ``unflowed'' sector, see ref.~\cite{Dei:2018mfl} for a discussion of the mode-number identification in the spectrally-flowed sectors. Bearing in mind the level-matching condition we hence have
\begin{equation}
P_+^*=- P_-^*,\qquad 
(R+ \tfrac{k}{2\pi} P_{+}^*)\, P_{+}^* = 2\pi \mathcal{N},\qquad
H_{\text{tot}}= H_0 +\frac{k}{\pi}P_{+}^*+ \hat{\mu}_{\text{tot}},
\end{equation}
which corrects the asymptotic Bethe ansatz result~\eqref{eq:ABAsol} simply by shifting the vacuum energy to the lowest-energy state, yielding a perfect agreement with the WZW result---see appendix~\ref{app:WZW}.

\section{Conclusions}
\label{sec:conclusions}

We saw that the bosonic tree-level worldsheet S~matrix for pure-NSNS $\AdSSS$ is proportional to the identity, and it takes the ``shockwave'' form~\cite{Dray:1984ha} already found for flat-space strings~\cite{Dubovsky:2012wk} and for pure-NSNS $\AdS{3}\times\S^3\times\T^4$~\cite{Hoare:2013ida}.
Based on this (and on a previous analysis of the Green-Schwarz fermion contributions~\cite{Dei:2018yth}) we made a proposal for the \textit{exact} worldsheet S~matrix. We took it to be diagonal, and simply given by the exponentiation of the tree-level result. To check our proposal, we wrote down the mirror TBA equations and solved them analytically in the sector corresponding to highest-weight representations of the worldsheet WZW model, \textit{cf.}\ ref.~\cite{Eberhardt:2017fsi,Eberhardt:2017pty}, finding a perfect match. In the process, we noted that the mirror TBA boils down to a set of equations which can be though of as spin-chain Bethe ansatz---like for $\AdS{3}\times\S^3\times\T^4$~\cite{Dei:2018mfl}. There is an important subtlety here, with respect to that case: generically, the spin-chain vacuum is only a \textit{pseudo-vacuum}, \textit{i.e.}\ it is not the lowest-energy state of the theory. Lower-energy states can by acting with two \textit{negative-energy} fermion zero-modes. Only in the case where the spin-chain pseudo-vacuum describes a BPS state we find that this is a true (albeit degenerate) vacuum.

One immediate question is whether our conjectured S~matrix, and therefore the mirror TBA, reproduces the whole spectrum of the theory as predicted by the worldsheet WZW model---not just the highest-weight representations. In particular, new representations can be obtained from the highest-weight ones by means of \textit{spectral flow}~\cite{Maldacena:2000hw}. We expect these to appear when we relax some assumptions that we have made in solving the mirror TBA equations. In ref.~\cite{Dei:2018mfl} we discussed at some length how, when the effective size of the worldsheet $R_{\text{eff}}$ is comparable with the WZW level~$k$, spectrally-flowed sectors emerge. More specifically, the $m$-th spectrally flowed sector appears when $m\leq R_{\text{eff}}/k <(m+1)$. In the case at hand, we expect a similar qualitative structure, with the important caveat that different bounds arise depending on the mass of each particle, $m\leq (\mu R_{\text{eff}})/k<(m+1)$. This is due to the fact that there are \textit{three} WZW levels, one for each Ka\v{c}-Moody algebra; indeed the structure of the spectral flow for the $\AdSSS$ WZW model is rather intricate, see \textit{e.g.}\ ref.~\cite{Eberhardt:2017pty} for a detailed discussion.

An additional family of states comes from \textit{continuous} representations~\cite{Maldacena:2000hw}, which seem to be qualitatively different from the discrete one (spectrally flowed or not). Indeed these representations have recently attracted much attention, both for $\AdS{3}\times\S^3\times\T^4$~\cite{Giribet:2018ada, Gaberdiel:2018rqv,Eberhardt:2018ouy} and $\AdSSS$~\cite{Gaberdiel:2018rqv}. The role of these representations in describing the theory's spectrum seems especially  crucial when the level of the WZW model is small; then the dual appears to be a symmetric-product orbifold CFT~\cite{Giribet:2018ada,Gaberdiel:2018rqv,Eberhardt:2018ouy}. At this stage it is not clear how these states emerge from the worldsheet S~matrix in light-cone gauge, even in the simpler case of~$\AdS{3}\times\S^3\times\T^4$. This is certainly something that should be understood, and we hope to return to this question in the near future.

Another interesting feature of stringy WZW models is that not only their spectrum, but also their correlation functions are substantially simpler than for RR string backgrounds~\cite{Teschner:1997ft,Teschner:1999ug, Maldacena:2001km, Giribet:2007wp, Cardona:2009hk,Cardona:2010qf}. Integrability also allows the computation of stringy correlation functions, at least in the planar limit: Basso, Komatsu, and Vieira recently proposed an ``hexagon'' formalism~\cite{Basso:2015zoa} for the computation of three-point functions, which was later extended to four-point~\cite{Eden:2016xvg, Fleury:2016ykk, Basso:2017khq} and well as higher-point~\cite{Fleury:2017eph} correlation functions and, at least in some special cases, to non-planar corrections~\cite{Eden:2017ozn, Bargheer:2017nne, Bargheer:2018jvq}. That proposal has been developed in the context of the $\AdS{5}/\CFT_4$ duality. Despite many impressive advances, a major stumbling block remains the systematic incorporation of finite-size (``wrapping'') corrections. For three-point functions, this can be done order-by-order~\cite{Eden:2015ija, Basso:2015eqa,Basso:2017muf}---by formulae similar to those developed by L\"uscher~\cite{Luscher:1985dn,Luscher:1986pf}---while for higher-point correlators many puzzles remain~\cite{Eden:2018vug}. In any case, nothing like a ``mirror TBA'' for the hexagon program has been developed yet. Perhaps the pure-NSNS backgrounds---with their almost-trivial wrapping structure---may prove to be the ideal playground for advancing the hexagon~program.

Finally, an important ingredient in recognising the integrable structure of pure-NSNS backgrounds was their relation to $T\bar{T}$ deformations~\cite{Smirnov:2016lqw,Cavaglia:2016oda} of free theories~\cite{Baggio:2018gct}. It would be interesting to study in greater detail this relation, and explore recently-proposed ideas concerning the worldsheet description of the $T\bar{T}$ deformation \textit{of the dual CFT${}_2$}~\cite{Giveon:2017nie,Giveon:2017myj}. We hope to return to some of these questions in the near future.

\section*{Acknowledgements}
We thank Sergei Dubovsky, Lorentz Eberhardt, Christian Ferko, Matthias Gaberdiel, Ben Hoare, Roberto Tateo and Linus Wulff for useful discussions related to this work. We are especially grateful to Gleb Arutyunov and Riccardo Borsato for comments on an preliminary version of this manuscript.
This work is partially supported through a research
grant of the Swiss National Science Foundation, as well as by the NCCR SwissMAP, funded by the Swiss National Science Foundation.

\newpage
\appendix 

\section{Details of the near BMN expansion}
\label{app:details-BMN-expansion}
Solving equation~\eqref{eq:lightcone-gauge} for the worldsheet metric, together with the constraint ${\text{det} \gamma = -1}$, we find up to cubic order 
\begin{equation}
\begin{aligned}
\gamma^{00} = &  -1  + (1-a)  \sin \varphi \sin \vartheta \ (x_6 \pri x_7 - x_7 \pri x_6 ) \\
& + (1-a)  \cos \varphi \cos \vartheta \ (y_3 \pri y_4 - y_4 \pri y_3 ) -a(z_1 \pri z_2 - z_2 \pri z_1) \\
&  - (\tfrac{1}{2}-a) \bigl(\dot w^2 + \dot \psi^2 + \dot x^2 + \dot y^2 + \dot z^2 + \pri w^2 + \pri \psi^2 + \pri x^2 + \pri y^2 + \pri z^2 \bigr)\\ 
& - \tfrac{1}{2} \bigl(  \sin^2 \varphi \sin^2 \vartheta \ x^2 + \cos^2 \varphi \cos^2 \vartheta \ y^2 - z^2 \bigr)\\
&  - \tfrac{a}{2} \sin (2 \vartheta)(\sin^2 \varphi \ x^2 - \cos^2 \varphi \ y^2) \dot \psi \,,
\end{aligned}
\end{equation}
\begin{equation}
\begin{aligned}
\gamma^{11} = & \ 1  + (1-a)  \sin \varphi \sin \vartheta \ (x_6 \pri x_7 - x_7 \pri x_6 ) \\
& + (1-a)  \cos \varphi \cos \vartheta \ (y_3 \pri y_4 - y_4 \pri y_3 ) -a(z_1 \pri z_2 - z_2 \pri z_1) \\
&  - (\tfrac{1}{2}-a) \bigl(\dot w^2 + \dot \psi^2 + \dot x^2 + \dot y^2 + \dot z^2 + \pri w^2 + \pri \psi^2 + \pri x^2 + \pri y^2 + \pri z^2 \bigr)\\ 
& - \tfrac{1}{2} \bigl(  \sin^2 \varphi \sin^2 \vartheta \ x^2 + \cos^2 \varphi \cos^2 \vartheta \ y^2 - z^2 \bigr)\\
&  - \tfrac{a}{2} \sin (2 \vartheta)(\sin^2 \varphi \ x^2 - \cos^2 \varphi \ y^2) \dot \psi \,, 
\end{aligned}
\end{equation}
\begin{equation}
\begin{aligned}
\gamma^{01} = & \ (1-a) \sin \varphi \sin \vartheta \ ( x_7 \dot x_6 - x_6 \dot x_7 ) \\
& + (1-a) \cos \varphi \cos \vartheta \ ( y_4 \dot y_3 - y_3 \dot y_4 ) - a(z_2 \dot z_1 - z_1 \dot z_2) \\
& + (1-2a) ( \dot w \pri w + \dot \psi \pri \psi + \dot x_6 \pri x_6 + \dot x_7 \pri x_7 + \dot y_3 \pri y_3  + \dot y_4 \pri y_4 + \dot z_1 \pri z_1 + \dot z_2 \pri z_2 )  \\
& +\tfrac{a}{2}\sin(2 \vartheta) (\sin^2 \varphi \ x^2 - \cos^2 \varphi \ y^2) \pri \psi\ \,, 
\end{aligned}
\end{equation} 
where we adopted the shorthand notation
\begin{equation}
x^2 \equiv \ x_6^2 + x_7^2 \,, \qquad \dot x^2 \equiv  \ \dot x_6^2 + \dot x_7^2  \,, \qquad \pri x^2 \equiv \ \pri x_6^2 + \pri x_7^2 \,, 
\end{equation}
and similarly for the fields $y_3, y_4$ and $z_1, z_2$. The Virasoro constraints~\eqref{eq:virasoro-constraints-bis} give, up to cubic order 
\begin{equation}
\begin{aligned}
\pri x^- = & - \tfrac{1}{2} \bigl( \dot w \pri w + \dot \psi \pri \psi + \dot x_6 \pri x_6 + \dot x_7 \pri x_7 + \dot y_3 \pri y_3  + \dot y_4 \pri y_4 + \dot x_1 \pri x_1 + \dot x_2 \pri x_2  \bigr) \\
&  + \tfrac{1}{4} \sin(2 \vartheta) \bigl( \sin^2 \varphi \ x^2 - \cos^2 \varphi \ y^2  \bigr) \pri \psi \,,
\end{aligned}
\end{equation}
\begin{equation}
\begin{aligned}
\dot x^- = & -\tfrac{1}{4}(\dot w^2 + \dot \psi^2 + \dot x^2 + \dot y^2 + \dot z^2 + \pri w^2 + \pri \psi^2 + \pri x^2 + \pri y^2 + \pri z^2) \\
& + \tfrac{1}{4} (\sin^2 \varphi \sin^2 \vartheta \ x^2 + \cos^2 \varphi \cos^2 \vartheta \ y^2 + z^2 ) \\
& + \tfrac{1}{4} \sin(2 \vartheta) \bigl( \sin^2 \varphi \ x^2 - \cos^2 \varphi \ y^2  \bigr) \dot \psi\,. 
\end{aligned}
\end{equation}
The results above agree with \cite{Borsato:2015mma}, where they had been derived for $\vartheta = \varphi$ and $a = \dfrac{1}{2}$. 

\section{Quartic Lagrangian}
\label{app:quartic-L}

The quartic Lagrangian is a sum of terms each containing at most two different fields:
\begin{equation}
\mathcal{L}_4 = \ \mathcal{L}_4^{\scriptscriptstyle{WW}} + \mathcal{L}_4^{\scriptscriptstyle{XX}} + \mathcal{L}_4^{\scriptscriptstyle{YY}}+ \mathcal{L}_4^{\scriptscriptstyle{ZZ}} + \mathcal{L}_4^{\scriptscriptstyle{WX}} + \mathcal{L}_4^{\scriptscriptstyle{WY}} + \mathcal{L}_4^{\scriptscriptstyle{WZ}} + \mathcal{L}_4^{\scriptscriptstyle{XY}} + \mathcal{L}_4^{\scriptscriptstyle{XZ}} + \mathcal{L}_4^{\scriptscriptstyle{YZ}} \,, 
\end{equation}
where using the short-hand notation $c\equiv \cos\varphi\cos\vartheta$ and $s\equiv \sin\varphi\sin\vartheta$
\begingroup
\allowdisplaybreaks
\begin{align}
\mathcal{L}_4^{\scriptscriptstyle{WW}} = & \ \dfrac{1-2a}{8} (\dot W^2 - \pri{W}^2)(\dot{\bar W}^2 - \pri{\bar W}^2) \,, \\
\mathcal{L}_4^{\scriptscriptstyle{XX}} = & \ \dfrac{1-2a}{8}\biggl[- s^4 \  X^2 \bar X^2\\
\nonumber
& \qquad \qquad \qquad + (\dot X^2 - \pri X^2 + 2 i s \ X \pri X)(\dot{\bar X}^2 - \pri{\bar X}^2 - 2 i s \ \bar X \pri{\bar X}) \\
\nonumber
& \qquad \qquad \qquad + 2 \sin^2 \varphi  \cos^2 \vartheta \ X \bar X (- \dot X \dot{\bar X} + \pri X \pri{\bar X} + i s \ \pri X \bar X \\
\nonumber
& \qquad \qquad \qquad \qquad \qquad \qquad \qquad \qquad \qquad \qquad - i s \ X \pri{\bar X} + s^2 \ X \bar X )\biggr] \\
\nonumber
& + \dfrac{1}{4} a \biggl[ i s \ X \pri X (\dot{\bar X}^2 + \sin^2 \varphi \ \bar X^2 - \pri{\bar X}^2 - 2 i s \ \bar X \pri{\bar X}) + c.c. \\
\nonumber
& \qquad \quad \  + \sin^2 \varphi \cos^2 \vartheta \ X \bar X (-2 \dot{X} \dot{\bar X} + 2 \pri{X} \pri{\bar X} + i s \ \pri X \bar X \\
\nonumber
&  \qquad \qquad \qquad \qquad \qquad \qquad \qquad \qquad \qquad \qquad -i s \ X \pri{\bar X} + 2 s^2 \ X \bar X ) \biggr] \,, \\
\mathcal{L}_4^{\scriptscriptstyle{YY}} = & \ \dfrac{1-2a}{8}\biggl[- c^4 \  Y^2 \bar Y^2\\
\nonumber
& \qquad \qquad \qquad + (\dot Y^2 - \pri Y^2 + 2 i c \ Y \pri Y)(\dot{\bar Y}^2 - \pri{\bar Y}^2 - 2 i c \ \bar Y \pri{\bar Y}) \\
\nonumber
& \qquad \qquad \qquad + 2 \cos^2 \varphi  \sin^2 \vartheta \ Y \bar Y (- \dot Y \dot{\bar Y} + \pri Y \pri{\bar Y} + i c \ \pri Y \bar Y \\
\nonumber
& \qquad \qquad \qquad \qquad \qquad \qquad \qquad \qquad \qquad \qquad - i c \ Y \pri{\bar Y} + c^2 \ Y \bar Y )\biggr] \\
\nonumber
& + \dfrac{1}{4} a \biggl[ i c \ Y \pri Y (\dot{\bar Y}^2 + \cos^2 \varphi \ \bar Y^2 - \pri{\bar Y}^2 - 2 i c \ \bar Y \pri{\bar Y}) + c.c. \\
\nonumber
& \qquad \quad \  + \cos^2 \varphi \sin^2 \vartheta \ Y \bar Y (-2 \dot{Y} \dot{\bar Y} + 2 \pri{Y} \pri{\bar Y} + i c \ \pri Y \bar Y \\
\nonumber
&  \qquad \qquad \qquad \qquad \qquad \qquad \qquad \qquad \qquad \qquad -i c \ Y \pri{\bar Y} + 2 c^2 \ Y \bar Y ) \biggr] \,, \\
\mathcal{L}_4^{\scriptscriptstyle{ZZ}} = & \ \dfrac{1-2a}{8} \biggl[(\dot Z^2 - \pri{Z}^2)(\dot{\bar Z}^2 - \pri{\bar Z}^2) - Z \bar Z (Z + 2 i \pri Z)(\bar Z - 2 i \pri{\bar Z})\biggr] \\
\nonumber
& \ -\dfrac{1}{4} a \biggl[ i Z \pri Z (\dot{\bar Z}^2 + \bar Z^2 - \pri{\bar Z}^2 - 2i \bar Z \pri{\bar Z}) + c.c.\biggr] \,, \\
\mathcal{L}_4^{\scriptscriptstyle{WX}} = & \ \dfrac{1-2a}{8}\biggl\{ -2( \pri{W} \dot{\bar W} + \dot W \pri{\bar W})(\pri X \dot{\bar X} + \dot X \pri{\bar X} + i s \ \dot X \bar X - i s \ X \dot{\bar X}) \\
\nonumber & \qquad \qquad \quad \  + 2 (\dot W \dot{\bar W} + \pri W \pri{\bar W})(\dot X \dot{\bar X} + \pri X \pri{\bar X} + i s \ \pri X \bar X \\
\nonumber
&\qquad \qquad \qquad \qquad \qquad \qquad \qquad \qquad  - i s \ X \pri{\bar X} + s^2 \ X \bar X) \biggr\} \\
\nonumber
& \ + \dfrac{1}{4} a \biggl\{ i s \ (\dot W \pri{\bar W} + \pri W \dot{\bar W})(X \dot{\bar X} - \dot X \bar X ) \\
\nonumber
& \qquad \quad \ \ + s \ (\dot W \dot{\bar W} + \pri{W} \pri{\bar W})(2 s \ X \bar X + i \pri X \bar X -i X \pri{\bar X}) \biggr\} \\
\nonumber
& \ + \dfrac{1}{8} \sin^2 \varphi \cos^2 \vartheta \ X \bar X \left[ (\dot W - \dot{\bar W})^2 - (\pri W - \pri{\bar W})^2 \right] \,, \\
\mathcal{L}_4^{\scriptscriptstyle{WY}} = & \ \dfrac{1-2a}{8}\biggl\{ -2( \pri{W} \dot{\bar W} + \dot W \pri{\bar W})(\pri Y \dot{\bar Y} + \dot Y \pri{\bar Y} + i c \ \dot Y \bar Y - i c \ Y \dot{\bar Y}) \\
\nonumber
& \qquad \qquad \quad \  + 2 (\dot W \dot{\bar W} + \pri W \pri{\bar W})(\dot Y \dot{\bar Y} + \pri Y \pri{\bar Y} + i c \ \pri Y \bar Y \\
\nonumber
&\qquad \qquad \qquad \qquad \qquad \qquad \qquad \qquad  - i c \ Y \pri{\bar Y} + c^2 \ Y \bar Y) \biggr\} \\
\nonumber
& \ + \dfrac{1}{4} a \biggl\{ i c \ (\dot W \pri{\bar W} + \pri W \dot{\bar W})(Y \dot{\bar Y} - \dot Y \bar Y ) \\
\nonumber
& \qquad \quad \ \ + c \ (\dot W \dot{\bar W} + \pri{W} \pri{\bar W})(2 c \ Y \bar Y + i \pri Y \bar Y -i Y \pri{\bar Y}) \biggr\} \\
\nonumber
& \ + \dfrac{1}{8} \cos^2 \varphi \sin^2 \vartheta \ Y \bar Y \left[ (\dot W - \dot{\bar W})^2 - (\pri W - \pri{\bar W})^2 \right]  \,, \\
\mathcal{L}_4^{\scriptscriptstyle{WZ}} = & \ -\dfrac{1-2a}{4}\biggl[ ( \pri{W} \dot{\bar W}+ \dot W \pri{\bar W})( \pri{Z} \dot{\bar Z} + \dot Z \pri{\bar Z}) -(\dot W \dot{\bar W} + \pri W \pri{\bar W})(\dot Z \dot{\bar Z} + \pri Z \pri{\bar Z} - Z \bar Z ) \biggr] \\ 
\nonumber
& \ - \dfrac{1}{4} a \biggl[ (\dot W \dot{\bar W} + \pri W \pri{\bar W})(2 Z \bar Z + i \pri Z \bar Z - i Z \pri{\bar Z}) + i( \pri{W} \dot{\bar W}+ \dot W \pri{\bar W})(\dot{\bar Z} Z - \dot Z \bar Z)  \biggr]  \,, \\
\mathcal{L}_4^{\scriptscriptstyle{XY}} = & \ - \dfrac{1-2a}{4} (\pri X \dot{\bar X} + \dot X \pri{\bar X} + i s \ \dot X \bar X - i s \ X \dot{\bar X}) \times \\
\nonumber
& \qquad \qquad \times(\pri Y \dot{\bar Y} + \dot Y \pri{\bar Y} + i c \ \dot Y \bar Y - i c \ Y \dot{\bar Y}) \\
\nonumber
& \ + \dfrac{1}{4}(1-2a)(\dot X \dot{\bar X} + \pri X \pri{\bar X} + i s \ \pri X \bar X - i s \ X \pri{\bar X} + s^2 \ X \bar X) \times \\
\nonumber
& \qquad \qquad \times (\dot Y \dot{\bar Y} + \pri Y \pri{\bar Y} + i c \ \pri Y \bar Y - i c \ Y \pri{\bar Y} + c^2 \ Y \bar Y) \\
\nonumber & \ + \dfrac{1}{4} (1-2a) c\,s \ (\pri X \bar X - X \pri{\bar X} - 2 i s \ X \bar X) \times \\
\nonumber
& \qquad \qquad \qquad \qquad \qquad \qquad \qquad \qquad \qquad \times(\pri Y \bar Y - Y \pri{\bar Y} - 2 i c \ Y \bar Y)  \\ 
\nonumber
& \ -\dfrac{1-2a}{4} c\,s \ (X \dot{\bar X} - \dot X \bar X)(Y \dot{\bar Y} - \dot Y \bar Y) \\
\nonumber
& \ + \dfrac{1}{4}a c \ (\dot X \dot{\bar X} + \pri X \pri{\bar X} - s^2 \ X \bar X)(2 c \ Y \bar Y + i \pri Y \bar Y - i Y \pri{\bar Y} ) \\
\nonumber
& \ + \dfrac{1}{4}a s \ (2 s \ X \bar X + i \pri X \bar X - i X \pri{\bar X} )(\dot Y \dot{\bar Y} + \pri Y \pri{\bar Y} - c^2 \ Y \bar Y) \\
\nonumber
& \ + \dfrac{i}{4}a c \ (\pri X \dot{\bar X} + \dot X \pri{\bar X})(Y \dot{\bar Y} - \dot Y \bar Y ) + \dfrac{i}{4}a s \ ( X \dot{\bar X} - \dot X \bar X)(\pri Y \dot{\bar Y} + \dot Y \pri{\bar Y} ) \,, \\
\mathcal{L}_4^{\scriptscriptstyle{XZ}} = & \ - \dfrac{1-2a}{4}\biggl[ (\pri X \dot{\bar X} + \dot X \pri{\bar X} + i s \ \dot X \bar X - i s \ X \dot{\bar X})( \pri{Z} \dot{\bar Z} + \dot Z \pri{\bar Z}) \\ 
\nonumber
& \qquad \qquad \qquad  \ -(\dot X \dot{\bar X} + \pri X \pri{\bar X} + i s \ \pri X \bar X - i s \ X \pri{\bar X} + s^2 \ X \bar X) \times \\
\nonumber
& \qquad \qquad \qquad \qquad \qquad \qquad \qquad \qquad \qquad \qquad  \times (\dot Z \dot{\bar Z} + \pri Z \pri{\bar Z} - Z \bar Z ) \biggr] \\
\nonumber
& \ - \dfrac{i}{4} a (\pri X \dot{\bar X} + \dot X \pri{\bar X})(Z \dot{\bar Z} - \dot Z \bar Z) + \dfrac{i}{4}a s \ (X \dot{\bar X} - \dot X \bar X)(\dot Z \pri{\bar Z} + \pri Z \dot{\bar Z}) \\
\nonumber
& \ - \dfrac{1}{4} a (\dot X \dot{\bar X} + \pri X \pri{\bar X} - s^2 \ X \bar X )(2 Z \bar Z + i \pri Z \bar Z  - i Z \pri{\bar Z}) \\
\nonumber
& \ + \dfrac{1}{4}a s \ (i \pri X \bar X - i X \pri{\bar X} + 2 s \ X \bar X) (\dot Z \dot{\bar Z} + \pri Z \pri{\bar Z} - Z \bar Z) \,, \\
\mathcal{L}_4^{\scriptscriptstyle{YZ}} = & \ - \dfrac{1-2a}{4}\biggl[ (\pri Y \dot{\bar Y} + \dot Y \pri{\bar Y} + i c \ \dot Y \bar Y - i c \ Y \dot{\bar Y})( \pri{Z} \dot{\bar Z} + \dot Z \pri{\bar Z}) \\ 
\nonumber
& \qquad \qquad \qquad  \ -(\dot Y \dot{\bar Y} + \pri Y \pri{\bar Y} + i c \ \pri Y \bar Y - i c \ Y \pri{\bar Y} + c^2 \ Y \bar Y) \times \\
\nonumber
& \qquad \qquad \qquad \qquad \qquad \qquad \qquad \qquad \qquad  \times (\dot Z \dot{\bar Z} + \pri Z \pri{\bar Z} - Z \bar Z ) \biggr] \\
\nonumber
& \ - \dfrac{i}{4} a (\pri Y \dot{\bar Y} + \dot Y \pri{\bar Y})(Z \dot{\bar Z} - \dot Z \bar Z) + \dfrac{i}{4}a c \ (Y \dot{\bar Y} - \dot Y \bar Y)(\dot Z \pri{\bar Z} + \pri Z \dot{\bar Z}) \\
\nonumber
& \ - \dfrac{1}{4} a (\dot Y \dot{\bar Y} + \pri Y \pri{\bar Y} - c^2 \ Y \bar Y )(2 Z \bar Z + i \pri Z \bar Z  - i Z \pri{\bar Z}) \\
\nonumber
& \ + \dfrac{1}{4}a c \ (i \pri Y \bar Y - i Y \pri{\bar Y} + 2 c \ Y \bar Y) (\dot Z \dot{\bar Z} + \pri Z \pri{\bar Z} - Z \bar Z) \,.
\end{align}
\endgroup

\section{Feynmal rules}
\label{app:feynman-rules}

From eqs. (\ref{eq:L2}-\ref{eq:L4}) one can read off the following Feynman rules. We have associated dashed black lines to the $W$ field and straight red, blue and green lines respectively for $X, Y$ and $Z$ fields. In the following we use the short-hand notation $c\equiv \cos\varphi\cos\vartheta$ and $s\equiv \sin\varphi\sin\vartheta$. For the propagators we find

\begin{fmffile}{Feynman-rule-propagators}
\begin{equation}
\begin{aligned}
& \begin{gathered}
\begin{fmfgraph*}(50,30)
  \fmfleft{i1}
  \fmfright{o1}
  \fmf{scalar}{i1,o1}
   \marrow{a}{up}{top}{$p$}{i1,o1}
  \end{fmfgraph*}
\end{gathered} = \dfrac{i}{\omega^2 - p^2 + i \varepsilon} \,,  & 
&  \begin{gathered}
\begin{fmfgraph*}(50,30)
  \fmfleft{i1}
  \fmfright{o1}
  \fmf{fermion,foreground=(1,,0.1,,0.1)}{i1,o1}
   \marrow{a}{up}{top}{$p$}{i1,o1}
  \end{fmfgraph*}
\end{gathered} =  \dfrac{i}{\omega^2 - (p + s)^2  + i \varepsilon} \,, \\
& \begin{gathered}
 \begin{fmfgraph*}(50,30)
  \fmfleft{i1}
  \fmfright{o1}
  \fmf{fermion,foreground=(0.1,,0.1,,1)}{i1,o1}
   \marrow{a}{up}{top}{$p$}{i1,o1}
  \end{fmfgraph*}
\end{gathered} = \dfrac{i}{\omega^2 - (p + c)^2  + i \varepsilon} \,, & \qquad 
&  \begin{gathered}
\begin{fmfgraph*}(50,30)
  \fmfleft{i1}
  \fmfright{o1}
  \fmf{fermion,foreground=(0.1,,1,,0.1)}{i1,o1}
  \marrow{a}{up}{top}{$p$}{i1,o1}
  \end{fmfgraph*}
  \end{gathered} = \dfrac{i}{\omega^2 - (p + 1)^2  + i \varepsilon} \,. 
\end{aligned}
\end{equation}
\end{fmffile}  

\noindent The cubic vertices are

\begin{fmffile}{Feynman-rule-cubic}
\begin{align}
\begin{gathered}
\begin{fmfgraph*}(60,60)
  \fmfleft{i1,i2}
  \fmfright{o1}
  \fmf{fermion,foreground=(1,,0.1,,0.1)}{i1,v1,i2}
  \fmf{scalar}{v1,o1}
    \Marrow{a}{right}{bot}{$\quad p_1$}{i1,v1}{7} 
    \Marrow{b}{right}{rt}{$\ p_2$}{v1,i2}{7}
    \Marrow{c}{down}{bot}{$p_3$}{v1,o1}{7}
	        \end{fmfgraph*}
\end{gathered} = 
\begin{gathered}
\begin{fmfgraph*}(60,60)
  \fmfleft{i1,i2}
  \fmfright{o1}
  \fmf{fermion,foreground=(1,,0.1,,0.1)}{i1,v1,i2}
  \fmf{scalar}{o1,v1}
    \Marrow{a}{right}{bot}{$\quad p_1$}{i1,v1}{7} 
    \Marrow{b}{right}{rt}{$\ p_2$}{v1,i2}{7}
    \Marrow{c}{down}{bot}{$p_3$}{o1,v1}{7}
	        \end{fmfgraph*}
\end{gathered} = & -\dfrac{i}{4}\sin \varphi \cos \vartheta \bigl[ p_3(\omega_1 + \omega_2) - \omega_3(p_1 + p_2 + 2 s)\bigr] \,, \\
\begin{gathered}
\begin{fmfgraph*}(60,60)
  \fmfleft{i1,i2}
  \fmfright{o1}
  \fmf{fermion,foreground=(0.1,,0.1,,1)}{i1,v1,i2}
  \fmf{scalar}{v1,o1}
    \Marrow{a}{right}{bot}{$\quad p_1$}{i1,v1}{7} 
    \Marrow{b}{right}{rt}{$\ p_2$}{v1,i2}{7}
    \Marrow{c}{down}{bot}{$p_3$}{v1,o1}{7}
	        \end{fmfgraph*}
\end{gathered} = 
\begin{gathered}
\begin{fmfgraph*}(60,60)
  \fmfleft{i1,i2}
  \fmfright{o1}
  \fmf{fermion,foreground=(0.1,,0.1,,1)}{i1,v1,i2}
  \fmf{scalar}{o1,v1}
    \Marrow{a}{right}{bot}{$\quad p_1$}{i1,v1}{7} 
    \Marrow{b}{right}{rt}{$\ p_2$}{v1,i2}{7}
    \Marrow{c}{down}{bot}{$p_3$}{o1,v1}{7}
	        \end{fmfgraph*}
\end{gathered} = & \  \dfrac{i}{4}\sin \vartheta \cos \varphi \bigl[ p_3(\omega_1 + \omega_2) - \omega_3(p_1 + p_2 + 2 c )\bigr] \,.
\end{align}
\end{fmffile}  

\noindent For the quartic vertices we find

\begingroup
\allowdisplaybreaks
\begin{fmffile}{Feynman-rule-quartic}
\begin{align}
\begin{gathered}
	\begin{fmfgraph*}(60,60)
  \fmfleft{i1,i2}
  \fmfright{o1,o2}
  \fmf{scalar}{v1,o1}
  \fmf{scalar}{v1,o2}
  \fmf{scalar}{i1,v1}
  \fmf{scalar}{i2,v1}
    \Marrow{a}{left}{bot}{$p_1 \ $}{i2,v1}{7} 
    \Marrow{b}{left}{top}{$p_2 \ \ \ $}{i1,v1}{7}
    \Marrow{c}{right}{bot}{$\ \ \ p_3$}{v1,o2}{7}
    \Marrow{d}{right}{top}{$\ \ \ p_4$}{v1,o1}{7}
	        \end{fmfgraph*}
  \end{gathered}
   \ \ = & \ \dfrac{i}{2} (1-2a) \ (p_1 p_2 - \omega_1 \omega_2)(p_3 p_4 - \omega_3 \omega_4) \,, \\ 
\begin{gathered}
	\begin{fmfgraph*}(60,60)
  \fmfleft{i1,i2}
  \fmfright{o1,o2}
  \fmf{fermion,foreground=(1,,0.1,,0.1)}{v1,o1}
  \fmf{fermion,foreground=(1,,0.1,,0.1)}{v1,o2}
  \fmf{fermion,foreground=(1,,0.1,,0.1)}{i1,v1}
  \fmf{fermion,foreground=(1,,0.1,,0.1)}{i2,v1}
    \Marrow{a}{left}{bot}{$p_1 \ $}{i2,v1}{7} 
    \Marrow{b}{left}{top}{$p_2 \ \ \ $}{i1,v1}{7}
    \Marrow{c}{right}{bot}{$\ \ \ p_3$}{v1,o2}{7}
    \Marrow{d}{right}{top}{$\ \ \ p_4$}{v1,o1}{7}
	        \end{fmfgraph*}
  \end{gathered}
   \ \  = &  - \dfrac{i}{2} (1-2a) \ s^4 \\ \nonumber
   \\[-1.4cm] \nonumber & + \dfrac{i}{2}(1-2a) \bigl[(p_1 + s)(p_2 + s) - \omega_1 \omega_2 - s^2 \bigr] \times \\ \nonumber
	& \qquad \qquad \qquad \qquad \qquad \qquad \times \bigl[(p_3 + s)(p_4 + s) - \omega_3 \omega_4 - s^2 \bigr] \\ \nonumber
    &  + \dfrac{i}{4} (1-2a) \sin^2 \varphi \cos^2 \vartheta \ \bigl[ (p_1+p_2)(p_3+p_4) - (\omega_1 + \omega_2)(\omega_3 + \omega_4) \\ \nonumber
    & \qquad \qquad \qquad \qquad \qquad \qquad + 4 s^2 + 2 s \ (p_1 + p_2 + p_3 + p_4) \bigr] \\ \nonumber
    &  +\dfrac{i a}{2} \ s \ \bigl\{ (p_1 + p_2) [(p_3 + s)(p_4 + s) - \omega_3 \omega_4] \\ \nonumber
    & \qquad \qquad \qquad \quad + (p_3 + p_4) [(p_1 + s)(p_2 + s) - \omega_1 \omega_2] \bigr\} \\ \nonumber
    &  + \dfrac{i a}{2} \sin^2 \varphi \cos^2 \vartheta \ [(p_1 + p_2 + 2 s)(p_3 + p_4 + 2 s) + \\ \nonumber 
		& \qquad \qquad \qquad \qquad \qquad \qquad \qquad \qquad \qquad - (\omega_1 + \omega_2)(\omega_3 + \omega_4)] \,, \\
\begin{gathered}
	\begin{fmfgraph*}(60,60)
  \fmfleft{i1,i2}
  \fmfright{o1,o2}
  \fmf{fermion,foreground=(0.1,,0.1,,1)}{v1,o1}
  \fmf{fermion,foreground=(0.1,,0.1,,1)}{v1,o2}
  \fmf{fermion,foreground=(0.1,,0.1,,1)}{i1,v1}
  \fmf{fermion,foreground=(0.1,,0.1,,1)}{i2,v1}
    \Marrow{a}{left}{bot}{$p_1 \ $}{i2,v1}{7} 
    \Marrow{b}{left}{top}{$p_2 \ \ \ $}{i1,v1}{7}
    \Marrow{c}{right}{bot}{$\ \ \ p_3$}{v1,o2}{7}
    \Marrow{d}{right}{top}{$\ \ \ p_4$}{v1,o1}{7}
	        \end{fmfgraph*}
  \end{gathered}
   \ \  = &  - \dfrac{i}{2} (1-2a) \ c^4 \\ \nonumber
   \\[-1.4cm] \nonumber & + \dfrac{i}{2}(1-2a) \bigl[(p_1 + c)(p_2 + c) - \omega_1 \omega_2 - c^2 \bigr] \times \\ \nonumber
    & \qquad \qquad \qquad \qquad \qquad \qquad \times	\bigl[(p_3 + c)(p_4 + c) - \omega_3 \omega_4 - c^2 \bigr] \\ \nonumber
    & \ + \dfrac{i}{4} (1-2a) \cos^2 \varphi \sin^2 \vartheta \ \bigl[ (p_1+p_2)(p_3+p_4) - (\omega_1 + \omega_2)(\omega_3 + \omega_4) \\ \nonumber
    & \qquad \qquad \qquad \qquad \qquad \qquad  + 4 c^2 + 2 c \ (p_1 + p_2 + p_3 + p_4) \bigr] \\ \nonumber
    & \ +\dfrac{i a}{2} \ c \ \bigl\{ (p_1 + p_2) [(p_3 + c)(p_4 + c) - \omega_3 \omega_4] \\ \nonumber
    & \qquad \qquad \qquad \quad \  + (p_3 + p_4) [(p_1 + c)(p_2 + c) - \omega_1 \omega_2] \bigr\} \\ \nonumber
    & \ + \dfrac{i a}{2} \cos^2 \varphi \sin^2 \vartheta \ [(p_1 + p_2 + 2 c)(p_3 + p_4 + 2 c) + \\ \nonumber 
		& \qquad \qquad \qquad \qquad \qquad \qquad \qquad \qquad \qquad- (\omega_1 + \omega_2)(\omega_3 + \omega_4)] \,, \\
\begin{gathered}
	\begin{fmfgraph*}(60,60)
  \fmfleft{i1,i2}
  \fmfright{o1,o2}
  \fmf{fermion,foreground=(0.1,,1,,0.1)}{v1,o1}
  \fmf{fermion,foreground=(0.1,,1,,0.1)}{v1,o2}
  \fmf{fermion,foreground=(0.1,,1,,0.1)}{i1,v1}
  \fmf{fermion,foreground=(0.1,,1,,0.1)}{i2,v1}
    \Marrow{a}{left}{bot}{$p_1 \ $}{i2,v1}{7} 
    \Marrow{b}{left}{top}{$p_2 \ \ \ $}{i1,v1}{7}
    \Marrow{c}{right}{bot}{$\ \ \ p_3$}{v1,o2}{7}
    \Marrow{d}{right}{top}{$\ \ \ p_4$}{v1,o1}{7}
	        \end{fmfgraph*}
  \end{gathered}
   \ \ = & \ \dfrac{i}{2}(1-2a)[(p_1 p_2 - \omega_1 \omega_2)(p_3 p_4 - \omega_3 \omega_4) + \\ \nonumber
	\\[-1.4cm] \nonumber & \qquad \qquad \qquad \qquad \qquad \qquad - (p_1 + p_2 +1)(p_3 + p_4 +1)] \\ \nonumber
	  &  - \dfrac{i}{2}a \bigl\{(p_1 + p_2)[(p_3 + 1)(p_4 + 1) - \omega_3 \omega_4] \\ \nonumber
	& \qquad \qquad + (p_3 + p_4)[(p_1 + 1)(p_2 + 1) - \omega_1 \omega_2] \bigr\} \,, \\
\begin{gathered}
	\begin{fmfgraph*}(60,60)
  \fmfleft{i1,i2}
  \fmfright{o1,o2}
  \fmf{fermion,foreground=(1,,0.1,,0.1)}{o1,v1,o2}
  \fmf{scalar}{i1,v1,i2}
    \Marrow{a}{left}{bot}{$p_1 \ $}{v1,i2}{7} 
    \Marrow{b}{left}{top}{$p_2 \ \ \ $}{i1,v1}{7}
    \Marrow{c}{right}{bot}{$\ \ \ p_3$}{v1,o2}{7}
    \Marrow{d}{right}{top}{$\ \ \ p_4$}{o1,v1}{7}
	        \end{fmfgraph*}
  \end{gathered}
   \ \ = & \ \dfrac{i}{4} (1-2a) \sin^2 \varphi \cos^2 \vartheta \ (p_1 p_2 - \omega_1 \omega_2) \\ \nonumber
  \\[-1.4cm] \nonumber & \ - \dfrac{i}{4}(1-2a) (p_1 \omega_2 + p_2 \omega_1) \bigl[ \omega_3 (p_4 + s) + \omega_4 (p_3 + s)  \bigr]   \\ \nonumber
  & \ + \dfrac{i}{4}(1-2a)(\omega_1 \omega_2 + p_1 p_2) \bigl[ \omega_3 \omega_4 + (p_3 + s)(p_4 + s) \bigr] \\ \nonumber
  & \ + \dfrac{i}{4} \ a \ s \ [(\omega_1 \omega_2 + p_1 p_2)(p_3 + p_4 + 2 s)- (p_1 \omega_2 + p_2 \omega_1)(\omega_3 + \omega_4)] \\ \nonumber
  & \ + \dfrac{i}{2} \ a  \sin^2 \varphi \cos^2 \vartheta \ (p_1 p_2 - \omega_1 \omega_2)  \,, \\ \nonumber
   & \\
\begin{gathered}
	\begin{fmfgraph*}(60,60)
  \fmfleft{i1,i2}
  \fmfright{o1,o2}
  \fmf{fermion,foreground=(1,,0.1,,0.1)}{o1,v1,o2}
  \fmf{scalar}{i1,v1}
  \fmf{scalar}{i2,v1}
    \Marrow{a}{left}{bot}{$p_1 \ $}{i2,v1}{7} 
    \Marrow{b}{left}{top}{$p_2 \ \ \ $}{i1,v1}{7}
    \Marrow{c}{right}{bot}{$\ \ \ p_3$}{v1,o2}{7}
    \Marrow{d}{right}{top}{$\ \ \ p_4$}{o1,v1}{7}
	        \end{fmfgraph*}
  \end{gathered} \ \  = & \ \
\begin{gathered}
	\begin{fmfgraph*}(60,60)
  \fmfleft{i1,i2}
  \fmfright{o1,o2}
  \fmf{fermion,foreground=(1,,0.1,,0.1)}{o1,v1,o2}
  \fmf{scalar}{v1,i1}
  \fmf{scalar}{v1,i2}
    \Marrow{a}{left}{bot}{$p_1 \ $}{v1,i2}{7} 
    \Marrow{b}{left}{top}{$p_2 \ \ \ $}{v1,i1}{7}
    \Marrow{c}{right}{bot}{$\ \ \ p_3$}{v1,o2}{7}
    \Marrow{d}{right}{top}{$\ \ \ p_4$}{o1,v1}{7}
	        \end{fmfgraph*}
  \end{gathered}
   \ \ = \ \dfrac{i}{4} \sin^2 \varphi \cos^2 \vartheta \ (p_1 p_2 - \omega_1 \omega_2) \,,    \\
\begin{gathered}
	\begin{fmfgraph*}(60,60)
  \fmfleft{i1,i2}
  \fmfright{o1,o2}
  \fmf{fermion,foreground=(0.1,,0.1,,1)}{o1,v1,o2}
  \fmf{scalar}{i1,v1,i2}
    \Marrow{a}{left}{bot}{$p_1 \ $}{v1,i2}{7} 
    \Marrow{b}{left}{top}{$p_2 \ \ \ $}{i1,v1}{7}
    \Marrow{c}{right}{bot}{$\ \ \ p_3$}{v1,o2}{7}
    \Marrow{d}{right}{top}{$\ \ \ p_4$}{o1,v1}{7}
	        \end{fmfgraph*}
  \end{gathered}
    \ \ = & \ \dfrac{i}{4} (1-2a) \cos^2 \varphi \sin^2 \vartheta \ (p_1 p_2 - \omega_1 \omega_2) \\ \nonumber
  \\[-1.4cm] \nonumber & \ - \dfrac{i}{4}(1-2a) (p_1 \omega_2 + p_2 \omega_1) \bigl[ \omega_3 (p_4 + c) + \omega_4 (p_3 + c)  \bigr]   \\ \nonumber
  & \ + \dfrac{i}{4}(1-2a)(\omega_1 \omega_2 + p_1 p_2) \bigl[ \omega_3 \omega_4 + (p_3 + c)(p_4 + c) \bigr] \\ \nonumber
  & \ + \dfrac{i}{4} \ a \ c \ [(\omega_1 \omega_2 + p_1 p_2)(p_3 + p_4 + 2 c)- (p_1 \omega_2 + p_2 \omega_1)(\omega_3 + \omega_4)] \\ \nonumber
  & \ + \dfrac{i}{2} \ a \cos^2 \varphi \sin^2 \vartheta \ (p_1 p_2 - \omega_1 \omega_2)  \,, \\ \nonumber
   & \\
\begin{gathered}
	\begin{fmfgraph*}(60,60)
  \fmfleft{i1,i2}
  \fmfright{o1,o2}
  \fmf{fermion,foreground=(0.1,,0.1,,1)}{o1,v1,o2}
  \fmf{scalar}{i1,v1}
  \fmf{scalar}{i2,v1}
    \Marrow{a}{left}{bot}{$p_1 \ $}{i2,v1}{7} 
    \Marrow{b}{left}{top}{$p_2 \ \ \ $}{i1,v1}{7}
    \Marrow{c}{right}{bot}{$\ \ \ p_3$}{v1,o2}{7}
    \Marrow{d}{right}{top}{$\ \ \ p_4$}{o1,v1}{7}
	        \end{fmfgraph*}
  \end{gathered} \ \  = & \ \
\begin{gathered}
	\begin{fmfgraph*}(60,60)
  \fmfleft{i1,i2}
  \fmfright{o1,o2}
  \fmf{fermion,foreground=(0.1,,0.1,,1)}{o1,v1,o2}
  \fmf{scalar}{v1,i1}
  \fmf{scalar}{v1,i2}
    \Marrow{a}{left}{bot}{$p_1 \ $}{v1,i2}{7} 
    \Marrow{b}{left}{top}{$p_2 \ \ \ $}{v1,i1}{7}
    \Marrow{c}{right}{bot}{$\ \ \ p_3$}{v1,o2}{7}
    \Marrow{d}{right}{top}{$\ \ \ p_4$}{o1,v1}{7}
	        \end{fmfgraph*}
  \end{gathered}
   \ \ = \ \dfrac{i}{4} \sin^2 \vartheta \cos^2 \varphi \ (p_1 p_2 - \omega_1 \omega_2) \,,   \\
\begin{gathered}
	\begin{fmfgraph*}(60,60)
  \fmfleft{i1,i2}
  \fmfright{o1,o2}
  \fmf{fermion,foreground=(0.1,,1,,0.1)}{o1,v1,o2}
  \fmf{scalar}{i1,v1,i2}
    \Marrow{a}{left}{bot}{$p_1 \ $}{v1,i2}{7} 
    \Marrow{b}{left}{top}{$p_2 \ \ \ $}{i1,v1}{7}
    \Marrow{c}{right}{bot}{$\ \ \ p_3$}{v1,o2}{7}
    \Marrow{d}{right}{top}{$\ \ \ p_4$}{o1,v1}{7}
	        \end{fmfgraph*}
  \end{gathered}
   \ \ = & \  \dfrac{i}{4}(1-2a)\bigl[ (\omega_1 \omega_2 + p_1 p_2)(\omega_3 \omega_4 + p_3 p_4 - 1) + \\ \nonumber 
	\\[-1.4cm] \nonumber & \qquad \qquad \qquad \qquad \qquad \qquad - (p_1 \omega_2 + \omega_1 p_2)(p_3 \omega_4 + \omega_3 p_4) \bigr]  \\ \nonumber
    & \ -\dfrac{i}{4}a \bigl[(\omega_1 \omega_2 + p_1 p_2)(p_3 + p_4 +2) - (p_1 \omega_2 + p_2 \omega_1)(\omega_3 + \omega_4)\bigr] \,, \\
\begin{gathered}
	\begin{fmfgraph*}(60,60)
  \fmfleft{i1,i2}
  \fmfright{o1,o2}
  \fmf{fermion,foreground=(0.1,,0.1,,1)}{o1,v1,o2}
  \fmf{fermion,foreground=(1,,0.1,,0.1)}{i1,v1,i2}
    \Marrow{a}{left}{bot}{$p_1 \ $}{v1,i2}{7} 
    \Marrow{b}{left}{top}{$p_2 \ \ \ $}{i1,v1}{7}
    \Marrow{c}{right}{bot}{$\ \ \ p_3$}{v1,o2}{7}
    \Marrow{d}{right}{top}{$\ \ \ p_4$}{o1,v1}{7}
	        \end{fmfgraph*}
  \end{gathered}
   \ \ = & \ -\dfrac{i}{4}(1-2a) \bigl[ \omega_1 (p_2 + s) +  \omega_2 (p_1 + s) \bigr] \times \\ \nonumber
	\\[-1.4cm] \nonumber & \qquad \qquad \qquad \qquad \qquad \qquad \times \bigl[ \omega_3 (p_4 + c) +  \omega_4 (p_3 + c) \bigr] \\ \nonumber
   & \ + \dfrac{i}{4}(1-2a) \bigl[ \omega_1 \omega_2 + (p_1 + s)(p_2 + s) \bigr]\bigl[ \omega_3 \omega_4 + (p_3 + c)(p_4 + c) \bigr] \\ \nonumber
   & \ + \dfrac{i}{4}(1-2a) \ s \ c \ \bigl[ (\omega_1 + \omega_2)(\omega_3 + \omega_4) + \\ \nonumber 
	& \qquad \qquad \qquad \qquad \qquad \qquad - (p_1 + p_2 + 2 s)(p_3 + p_4 + 2 c) \bigr] \\ \nonumber
   & \ + \dfrac{i}{4}\ a \ c \ [(\omega_1 \omega_2 + p_1p_2 - s^2)(p_3 + p_4 + 2 c) + \\ \nonumber 
	& \qquad \qquad \qquad \qquad \qquad \qquad - (p_1 \omega_2 + p_2 \omega_1)(\omega_3 + \omega_4)]\\ \nonumber
   & \ + \dfrac{i}{4} \ a \ s \ [(p_1 + p_2 + 2 s)(\omega_3 \omega_4 + p_3 p_4 - c^2) + \\ \nonumber 
	& \qquad \qquad \qquad \qquad \qquad \qquad - (\omega_1 + \omega_2)(p_3 \omega_4 + p_4 \omega_3)] \,, \\
\begin{gathered}
	\begin{fmfgraph*}(60,60)
  \fmfleft{i1,i2}
  \fmfright{o1,o2}
  \fmf{fermion,foreground=(0.1,,1,,0.1)}{o1,v1,o2}
  \fmf{fermion,foreground=(1,,0.1,,0.1)}{i1,v1,i2}
    \Marrow{a}{left}{bot}{$p_1 \ $}{v1,i2}{7} 
    \Marrow{b}{left}{top}{$p_2 \ \ \ $}{i1,v1}{7}
    \Marrow{c}{right}{bot}{$\ \ \ p_3$}{v1,o2}{7}
    \Marrow{d}{right}{top}{$\ \ \ p_4$}{o1,v1}{7}
	        \end{fmfgraph*}
  \end{gathered}
   \ \ = & \ - \dfrac{i}{4}(1-2a) \bigl[ \omega_1 (p_2 + s) +  \omega_2 (p_1 + s) \bigr]\bigl[ \omega_3 p_4 + \omega_4 p_3 \bigr] \\ \nonumber
 \\[-1.4cm] \nonumber  & \ + \dfrac{i}{4}(1-2a) \bigl[ \omega_1 \omega_2 + (p_1 + s)(p_2 + s) \bigr]\bigl[ \omega_3 \omega_4 + p_3 p_4 -1 ) \bigr] \\ \nonumber
& \ + \dfrac{i}{4}\ a \ \bigl[(\omega_1 p_2 + \omega_2 p_1)(\omega_3 + \omega_4) + \\ \nonumber 
& \qquad \qquad \qquad \qquad \qquad \qquad - (\omega_1 \omega_2 + p_1 p_2 - s^2)(p_3 + p_4 +2)\bigr] \\ \nonumber
& \ + \dfrac{i}{4} \ a \ s \ [(p_1 + p_2 + 2 s)(\omega_3 \omega_4 + p_3 p_4 -1) + \\ \nonumber
& \qquad \qquad \qquad \qquad \qquad \qquad - (\omega_1 + \omega_2)(\omega_3 p_4 + \omega_4 p_3)]  \,, \\
\begin{gathered}
	\begin{fmfgraph*}(60,60)
  \fmfleft{i1,i2}
  \fmfright{o1,o2}
  \fmf{fermion,foreground=(0.1,,1,,0.1)}{o1,v1,o2}
  \fmf{fermion,foreground=(0.1,,0.1,,1)}{i1,v1,i2}
    \Marrow{a}{left}{bot}{$p_1 \ $}{v1,i2}{7} 
    \Marrow{b}{left}{top}{$p_2 \ \ \ $}{i1,v1}{7}
    \Marrow{c}{right}{bot}{$\ \ \ p_3$}{v1,o2}{7}
    \Marrow{d}{right}{top}{$\ \ \ p_4$}{o1,v1}{7}
	        \end{fmfgraph*}
  \end{gathered}
   \ \ = & \ - \dfrac{i}{4}(1-2a) \bigl[ \omega_1 (p_2 + c) +  \omega_2 (p_1 +  c) \bigr]\bigl[ \omega_3 p_4 + \omega_4 p_3 \bigr] \\ \nonumber
 \\[-1.4cm] \nonumber  & \ + \dfrac{i}{4}(1-2a) \bigl[ \omega_1 \omega_2 + (p_1 + c)(p_2 +  c) \bigr]\bigl[ \omega_3 \omega_4 + p_3 p_4 -1 ) \bigr] \\ \nonumber
& \ + \dfrac{i}{4}\ a \ \bigl[(\omega_1 p_2 + \omega_2 p_1)(\omega_3 + \omega_4) + \\ \nonumber 
& \qquad \qquad \qquad \qquad - (\omega_1 \omega_2 + p_1 p_2 - c^2)(p_3 + p_4 +2)\bigr] \\ \nonumber
& \ + \dfrac{i}{4} \ a \ c \ [(p_1 + p_2 + 2 c)(\omega_3 \omega_4 + p_3 p_4 -1) + \\ \nonumber 
& \qquad \qquad \qquad \qquad \qquad \qquad - (\omega_1 + \omega_2)(\omega_3 p_4 + \omega_4 p_3)]  \,. \\ \nonumber 
\end{align} 
\end{fmffile}  
\endgroup

\section{Wess-Zumino-Witten description of \texorpdfstring{$\AdSSS$}{AdS3xS3xS3xS1} strings}
\label{app:WZW}
Closed strings propagating on $\text{AdS}_3 \times \text{S}^3 \times \text{S}^3 \times \text{S}^1$ with pure NS-NS flux can be described by a Wess-Zumino-Witten model based on the $\mathcal{N} = 1$ Ka\v{c}-Moody algebras $\mathfrak{sl}(2)_k^{(1)} \oplus \mathfrak{su}(2)_{k_1}^{(1)} \oplus \mathfrak{su}(2)_{k_2}^{(1)} \oplus \mathfrak{u}(1)$ \cite{Elitzur:1998mm,Eberhardt:2017fsi}. The following commutation relations characterize the $\mathfrak{sl}(2)_k^{(1)}$ algebra
\begin{equation}
\begin{aligned}
& [\text{L}^+_m, \text{L}^-_n] = - 2 \text{L}_{m+n}^3 + k m \delta_{m,-n} &  & [\text{L}^3_m, \text{L}^\pm_n] = \pm \text{L}^\pm_{m+n} & & [\text{L}^3_m, \text{L}^3_n] = -\frac{k}{2} m \delta_{m, -n}  \\
& [\text{L}_m^\pm, \psi_r^3] = \mp \psi_{r+m}^\pm & & [\text{L}_m^3, \psi_r^\pm] = \pm \psi_{r+m}^\pm & & [\text{L}_m^\pm, \psi_r^\mp] = \mp 2 \psi^3_{m+r} \\
& \{ \psi^+_r, \psi^-_s \} = k \delta_{r, -s} & & \{ \psi^3_r, \psi^3_s \} = -\frac{k}{2} \delta_{r, -s} \,,  
\end{aligned}
\label{eq:sl(2)-kac-moody-algebra}
\end{equation}
while for the $\mathfrak{su}(2)_{k_I}^{(1)}$ algebra we have 
\begin{equation}
\begin{aligned}
& [\text{J}^{I,+}_m, \text{J}^{I,-}_n] = 2 \text{J}_{m+n}^{I,3} + k m \delta_{m,-n} &  & [\text{J}^{I,3}_m, \text{J}^{I,\pm}_n] = \pm \text{J}^{I,\pm}_{m+n} & & [\text{J}^{I,3}_m, \text{J}^{I,3}_n] = \frac{k_I}{2} m \delta_{m, -n}  \\
& [\text{J}
_m^{I,\pm}, \theta_r^{I,3}] = \mp \theta_{r+m}^{I,\pm} & & [\text{J}_m^{I,3}, \theta_r^{I,\pm}] = \pm \theta_{r+m}^{I,\pm} & & [\text{J}_m^{I,\pm}, \theta_r^{I,\mp}] = \pm 2 \theta^{I,3}_{m+r} \\
& \{ \theta^{I,+}_r, \theta^{I,-}_s \} = k_I \delta_{r, -s} & & \{ \theta^{I,3}_r, \theta^{I,3}_s \} = \frac{k_I}{2} \delta_{r, -s} \,,  
\end{aligned}
\label{eq:su(2)-kac-moody-algebra}
\end{equation}
where $I=1,2$ labels respectively $\mathfrak{su}(2)_{k_1}^{(1)}$ and $\mathfrak{su}(2)_{k_2}^{(1)}$. As explained \emph{e.g.}~in \cite{Ferreira:2017pgt}, it is convenient to define decoupled bosonic modes $L^a_n$ and $J^{I,a}_n$ commuting with the worldsheet fermions. The resulting Ka\v{c}-Moody algebras have levels shifted respectively by +2 and $-2$ for $\mathfrak{sl}(2)$ and $\mathfrak{su}(2)$. One can schematically write
\begin{equation}
\begin{aligned}
& \mathfrak{sl}(2)^{(1)}_k = \mathfrak{sl}(2)_{k+2} \oplus \text{free fermions} \,, \\
& \mathfrak{su}(2)^{(1)}_{k_I} = \mathfrak{su}(2)_{k_I-2} \oplus \text{free fermions} \,.
\end{aligned}
\end{equation}
Finally, the $\mathfrak{u}(1)$ algebra contains a free bosonic mode $\alpha_n$ and its superpartner $\gamma_n$ arising from the one-sphere. The one we described so far is the holomorphic sector of the theory. The anti-holomorphic sector is defined in complete analogy and will be denoted in the following by tildes, \emph{e.g.} $\tilde L^a_n, J^{I,a}_n$, etc. As usual in RNS formalism, worldsheet fermions are integer-moded in the R sector and hal-integer-moded in the NS sector. The spectrum is realised by tensoring holomorphic and anti-holomorphic representations and target space supersymmetry is obtained by imposing GSO projection. We are now going to revise the construction of the unflowed representations of the spectrum. The spectrum also contains  so-called \emph{spectrally flowed} representations \cite{Maldacena:2000hw}, which we will not discuss here---see ref.~\cite{Dei:2018mfl} for a discussion of the spectral flow in the spin-chain description, and ref.~\cite{Eberhardt:2017fsi} for a detailed description of spectral flow for $\AdSSS$.

Spectrally unflowed representations of the Kac-Moody algebras are built by acting with negative-moded modes on a highest-weight Ka\v{c}-Moody module, which we will denote by $\ket{\ell_0, j_{0,1}, j_{0,2}}$. The latter is a lowest-weight state of $\mathfrak{sl}(2)$ and a highest-weight state of $\mathfrak{su}(2)_I$:  
\begin{equation}
\begin{aligned}
L_0^- \ket{\ell_0, j_{0,1}, j_{0,2}} &= 0 \,, \qquad &  L_0^3 \ket{\ell_0, j_{0,1}, j_{0,2}} &= \ell_0 \ket{\ell_0, j_{0,1}, j_{0,2}} \,,  \\
J_0^{I,+} \ket{\ell_0, j_{0,1}, j_{0,2}} &= 0 \,, \qquad & J_0^{I,3} \ket{\ell_0, j_{0,1}, j_{0,2}} &= j_{0,I} \ket{\ell_0, j_{0,1}, j_{0,2}} \,, 
\end{aligned}
\end{equation}
and
\begin{equation}
L_n^a \ket{\ell_0, j_{0,1}, j_{0,2}} = 0 \,, \qquad   J_n^{I,a} \ket{\ell_0, j_{0,1}, j_{0,2}} = \ 0 \,,  \quad \text{for} \ n>0 \,. 
\end{equation}
The ground state representation is the same in the holomorphic and anti-holomorphic sectors: $\ell_0 = \tilde \ell_0$ and $j_{0,I} = \tilde j_{0,I}$ \cite{Maldacena:2000hw}. The possible values of the $\mathfrak{sl}(2)$ and $\mathfrak{su}(2)$ spins are restricted by the unitarity and Maldacena-Ooguri bounds \cite{Maldacena:2000hw, Eberhardt:2018vho}
\begin{equation}
\dfrac{1}{2} < \ell_0 < \dfrac{k + 1}{2} \,, \qquad 0 \leq j_{0,I} \leq \dfrac{k_I-2}{2} \,. 
\label{eq:unitarity-bounds}
\end{equation}
The physicality condition for states to be annihilated by positive modes of the Virasoro algebra reduces to the so-called \emph{mass-shell condition}
\begin{equation}
-\dfrac{\ell_0(\ell_0 - 1)}{k} + \dfrac{j_{0,1}(j_{0,1} + 1)}{k_1} + \dfrac{j_{0,2}(j_{0,2} + 1)}{k_2} + N_{\rm{eff}} = 0 \ , 
\label{eq:unflowed-mass-shell-condition}
\end{equation}
where 
\begin{equation}
\label{eq:Neff}
N_{\rm{eff}} = \begin{cases} N - \tfrac{1}{2} & \rm{NS \ sector} \\ N & \rm{R \ sector}  \end{cases}
\end{equation}
and $N$ is the eigenvalue of the total number operator. Level-matching requires 
\begin{equation}
N_{\rm{eff}} = \tilde N_{\rm{eff}} \,. 
\label{eq:unflowed-level-matching}
\end{equation}
The supersymmetric spectrum is built by tensoring holomorphic and anti-holomorphic sectors and imposing the GSO projection. The latter amounts to requiring $N_{\rm{eff}}$ to be integer in the NS sector. The spacetime charges are related to the quantum numbers $\ell_0$ and $j_{0,I}$ by 
\begin{equation}
\begin{array}{lll}
\ell = \ell_0 + \delta \ell \ , \qquad & j_I = j_{0,I}  - \delta j_I \qquad  &\rm{NS \ sector} \\
\ell = \ell_0 + \delta \ell + s_\ell \ , \qquad  \qquad & j_I = j_{0,I}  - \delta j_I - s_{j,I} 
\qquad \qquad & \rm{R \ sector.}
\label{eq:WZW-spins}
\end{array}
\end{equation}
In eq.~\eqref{eq:WZW-spins} $\delta \ell$ denotes the $\mathfrak{sl}(2)$ charge of the state with respect to the ground state $\ket{\ell_0, j_{0,1}, j_{0,2}}$;  the modes $L_{-n}^\pm$ contribute to $\delta \ell$ with $\pm 1$. Similarly, but with a different sign convention, the modes $J_{-n}^{I,\pm}$ contribute to $\delta j$ with $\mp 1$. In the R sector, the additional labels $s_\ell, s_{j,I} = \pm \frac{1}{2}$ account for the different possible choices of fermionic zero-modes \cite{Ferreira:2017pgt}. Solving the mass-shell condition \eqref{eq:unflowed-mass-shell-condition} for $\ell_0$ and exploiting eq~\eqref{eq:WZW-spins} we find 
\begin{equation}
\ell = 1 + \delta \ell + s_l +   \sqrt{\cos^2 \varphi \ (2j_{0,1} + 1)^2 + \sin^2 \varphi \ (2j_{0,2} + 1)^2 + 4 k N_{\text{eff}}}   \,, 
\end{equation}
where with a small abuse of notation, in the NS sector $s_\ell=s_{j,1}=s_{j,2} = 0$. Similarly, making use of eq.~\eqref{eq:unflowed-level-matching}, in the antiholomorphic sector we find 
\begin{equation}
\tilde \ell = 1 + \delta \tilde \ell + \tilde s_l +   \sqrt{\cos^2 \varphi \ (2j_{0,1} + 1)^2 + \sin^2 \varphi \ (2j_{0,2} + 1)^2 + 4 k N_{\text{eff}}}   \,.  
\end{equation}
The worldsheet Hamiltonian is then
\begin{equation}
\begin{aligned}
H_{\text{tot}} = \ell + \tilde \ell - \mathcal{J} = &  \sqrt{\cos^2 \varphi \ (2j_{0,1} + 1)^2 + \sin^2 \varphi \ (2j_{0,2} + 1)^2 + 4 k N_{\text{eff}}} +  \delta + \tilde \delta \,, 
\end{aligned}
\end{equation}
where
\begin{equation}
\begin{aligned}
\delta & = \delta \ell + s_l + \cos \varphi \cos \vartheta \ (\delta j_1 + s_{j,1} ) + \sin \varphi \sin \vartheta \ (\delta j_2 + s_{j,2}) + 1 \,, \\
\tilde \delta & = \delta \tilde \ell + \tilde s_l + \cos \varphi \cos \vartheta \ (\delta \tilde j_1 + \tilde s_{j,1} ) + \sin \varphi \sin \vartheta \ (\delta \tilde j_2 + \tilde s_{j,2}) + 1 \,. 
\end{aligned}
\end{equation}
In the unflowed sector BPS states occur for   $j_{0,1} = j_{0,2} = \ell_0 - 1$, $N_{\text{eff}}=0$ and $\delta = \tilde \delta = 0$ \cite{Eberhardt:2017fsi,Eberhardt:2017pty}. They are realised by tensoring the following BPS states: 
\begin{equation}
\begin{aligned}
& \psi^-_{-\frac{1}{2}} \ket{\ell_0, \ell_0 -1, \ell_0-1} \,,  & \qquad & \text{NS sector} \\
& \ket{\ell_0, \ell_0 -1, \ell_0-1} \ \ \ \text{with} \ s_\ell = s_{j,1} = s_{j,2} = -\dfrac{1}{2} & \qquad & \text{R sector}
\end{aligned}
\end{equation} 
for a total of 4 BPS states for each value of $\ell_0$ and $j_{0,1}=j_{0,2} = \ell_0-1$ respecting the bounds in eq.~\eqref{eq:unitarity-bounds}. In particular, in the R-R sector we have
\begin{equation}
j_1=\tilde{\jmath}_1=j_{0,1}+\frac{1}{2}\,,\qquad
j_2=\tilde{\jmath}_2=j_{0,2}+\frac{1}{2}\,.
\end{equation}
We conclude by observing that
\begin{equation}
\delta+\tilde{\delta}= \hat{\mu}_{\text{tot}}\,,
\end{equation}
\textit{i.e.}\ the charge shift due to momentum-independent part of $\ell+\tilde{\ell}-\mathcal{J}$. The charges of the Ka\v{c}-Moody currents can be matched to the ones of the bosons of table~\ref{tab:S-matrix-basis}. The matching of the fermions is a little more involved, as it requires bearing in mind the GSO projection. This has been done in detail for this model in ref.~\cite{Dei:2018yth}.

\section{TBA integrals}
\label{app:integrals}

Here we will compute the integral $K$ and the sum $\kappa$ defined as
\begin{equation}
\begin{aligned}
K\,&=\sum_{a}^{(8|8)} \int\limits_0^{+\infty}\text{d}\bar{p} (-1)^{F_a+1} \log\Big(1-(-1)^{F_a}e^{-\epsilon_b(\bar{p})}\Big),\\
\kappa\,&=-i\sum_{a}^{(8|8)}\mu_a\,(-1)^{F_a} \log\Big(1-(-1)^{F_a}e^{-\epsilon_b(0)}\Big),
\end{aligned}
\end{equation}
where
\begin{equation}
\epsilon_b(\bar{p})= (\bar{p}+i\mu_a)c+\psi_a.
\end{equation}
Let us begin with the integral, which we can simplify as
\begin{equation}
K=-\sum_{a}^{(8|8)} (-1)^{F_a}\int\limits_0^{+\infty}\text{d}\bar{p}\,\log\Big(1-e^{-c(\bar{p}+i\mu_a)}\Big)=
\sum_{a}^{(8|8)} (-1)^{F_a} \frac{1}{c}\Li{e^{-i c\mu_a}},
\end{equation}
where we assumed that $c\geq 0 $ (which will turn out to be the case). This is immediately zero in presence of manifest supersymmetry. More generally, recall that the masses $\{\mu_a\}$ come in pairs of opposite signs, as a consequence of crossing invariance. Note the identity
\begin{equation}
\Li{e^{i t}}+\Li{e^{-i t}}=\frac{\pi^2}{3}-\frac{t(2\pi-t)}{2},\qquad 0\leq t\leq 2\pi.
\end{equation}
As long as we are not in a spectrally-flowed sector, we can straightforwardly apply it and find
\begin{equation}
K=-\pi \sum_a^{(4|4)} (-1)^{F_a} \mu_a+\frac{c}{2}\sum_a^{(4|4)} (-1)^{F_a} (\mu_a)^2\,.
\end{equation}
As already observed in ref.~\cite{Dei:2018yth}, the second sum vanishes; this is due to $\mathfrak{so}(8)$ triality%
\footnote{We thank L.~Eberhardt for this remark.}. Hence the integral~$K$ does not depends on $c$ at all and takes the value
\begin{equation}
K=-\pi \sum_a^{(4|4)} (-1)^{F_a} \mu_a\,.
\end{equation}
We can similarly compute the sum $\kappa$, which gives
\begin{equation}
\kappa = i\sum_{a}^{(8|8)}\mu_a\,(-1)^{F_a} \log\Big(1-e^{i\,c\,\mu_a}\Big)=i\sum_{a}^{(4|4)}\mu_a\,(-1)^{F_a}(ic\mu_a-i \pi),
\end{equation}
where again we have used that the masses come in pairs of opposite signs and assumed $0\leq c\mu_a\leq 2\pi$. Like before the $c$-dependent term cancels and we have
\begin{equation}
\kappa=+\pi \sum_a^{(4|4)} (-1)^{F_a} \mu_a\,.
\end{equation}
Using the values of the masses $\{\mu_a\}$ we have
\begin{equation}
K=-\kappa=-\pi \big(
\sin\varphi\sin\vartheta+\cos\varphi\cos\vartheta-1
\big).
\end{equation}

\bibliographystyle{JHEP}
\bibliography{refs}

\makeatletter \@ifundefined{Sphere}{\newcommand{\Sphere}{\text{S}}}{}
  \@ifundefined{AdS}{\newcommand{\AdS}{\text{AdS}}}{}
  \@ifundefined{CFT}{\newcommand{\CFT}{\text{CFT}}}{}
  \@ifundefined{CP}{\newcommand{\CP}{\text{CP}}}{}
  \@ifundefined{Torus}{\newcommand{\Torus}{\text{T}}}{}
  \@ifundefined{superN}{\newcommand{\superN}{\mathcal{N}}}{}
  \@ifundefined{grpOSp}{\newcommand{\grpOSp}{\mathrm{OSp}}}{}
  \@ifundefined{grpPSU}{\newcommand{\grpPSU}{\mathrm{PSU}}}{}
  \@ifundefined{grpSU}{\newcommand{\grpSU}{\mathrm{SU}}}{}
  \@ifundefined{grpU}{\newcommand{\grpU}{\mathrm{U}}}{}
  \@ifundefined{grpD}{\newcommand{\grpD}{\mathrm{D}}}{}
  \@ifundefined{grpSL}{\newcommand{\grpSL}{\mathrm{SL}}}{}
  \@ifundefined{grpSp}{\newcommand{\grpSp}{\mathrm{Sp}}}{}
  \@ifundefined{grpUSp}{\newcommand{\grpUSp}{\mathrm{USp}}}{}
  \@ifundefined{grpSO}{\newcommand{\grpSO}{\mathrm{SO}}}{}
  \@ifundefined{grpO}{\newcommand{\grpO}{\mathrm{O}}}{}
  \@ifundefined{algOSp}{\newcommand{\algOSp}{\mathfrak{osp}}}{}
  \@ifundefined{algPSU}{\newcommand{\algPSU}{\mathfrak{psu}}}{}
  \@ifundefined{algSU}{\newcommand{\algSU}{\mathfrak{su}}}{}
  \@ifundefined{algSp}{\newcommand{\algSp}{\mathfrak{sp}}}{}
  \@ifundefined{algSL}{\newcommand{\algSL}{\mathfrak{sl}}}{}
  \@ifundefined{algGL}{\newcommand{\algGL}{\mathfrak{gl}}}{}
  \@ifundefined{algU}{\newcommand{\algU}{\mathfrak{u}}}{}
  \@ifundefined{algSO}{\newcommand{\algSO}{\mathfrak{so}}}{}
  \@ifundefined{algO}{\newcommand{\algO}{\mathfrak{o}}}{}
  \@ifundefined{Integers}{\newcommand{\Integers}{\mathbb{Z}}}{}
  \@ifundefined{Reals}{\newcommand{\Reals}{\mathbb{R}}}{} \makeatother

\providecommand{\href}[2]{#2}\begingroup\raggedright\begin{thebibliography}{10}

\bibitem{Maldacena:1997re}
J.~M. Maldacena, \emph{{The Large N limit of superconformal field theories and
  supergravity}}, \href{https://doi.org/10.1023/A:1026654312961,
  10.4310/ATMP.1998.v2.n2.a1}{\emph{Int. J. Theor. Phys.} {\bfseries 38} (1999)
  1113} [\href{https://arxiv.org/abs/hep-th/9711200}{{\ttfamily
  hep-th/9711200}}].

\bibitem{Gubser:1998bc}
S.~S. Gubser, I.~R. Klebanov and A.~M. Polyakov, \emph{Gauge theory correlators
  from non-critical string theory},
  \href{https://doi.org/10.1016/S0370-2693(98)00377-3}{\emph{Phys. Lett.}
  {\bfseries B428} (1998) 105}
  [\href{https://arxiv.org/abs/hep-th/9802109}{{\ttfamily hep-th/9802109}}].

\bibitem{Witten:1998qj}
E.~Witten, \emph{{Anti-de Sitter space and holography}},
  \href{https://doi.org/10.4310/ATMP.1998.v2.n2.a2}{\emph{Adv. Theor. Math.
  Phys.} {\bfseries 2} (1998) 253}
  [\href{https://arxiv.org/abs/hep-th/9802150}{{\ttfamily hep-th/9802150}}].

\bibitem{David:2002wn}
J.~R. David, G.~Mandal and S.~R. Wadia, \emph{Microscopic formulation of black
  holes in string theory},
  \href{https://doi.org/10.1016/S0370-1573(02)00271-5}{\emph{Phys. Rept.}
  {\bfseries 369} (2002) 549}
  [\href{https://arxiv.org/abs/hep-th/0203048}{{\ttfamily hep-th/0203048}}].

\bibitem{Giveon:1998ns}
A.~Giveon, D.~Kutasov and N.~Seiberg, \emph{{Comments on string theory on
  AdS(3)}}, \href{https://doi.org/10.4310/ATMP.1998.v2.n4.a3}{\emph{Adv. Theor.
  Math. Phys.} {\bfseries 2} (1998) 733}
  [\href{https://arxiv.org/abs/hep-th/9806194}{{\ttfamily hep-th/9806194}}].

\bibitem{Elitzur:1998mm}
S.~Elitzur, O.~Feinerman, A.~Giveon and D.~Tsabar, \emph{String theory on
  {$\mathrm{AdS}_3\times \mathrm{S}^3\times \mathrm{S}^3\times\mathrm{S}^1$}},
  \href{https://doi.org/10.1016/S0370-2693(99)00101-X}{\emph{Phys. Lett.}
  {\bfseries B449} (1999) 180}
  [\href{https://arxiv.org/abs/hep-th/9811245}{{\ttfamily hep-th/9811245}}].

\bibitem{Sevrin:1988ew}
A.~Sevrin, W.~Troost and A.~Van~Proeyen, \emph{Superconformal algebras in
  two-dimensions with {$\superN = 4$}},
  \href{https://doi.org/10.1016/0370-2693(88)90645-4}{\emph{Phys. Lett.}
  {\bfseries B208} (1988) 447}.

\bibitem{OhlssonSax:2018hgc}
O.~Ohlsson~Sax and B.~Stefa{\'n}ski, \emph{{Closed strings and moduli in
  AdS$_{3}$/CFT$_{2}$}},
  \href{https://doi.org/10.1007/JHEP05(2018)101}{\emph{JHEP} {\bfseries 05}
  (2018) 101} [\href{https://arxiv.org/abs/1804.02023}{{\ttfamily
  1804.02023}}].

\bibitem{Maldacena:2000hw}
J.~M. Maldacena and H.~Ooguri, \emph{Strings in {$\AdS{3}$} and {$\grpSL(2,R)$}
  {WZW} model. {I}}, \href{https://doi.org/10.1063/1.1377273}{\emph{J. Math.
  Phys.} {\bfseries 42} (2001) 2929}
  [\href{https://arxiv.org/abs/hep-th/0001053}{{\ttfamily hep-th/0001053}}].

\bibitem{Maldacena:2000kv}
J.~M. Maldacena, H.~Ooguri and J.~Son, \emph{Strings in {$\AdS{3}$} and the
  {$\grpSL(2,R)$} {WZW} model. {II}: {E}uclidean black hole},
  \href{https://doi.org/10.1063/1.1377039}{\emph{J. Math. Phys.} {\bfseries 42}
  (2001) 2961} [\href{https://arxiv.org/abs/hep-th/0005183}{{\ttfamily
  hep-th/0005183}}].

\bibitem{Maldacena:2001km}
J.~M. Maldacena and H.~Ooguri, \emph{Strings in {$\AdS{3}$} and the
  {$\grpSL(2,R)$} {WZW} model.~{III}: Correlation functions},
  \href{https://doi.org/10.1103/PhysRevD.65.106006}{\emph{Phys. Rev.}
  {\bfseries D65} (2002) 106006}
  [\href{https://arxiv.org/abs/hep-th/0111180}{{\ttfamily hep-th/0111180}}].

\bibitem{Pakman:2003cu}
A.~Pakman, \emph{{Unitarity of supersymmetric SL(2,R) / U(1) and no ghost
  theorem for fermionic strings in AdS(3) x N}},
  \href{https://doi.org/10.1088/1126-6708/2003/01/077}{\emph{JHEP} {\bfseries
  01} (2003) 077} [\href{https://arxiv.org/abs/hep-th/0301110}{{\ttfamily
  hep-th/0301110}}].

\bibitem{Israel:2003ry}
D.~Israel, C.~Kounnas and M.~P. Petropoulos, \emph{{Superstrings on NS5
  backgrounds, deformed AdS(3) and holography}},
  \href{https://doi.org/10.1088/1126-6708/2003/10/028}{\emph{JHEP} {\bfseries
  10} (2003) 028} [\href{https://arxiv.org/abs/hep-th/0306053}{{\ttfamily
  hep-th/0306053}}].

\bibitem{Raju:2007uj}
S.~Raju, \emph{{Counting giant gravitons in AdS(3)}},
  \href{https://doi.org/10.1103/PhysRevD.77.046012}{\emph{Phys. Rev.}
  {\bfseries D77} (2008) 046012}
  [\href{https://arxiv.org/abs/0709.1171}{{\ttfamily 0709.1171}}].

\bibitem{Giribet:2007wp}
G.~Giribet, A.~Pakman and L.~Rastelli, \emph{{Spectral Flow in AdS(3)/CFT(2)}},
  \href{https://doi.org/10.1088/1126-6708/2008/06/013}{\emph{JHEP} {\bfseries
  06} (2008) 013} [\href{https://arxiv.org/abs/0712.3046}{{\ttfamily
  0712.3046}}].

\bibitem{Ferreira:2017pgt}
K.~Ferreira, M.~R. Gaberdiel and J.~I. Jottar, \emph{{Higher spins on AdS$_{3}$
  from the worldsheet}},
  \href{https://doi.org/10.1007/JHEP07(2017)131}{\emph{JHEP} {\bfseries 07}
  (2017) 131} [\href{https://arxiv.org/abs/1704.08667}{{\ttfamily
  1704.08667}}].

\bibitem{Berkovits:1999im}
N.~Berkovits, C.~Vafa and E.~Witten, \emph{{Conformal field theory of AdS
  background with Ramond-Ramond flux}},
  \href{https://doi.org/10.1088/1126-6708/1999/03/018}{\emph{JHEP} {\bfseries
  03} (1999) 018} [\href{https://arxiv.org/abs/hep-th/9902098}{{\ttfamily
  hep-th/9902098}}].

\bibitem{Eberhardt:2018exh}
L.~Eberhardt and K.~Ferreira, \emph{{The plane-wave spectrum from the
  worldsheet}}, \href{https://doi.org/10.1007/JHEP10(2018)109}{\emph{JHEP}
  {\bfseries 10} (2018) 109}
  [\href{https://arxiv.org/abs/1805.12155}{{\ttfamily 1805.12155}}].

\bibitem{Hoare:2013pma}
B.~Hoare and A.~A. Tseytlin, \emph{On string theory on {$\AdS{3} \times S^3
  \times T^4$} with mixed 3-form flux: tree-level {S}-matrix},
  \href{https://doi.org/10.1016/j.nuclphysb.2013.05.005}{\emph{Nucl. Phys.}
  {\bfseries B873} (2013) 682}
  [\href{https://arxiv.org/abs/1303.1037}{{\ttfamily 1303.1037}}].

\bibitem{Babichenko:2014yaa}
A.~Babichenko, A.~Dekel and O.~Ohlsson~Sax, \emph{Finite-gap equations for
  strings on {$\AdS{3} \times \Sphere^3 \times \Torus^4$} with mixed 3-form
  flux}, \href{https://doi.org/10.1007/JHEP11(2014)122}{\emph{JHEP} {\bfseries
  1411} (2014) 122} [\href{https://arxiv.org/abs/1405.6087}{{\ttfamily
  1405.6087}}].

\bibitem{Sax:2014mea}
O.~Ohlsson~Sax, A.~Sfondrini and B.~Stefa{\'n}ski, jr., \emph{Integrability and
  the conformal field theory of the {H}iggs branch},
  \href{https://doi.org/10.1007/JHEP06(2015)103}{\emph{JHEP} {\bfseries 06}
  (2015) 103} [\href{https://arxiv.org/abs/1411.3676}{{\ttfamily 1411.3676}}].

\bibitem{Hernandez:2014eta}
R.~Hern{\'a}ndez and J.~M. Nieto, \emph{Spinning strings in {$\AdS{3} \times
  S^3$} with {NS-NS} flux},
  \href{https://doi.org/10.1016/j.nuclphysb.2015.04.011,
  10.1016/j.nuclphysb.2014.10.001}{\emph{Nucl. Phys.} {\bfseries B888} (2014)
  236} [\href{https://arxiv.org/abs/1407.7475}{{\ttfamily 1407.7475}}].

\bibitem{Nieto:2018jzi}
J.~M. Nieto and R.~Ruiz, \emph{{One-loop quantization of rigid spinning strings
  in $AdS_3 \times S^3 \times T^4$ with mixed flux}},
  \href{https://doi.org/10.1007/JHEP07(2018)141}{\emph{JHEP} {\bfseries 07}
  (2018) 141} [\href{https://arxiv.org/abs/1804.10477}{{\ttfamily
  1804.10477}}].

\bibitem{Babichenko:2009dk}
A.~Babichenko, B.~Stefa{\'n}ski, jr. and K.~Zarembo, \emph{Integrability and
  the {$\AdS{3}/\CFT_{2}$} correspondence},
  \href{https://doi.org/10.1007/JHEP03(2010)058}{\emph{JHEP} {\bfseries 1003}
  (2010) 058} [\href{https://arxiv.org/abs/0912.1723}{{\ttfamily 0912.1723}}].

\bibitem{Sundin:2012gc}
P.~Sundin and L.~Wulff, \emph{Classical integrability and quantum aspects of
  the {$\AdS{3} \times S^3 \times S^3 \times S^1$} superstring},
  \href{https://doi.org/10.1007/JHEP10(2012)109}{\emph{JHEP} {\bfseries 1210}
  (2012) 109} [\href{https://arxiv.org/abs/1207.5531}{{\ttfamily 1207.5531}}].

\bibitem{Cagnazzo:2012se}
A.~Cagnazzo and K.~Zarembo, \emph{{B}-field in {$\AdS{3}/\CFT_{2}$}
  correspondence and integrability},
  \href{https://doi.org/10.1007/JHEP11(2012)133,
  10.1007/JHEP04(2013)003}{\emph{JHEP} {\bfseries 1211} (2012) 133}
  [\href{https://arxiv.org/abs/1209.4049}{{\ttfamily 1209.4049}}].

\bibitem{Arutyunov:2009ga}
G.~Arutyunov and S.~Frolov, \emph{Foundations of the {$\AdS{5} \times S^5$}
  superstring. part {I}},
  \href{https://doi.org/10.1088/1751-8113/42/25/254003}{\emph{J. Phys. A}
  {\bfseries A42} (2009) 254003}
  [\href{https://arxiv.org/abs/0901.4937}{{\ttfamily 0901.4937}}].

\bibitem{Beisert:2010jr}
N.~Beisert et~al., \emph{Review of {AdS/CFT} integrability: An overview},
  \href{https://doi.org/10.1007/s11005-011-0529-2}{\emph{Lett. Math. Phys.}
  {\bfseries 99} (2012) 3} [\href{https://arxiv.org/abs/1012.3982}{{\ttfamily
  1012.3982}}].

\bibitem{Sfondrini:2014via}
A.~Sfondrini, \emph{Towards integrability for {$\AdS{3}/\CFT_{2}$}},
  \href{https://doi.org/10.1088/1751-8113/48/2/023001}{\emph{J. Phys.}
  {\bfseries A48} (2015) 023001}
  [\href{https://arxiv.org/abs/1406.2971}{{\ttfamily 1406.2971}}].

\bibitem{Borsato:2012ud}
R.~Borsato, O.~Ohlsson~Sax and A.~Sfondrini, \emph{A dynamic {$\algSU(1|1)^2$}
  {S}-matrix for {$\AdS{3}/\CFT_{2}$}},
  \href{https://doi.org/10.1007/JHEP04(2013)113}{\emph{JHEP} {\bfseries 1304}
  (2013) 113} [\href{https://arxiv.org/abs/1211.5119}{{\ttfamily 1211.5119}}].

\bibitem{Borsato:2013qpa}
R.~Borsato, O.~Ohlsson~Sax, A.~Sfondrini, B.~Stefa{\'n}ski, jr. and
  A.~Torrielli, \emph{The all-loop integrable spin-chain for strings on
  {$\AdS{3} \times S^3 \times T^4$}: the massive sector},
  \href{https://doi.org/10.1007/JHEP08(2013)043}{\emph{JHEP} {\bfseries 1308}
  (2013) 043} [\href{https://arxiv.org/abs/1303.5995}{{\ttfamily 1303.5995}}].

\bibitem{Hoare:2013ida}
B.~Hoare and A.~Tseytlin, \emph{Massive {S}-matrix of {$\AdS{3} \times S^3
  \times T^4$} superstring theory with mixed 3-form flux},
  \href{https://doi.org/10.1016/j.nuclphysb.2013.04.024}{\emph{Nucl. Phys.}
  {\bfseries B873} (2013) 395}
  [\href{https://arxiv.org/abs/1304.4099}{{\ttfamily 1304.4099}}].

\bibitem{Borsato:2014exa}
R.~Borsato, O.~Ohlsson~Sax, A.~Sfondrini and B.~Stefa{\'n}ski, jr.,
  \emph{Towards the all-loop worldsheet {S} matrix for {$\AdS{3} \times S^3
  \times T^4$}},
  \href{https://doi.org/10.1103/PhysRevLett.113.131601}{\emph{Phys. Rev. Lett.}
  {\bfseries 113} (2014) 131601}
  [\href{https://arxiv.org/abs/1403.4543}{{\ttfamily 1403.4543}}].

\bibitem{Borsato:2014hja}
R.~Borsato, O.~Ohlsson~Sax, A.~Sfondrini and B.~Stefa{\'n}ski, jr, \emph{The
  complete {$\AdS{3} \times S^3 \times T^4$} worldsheet {S}-matrix},
  \href{https://doi.org/10.1007/JHEP10(2014)066}{\emph{JHEP} {\bfseries 1410}
  (2014) 66} [\href{https://arxiv.org/abs/1406.0453}{{\ttfamily 1406.0453}}].

\bibitem{Lloyd:2014bsa}
T.~Lloyd, O.~Ohlsson~Sax, A.~Sfondrini and B.~Stefa{\'n}ski, jr., \emph{The
  complete worldsheet {S} matrix of superstrings on {$\AdS{3} \times S^3 \times
  T^4$} with mixed three-form flux},
  \href{https://doi.org/10.1016/j.nuclphysb.2014.12.019}{\emph{Nucl. Phys.}
  {\bfseries B891} (2015) 570}
  [\href{https://arxiv.org/abs/1410.0866}{{\ttfamily 1410.0866}}].

\bibitem{Borsato:2015mma}
R.~Borsato, O.~Ohlsson~Sax, A.~Sfondrini and B.~Stefa{\'n}ski, jr., \emph{The
  {$\AdS{3} \times S^3 \times S^3 \times S^1$} worldsheet {S} matrix},
  \href{https://doi.org/10.1088/1751-8113/48/41/415401}{\emph{J. Phys.}
  {\bfseries A48} (2015) 415401}
  [\href{https://arxiv.org/abs/1506.00218}{{\ttfamily 1506.00218}}].

\bibitem{Rughoonauth:2012qd}
N.~Rughoonauth, P.~Sundin and L.~Wulff, \emph{Near {BMN} dynamics of the
  {$\AdS{3} \times S^3 \times S^3 \times S^1$} superstring},
  \href{https://doi.org/10.1007/JHEP07(2012)159}{\emph{JHEP} {\bfseries 1207}
  (2012) 159} [\href{https://arxiv.org/abs/1204.4742}{{\ttfamily 1204.4742}}].

\bibitem{Sundin:2013ypa}
P.~Sundin and L.~Wulff, \emph{Worldsheet scattering in {$\AdS{3}/\CFT_{2}$}},
  \href{https://doi.org/10.1007/JHEP07(2013)007}{\emph{JHEP} {\bfseries 1307}
  (2013) 007} [\href{https://arxiv.org/abs/1302.5349}{{\ttfamily 1302.5349}}].

\bibitem{Hoare:2013lja}
B.~Hoare, A.~Stepanchuk and A.~Tseytlin, \emph{Giant magnon solution and
  dispersion relation in string theory in {$\AdS{3} \times S^3 \times T^4$}
  with mixed flux},
  \href{https://doi.org/10.1016/j.nuclphysb.2013.12.011}{\emph{Nucl. Phys.}
  {\bfseries B879} (2014) 318}
  [\href{https://arxiv.org/abs/1311.1794}{{\ttfamily 1311.1794}}].

\bibitem{Sundin:2014ema}
P.~Sundin and L.~Wulff, \emph{One- and two-loop checks for the {$\AdS{3} \times
  \Sphere^3 \times \Torus^4$} superstring with mixed flux},
  \href{https://doi.org/10.1088/1751-8113/48/10/105402}{\emph{J. Phys.}
  {\bfseries A48} (2015) 105402}
  [\href{https://arxiv.org/abs/1411.4662}{{\ttfamily 1411.4662}}].

\bibitem{Sundin:2015uva}
P.~Sundin and L.~Wulff, \emph{The {$\AdS{n} \times \Sphere^n \times
  {\Torus}^{10-2n}$} {BMN} string at two loops},
  \href{https://doi.org/10.1007/JHEP11(2015)154}{\emph{JHEP} {\bfseries 11}
  (2015) 154} [\href{https://arxiv.org/abs/1508.04313}{{\ttfamily
  1508.04313}}].

\bibitem{Sundin:2016gqe}
P.~Sundin and L.~Wulff, \emph{{The complete one-loop BMN S-matrix in $
  \text{AdS}_{3} \times \text{S}^{3} \times \text{T}^{4}$}},
  \href{https://doi.org/10.1007/JHEP06(2016)062}{\emph{JHEP} {\bfseries 06}
  (2016) 062} [\href{https://arxiv.org/abs/1605.01632}{{\ttfamily
  1605.01632}}].

\bibitem{Borsato:2013hoa}
R.~Borsato, O.~Ohlsson~Sax, A.~Sfondrini, B.~Stefa{\'n}ski, jr. and
  A.~Torrielli, \emph{Dressing phases of {$\AdS{3}/\CFT_{2}$}},
  \href{https://doi.org/10.1103/PhysRevD.88.066004}{\emph{Phys. Rev.}
  {\bfseries D88} (2013) 066004}
  [\href{https://arxiv.org/abs/1306.2512}{{\ttfamily 1306.2512}}].

\bibitem{Borsato:2016xns}
R.~Borsato, O.~Ohlsson~Sax, A.~Sfondrini, B.~Stefa{\'n}ski, A.~Torrielli and
  O.~Ohlsson~Sax, \emph{{On the dressing factors, Bethe equations and Yangian
  symmetry of strings on $\AdS{3} \times S^3 \times T^4$}},
  \href{https://doi.org/10.1088/1751-8121/50/2/024004}{\emph{J. Phys.}
  {\bfseries A50} (2017) 024004}
  [\href{https://arxiv.org/abs/1607.00914}{{\ttfamily 1607.00914}}].

\bibitem{Borsato:2016kbm}
R.~Borsato, O.~Ohlsson~Sax, A.~Sfondrini and B.~Stefa{\'n}ski, \emph{{On the
  spectrum of $\AdS{3} \times S^3 \times T^4$ strings with Ramond-Ramond
  flux}}, \href{https://doi.org/10.1088/1751-8113/49/41/41LT03}{\emph{J. Phys.}
  {\bfseries A49} (2016) 41LT03}
  [\href{https://arxiv.org/abs/1605.00518}{{\ttfamily 1605.00518}}].

\bibitem{Janik:2006dc}
R.~A. Janik, \emph{The {$\AdS{5} \times \Sphere^5$} superstring worldsheet
  {S}-matrix and crossing symmetry},
  \href{https://doi.org/10.1103/PhysRevD.73.086006}{\emph{Phys. Rev.}
  {\bfseries D73} (2006) 086006}
  [\href{https://arxiv.org/abs/hep-th/0603038}{{\ttfamily hep-th/0603038}}].

\bibitem{Berenstein:2002jq}
D.~E. Berenstein, J.~M. Maldacena and H.~S. Nastase, \emph{Strings in flat
  space and {pp} waves from {$\superN = 4$} super {Y}ang {M}ills},
  \href{https://doi.org/10.1088/1126-6708/2002/04/013}{\emph{JHEP} {\bfseries
  0204} (2002) 013} [\href{https://arxiv.org/abs/hep-th/0202021}{{\ttfamily
  hep-th/0202021}}].

\bibitem{Baggio:2018gct}
M.~Baggio and A.~Sfondrini, \emph{{Strings on NS-NS Backgrounds as Integrable
  Deformations}}, \href{https://doi.org/10.1103/PhysRevD.98.021902}{\emph{Phys.
  Rev.} {\bfseries D98} (2018) 021902}
  [\href{https://arxiv.org/abs/1804.01998}{{\ttfamily 1804.01998}}].

\bibitem{Dei:2018mfl}
A.~Dei and A.~Sfondrini, \emph{{Integrable spin chain for stringy
  Wess-Zumino-Witten models}},
  \href{https://doi.org/10.1007/JHEP07(2018)109}{\emph{JHEP} {\bfseries 07}
  (2018) 109} [\href{https://arxiv.org/abs/1806.00422}{{\ttfamily
  1806.00422}}].

\bibitem{Arutyunov:2004yx}
G.~Arutyunov and S.~Frolov, \emph{Integrable {H}amiltonian for classical
  strings on {$\AdS{5} \times S^5$}},
  \href{https://doi.org/10.1088/1126-6708/2005/02/059}{\emph{JHEP} {\bfseries
  0502} (2005) 059} [\href{https://arxiv.org/abs/hep-th/0411089}{{\ttfamily
  hep-th/0411089}}].

\bibitem{Arutyunov:2005hd}
G.~Arutyunov and S.~Frolov, \emph{Uniform light-cone gauge for strings in
  {$\AdS{5} \times S^5$}: {S}olving {$\algSU(1|1)$} sector},
  \href{https://doi.org/10.1088/1126-6708/2006/01/055}{\emph{JHEP} {\bfseries
  0601} (2006) 055} [\href{https://arxiv.org/abs/hep-th/0510208}{{\ttfamily
  hep-th/0510208}}].

\bibitem{Arutyunov:2006gs}
G.~Arutyunov, S.~Frolov and M.~Zamaklar, \emph{Finite-size effects from giant
  magnons}, \href{https://doi.org/10.1016/j.nuclphysb.2006.12.026}{\emph{Nucl.
  Phys.} {\bfseries B778} (2007) 1}
  [\href{https://arxiv.org/abs/hep-th/0606126}{{\ttfamily hep-th/0606126}}].

\bibitem{Dray:1984ha}
T.~Dray and G.~'t~Hooft, \emph{{The Gravitational Shock Wave of a Massless
  Particle}}, \href{https://doi.org/10.1016/0550-3213(85)90525-5}{\emph{Nucl.
  Phys.} {\bfseries B253} (1985) 173}.

\bibitem{Dubovsky:2012wk}
S.~Dubovsky, R.~Flauger and V.~Gorbenko, \emph{{Solving the Simplest Theory of
  Quantum Gravity}}, \href{https://doi.org/10.1007/JHEP09(2012)133}{\emph{JHEP}
  {\bfseries 09} (2012) 133} [\href{https://arxiv.org/abs/1205.6805}{{\ttfamily
  1205.6805}}].

\bibitem{Smirnov:2016lqw}
F.~A. Smirnov and A.~B. Zamolodchikov, \emph{{On space of integrable quantum
  field theories}},
  \href{https://doi.org/10.1016/j.nuclphysb.2016.12.014}{\emph{Nucl. Phys.}
  {\bfseries B915} (2017) 363}
  [\href{https://arxiv.org/abs/1608.05499}{{\ttfamily 1608.05499}}].

\bibitem{Cavaglia:2016oda}
A.~Cavagli{\`a}, S.~Negro, I.~M. Sz{\'e}cs{\'e}nyi and R.~Tateo, \emph{{$T
  \bar{T}$-deformed 2D Quantum Field Theories}},
  \href{https://doi.org/10.1007/JHEP10(2016)112}{\emph{JHEP} {\bfseries 10}
  (2016) 112} [\href{https://arxiv.org/abs/1608.05534}{{\ttfamily
  1608.05534}}].

\bibitem{Yang:1968rm}
C.-N. Yang and C.~P. Yang, \emph{{Thermodynamics of one-dimensional system of
  bosons with repulsive delta function interaction}},
  \href{https://doi.org/10.1063/1.1664947}{\emph{J. Math. Phys.} {\bfseries 10}
  (1969) 1115}.

\bibitem{Zamolodchikov:1989cf}
A.~B. Zamolodchikov, \emph{Thermodynamic {B}ethe ansatz in relativistic models.
  {S}caling three state {P}otts and {L}ee-{Y}ang models},
  \href{https://doi.org/10.1016/0550-3213(90)90333-9}{\emph{Nucl. Phys.}
  {\bfseries B342} (1990) 695}.

\bibitem{Ambjorn:2005wa}
J.~Ambj{\o}rn, R.~A. Janik and C.~Kristjansen, \emph{Wrapping interactions and
  a new source of corrections to the spin-chain/string duality},
  \href{https://doi.org/10.1016/j.nuclphysb.2005.12.007}{\emph{Nucl. Phys.}
  {\bfseries B736} (2006) 288}
  [\href{https://arxiv.org/abs/hep-th/0510171}{{\ttfamily hep-th/0510171}}].

\bibitem{Alday:2005jm}
L.~F. Alday, G.~Arutyunov and S.~Frolov, \emph{New integrable system of 2dim
  fermions from strings on {$\AdS{5} \times \Sphere^5$}},
  \href{https://doi.org/10.1088/1126-6708/2006/01/078}{\emph{JHEP} {\bfseries
  0601} (2006) 078} [\href{https://arxiv.org/abs/hep-th/0508140}{{\ttfamily
  hep-th/0508140}}].

\bibitem{OhlssonSax:2011ms}
O.~Ohlsson~Sax and B.~Stefa{\'n}ski, jr., \emph{Integrability, spin-chains and
  the {$\AdS{3}/\CFT_{2}$} correspondence},
  \href{https://doi.org/10.1007/JHEP08(2011)029}{\emph{JHEP} {\bfseries 1108}
  (2011) 029} [\href{https://arxiv.org/abs/1106.2558}{{\ttfamily 1106.2558}}].

\bibitem{Borsato:2012ss}
R.~Borsato, O.~Ohlsson~Sax and A.~Sfondrini, \emph{All-loop {B}ethe ansatz
  equations for {$\AdS{3}/\CFT_{2}$}},
  \href{https://doi.org/10.1007/JHEP04(2013)116}{\emph{JHEP} {\bfseries 1304}
  (2013) 116} [\href{https://arxiv.org/abs/1212.0505}{{\ttfamily 1212.0505}}].

\bibitem{Dei:2018yth}
A.~Dei, M.~R. Gaberdiel and A.~Sfondrini, \emph{{The plane-wave limit of ${\rm
  AdS}_3 \times {\rm S}^3 \times {\rm S}^3 \times {\rm S}^1$}},
  \href{https://doi.org/10.1007/JHEP08(2018)097}{\emph{JHEP} {\bfseries 08}
  (2018) 097} [\href{https://arxiv.org/abs/1805.09154}{{\ttfamily
  1805.09154}}].

\bibitem{Klose:2006zd}
T.~Klose, T.~McLoughlin, R.~Roiban and K.~Zarembo, \emph{Worldsheet scattering
  in {$\AdS{5} \times \Sphere^5$}}, {\emph{JHEP} {\bfseries 0703} (2007) 094}
  [\href{https://arxiv.org/abs/hep-th/0611169}{{\ttfamily hep-th/0611169}}].

\bibitem{Arutyunov:2007tc}
G.~Arutyunov and S.~Frolov, \emph{On string {S}-matrix, bound states and
  {TBA}}, \href{https://doi.org/10.1088/1126-6708/2007/12/024}{\emph{JHEP}
  {\bfseries 0712} (2007) 024}
  [\href{https://arxiv.org/abs/0710.1568}{{\ttfamily 0710.1568}}].

\bibitem{Baggio:2017kza}
M.~Baggio, O.~Ohlsson~Sax, A.~Sfondrini, B.~Stefa{\'n}ski and A.~Torrielli,
  \emph{{Protected string spectrum in AdS$_{3}$/CFT$_{2}$ from worldsheet
  integrability}}, \href{https://doi.org/10.1007/JHEP04(2017)091}{\emph{JHEP}
  {\bfseries 04} (2017) 091}
  [\href{https://arxiv.org/abs/1701.03501}{{\ttfamily 1701.03501}}].

\bibitem{Eberhardt:2017fsi}
L.~Eberhardt, M.~R. Gaberdiel, R.~Gopakumar and W.~Li, \emph{{BPS spectrum on
  AdS$_3\times $S$^3 \times $S$^3 \times $S$^1$}},
  \href{https://doi.org/10.1007/JHEP03(2017)124}{\emph{JHEP} {\bfseries 03}
  (2017) 124} [\href{https://arxiv.org/abs/1701.03552}{{\ttfamily
  1701.03552}}].

\bibitem{Baggio:2018rpv}
M.~Baggio, A.~Sfondrini, G.~Tartaglino-Mazzucchelli and H.~Walsh, \emph{{On
  $T\bar{T}$ deformations and supersymmetry}},
  \href{https://arxiv.org/abs/1811.00533}{{\ttfamily 1811.00533}}.

\bibitem{Staudacher:2004tk}
M.~Staudacher, \emph{{The Factorized S-matrix of CFT/AdS}},
  \href{https://doi.org/10.1088/1126-6708/2005/05/054}{\emph{JHEP} {\bfseries
  05} (2005) 054} [\href{https://arxiv.org/abs/hep-th/0412188}{{\ttfamily
  hep-th/0412188}}].

\bibitem{vanTongeren:2016hhc}
S.~J. van Tongeren, \emph{{Introduction to the thermodynamic Bethe ansatz}},
  \href{https://arxiv.org/abs/1606.02951}{{\ttfamily 1606.02951}}.

\bibitem{Arutyunov:2009zu}
G.~Arutyunov and S.~Frolov, \emph{String hypothesis for the {$\AdS{5} \times
  S^5$} mirror},
  \href{https://doi.org/10.1088/1126-6708/2009/03/152}{\emph{JHEP} {\bfseries
  0903} (2009) 152} [\href{https://arxiv.org/abs/0901.1417}{{\ttfamily
  0901.1417}}].

\bibitem{vanTongeren:2015uha}
S.~J. van Tongeren, \emph{{Y}ang-{B}axter deformations, {$\AdS/\CFT$}, and
  twist-noncommutative gauge theory},
  \href{https://doi.org/10.1016/j.nuclphysb.2016.01.012}{\emph{Nucl. Phys.}
  {\bfseries B904} (2016) 148}
  [\href{https://arxiv.org/abs/1506.01023}{{\ttfamily 1506.01023}}].

\bibitem{Dorey:1996re}
P.~Dorey and R.~Tateo, \emph{Excited states by analytic continuation of {TBA}
  equations}, \href{https://doi.org/10.1016/S0550-3213(96)00516-0}{\emph{Nucl.
  Phys.} {\bfseries B482} (1996) 639}
  [\href{https://arxiv.org/abs/hep-th/9607167}{{\ttfamily hep-th/9607167}}].

\bibitem{Eberhardt:2017pty}
L.~Eberhardt, M.~R. Gaberdiel and W.~Li, \emph{{A holographic dual for string
  theory on $ \mathrm{AdS}_3\times \mathrm{S}^3\times
  \mathrm{S}^3\times\mathrm{S}^1 $ }},
  \href{https://doi.org/10.1007/JHEP08(2017)111}{\emph{JHEP} {\bfseries 08}
  (2017) 111} [\href{https://arxiv.org/abs/1707.02705}{{\ttfamily
  1707.02705}}].

\bibitem{Giribet:2018ada}
G.~Giribet, C.~Hull, M.~Kleban, M.~Porrati and E.~Rabinovici,
  \emph{{Superstrings on AdS$_{3}$ at $ k = $ 1}},
  \href{https://doi.org/10.1007/JHEP08(2018)204}{\emph{JHEP} {\bfseries 08}
  (2018) 204} [\href{https://arxiv.org/abs/1803.04420}{{\ttfamily
  1803.04420}}].

\bibitem{Gaberdiel:2018rqv}
M.~R. Gaberdiel and R.~Gopakumar, \emph{{Tensionless string spectra on
  AdS$_{3}$}}, \href{https://doi.org/10.1007/JHEP05(2018)085}{\emph{JHEP}
  {\bfseries 05} (2018) 085}
  [\href{https://arxiv.org/abs/1803.04423}{{\ttfamily 1803.04423}}].

\bibitem{Eberhardt:2018ouy}
L.~Eberhardt, M.~R. Gaberdiel and R.~Gopakumar, \emph{{The Worldsheet Dual of
  the Symmetric Product CFT}},
  \href{https://arxiv.org/abs/1812.01007}{{\ttfamily 1812.01007}}.

\bibitem{Teschner:1997ft}
J.~Teschner, \emph{{On structure constants and fusion rules in the
  $SL(2,\mathbb{C}) / SU(2)$ WZNW model}},
  \href{https://doi.org/10.1016/S0550-3213(99)00072-3}{\emph{Nucl. Phys.}
  {\bfseries B546} (1999) 390}
  [\href{https://arxiv.org/abs/hep-th/9712256}{{\ttfamily hep-th/9712256}}].

\bibitem{Teschner:1999ug}
J.~Teschner, \emph{{Operator product expansion and factorization in the
  $H_{3}^{+}$ WZNW model}},
  \href{https://doi.org/10.1016/S0550-3213(99)00785-3}{\emph{Nucl. Phys.}
  {\bfseries B571} (2000) 555}
  [\href{https://arxiv.org/abs/hep-th/9906215}{{\ttfamily hep-th/9906215}}].

\bibitem{Cardona:2009hk}
C.~A. Cardona and C.~A. Nunez, \emph{{Three-point functions in superstring
  theory on AdS(3) x S**3 x T**4}},
  \href{https://doi.org/10.1088/1126-6708/2009/06/009}{\emph{JHEP} {\bfseries
  06} (2009) 009} [\href{https://arxiv.org/abs/0903.2001}{{\ttfamily
  0903.2001}}].

\bibitem{Cardona:2010qf}
C.~A. Cardona and I.~Kirsch, \emph{{Worldsheet four-point functions in
  $\AdS{3}/\CFT_{2}$}},
  \href{https://doi.org/10.1007/JHEP01(2011)015}{\emph{JHEP} {\bfseries 01}
  (2011) 015} [\href{https://arxiv.org/abs/1007.2720}{{\ttfamily 1007.2720}}].

\bibitem{Basso:2015zoa}
B.~Basso, S.~Komatsu and P.~Vieira, \emph{Structure constants and integrable
  bootstrap in planar {$\superN = 4$} {SYM} theory},
  \href{https://arxiv.org/abs/1505.06745}{{\ttfamily 1505.06745}}.

\bibitem{Eden:2016xvg}
B.~Eden and A.~Sfondrini, \emph{{Tessellating cushions: four-point functions in
  $\mathcal{N} $ = 4 SYM}},
  \href{https://doi.org/10.1007/JHEP10(2017)098}{\emph{JHEP} {\bfseries 10}
  (2017) 098} [\href{https://arxiv.org/abs/1611.05436}{{\ttfamily
  1611.05436}}].

\bibitem{Fleury:2016ykk}
T.~Fleury and S.~Komatsu, \emph{{Hexagonalization of Correlation Functions}},
  \href{https://doi.org/10.1007/JHEP01(2017)130}{\emph{JHEP} {\bfseries 01}
  (2017) 130} [\href{https://arxiv.org/abs/1611.05577}{{\ttfamily
  1611.05577}}].

\bibitem{Basso:2017khq}
B.~Basso, F.~Coronado, S.~Komatsu, H.~T. Lam, P.~Vieira and D.-l. Zhong,
  \emph{{Asymptotic Four Point Functions}},
  \href{https://arxiv.org/abs/1701.04462}{{\ttfamily 1701.04462}}.

\bibitem{Fleury:2017eph}
T.~Fleury and S.~Komatsu, \emph{{Hexagonalization of Correlation Functions II:
  Two-Particle Contributions}},
  \href{https://doi.org/10.1007/JHEP02(2018)177}{\emph{JHEP} {\bfseries 02}
  (2018) 177} [\href{https://arxiv.org/abs/1711.05327}{{\ttfamily
  1711.05327}}].

\bibitem{Eden:2017ozn}
B.~Eden, Y.~Jiang, D.~le~Plat and A.~Sfondrini, \emph{{Colour-dressed hexagon
  tessellations for correlation functions and non-planar corrections}},
  \href{https://doi.org/10.1007/JHEP02(2018)170}{\emph{JHEP} {\bfseries 02}
  (2018) 170} [\href{https://arxiv.org/abs/1710.10212}{{\ttfamily
  1710.10212}}].

\bibitem{Bargheer:2017nne}
T.~Bargheer, J.~Caetano, T.~Fleury, S.~Komatsu and P.~Vieira, \emph{{Handling
  Handles: Nonplanar Integrability in $\mathcal{N}=4$ Supersymmetric Yang-Mills
  Theory}}, \href{https://doi.org/10.1103/PhysRevLett.121.231602}{\emph{Phys.
  Rev. Lett.} {\bfseries 121} (2018) 231602}
  [\href{https://arxiv.org/abs/1711.05326}{{\ttfamily 1711.05326}}].

\bibitem{Bargheer:2018jvq}
T.~Bargheer, J.~Caetano, T.~Fleury, S.~Komatsu and P.~Vieira, \emph{{Handling
  handles. Part II. Stratification and data analysis}},
  \href{https://doi.org/10.1007/JHEP11(2018)095}{\emph{JHEP} {\bfseries 11}
  (2018) 095} [\href{https://arxiv.org/abs/1809.09145}{{\ttfamily
  1809.09145}}].

\bibitem{Eden:2015ija}
B.~Eden and A.~Sfondrini, \emph{{Three-point functions in ${\cal N}=4$ SYM: the
  hexagon proposal at three loops}},
  \href{https://doi.org/10.1007/JHEP02(2016)165}{\emph{JHEP} {\bfseries 02}
  (2016) 165} [\href{https://arxiv.org/abs/1510.01242}{{\ttfamily
  1510.01242}}].

\bibitem{Basso:2015eqa}
B.~Basso, V.~Goncalves, S.~Komatsu and P.~Vieira, \emph{{Gluing Hexagons at
  Three Loops}},
  \href{https://doi.org/10.1016/j.nuclphysb.2016.04.020}{\emph{Nucl. Phys.}
  {\bfseries B907} (2016) 695}
  [\href{https://arxiv.org/abs/1510.01683}{{\ttfamily 1510.01683}}].

\bibitem{Basso:2017muf}
B.~Basso, V.~Goncalves and S.~Komatsu, \emph{{Structure constants at wrapping
  order}}, \href{https://doi.org/10.1007/JHEP05(2017)124}{\emph{JHEP}
  {\bfseries 05} (2017) 124}
  [\href{https://arxiv.org/abs/1702.02154}{{\ttfamily 1702.02154}}].

\bibitem{Luscher:1985dn}
M.~L{\"u}scher, \emph{Volume dependence of the energy spectrum in massive
  quantum field theories. 1. {S}table particle states},
  \href{https://doi.org/10.1007/BF01211589}{\emph{Commun. Math. Phys.}
  {\bfseries 104} (1986) 177}.

\bibitem{Luscher:1986pf}
M.~L{\"u}scher, \emph{Volume dependence of the energy spectrum in massive
  quantum field theories. 2. {S}cattering states},
  \href{https://doi.org/10.1007/BF01211097}{\emph{Commun. Math. Phys.}
  {\bfseries 105} (1986) 153}.

\bibitem{Eden:2018vug}
B.~Eden, Y.~Jiang, M.~de~Leeuw, T.~Meier, D.~le~Plat and A.~Sfondrini,
  \emph{{Positivity of hexagon perturbation theory}},
  \href{https://doi.org/10.1007/JHEP11(2018)097}{\emph{JHEP} {\bfseries 11}
  (2018) 097} [\href{https://arxiv.org/abs/1806.06051}{{\ttfamily
  1806.06051}}].

\bibitem{Giveon:2017nie}
A.~Giveon, N.~Itzhaki and D.~Kutasov, \emph{{$ \mathrm{T}\overline{\mathrm{T}}
  $ and LST}}, \href{https://doi.org/10.1007/JHEP07(2017)122}{\emph{JHEP}
  {\bfseries 07} (2017) 122}
  [\href{https://arxiv.org/abs/1701.05576}{{\ttfamily 1701.05576}}].

\bibitem{Giveon:2017myj}
A.~Giveon, N.~Itzhaki and D.~Kutasov, \emph{{A solvable irrelevant deformation
  of AdS$_{3}$/CFT$_{2}$}},
  \href{https://doi.org/10.1007/JHEP12(2017)155}{\emph{JHEP} {\bfseries 12}
  (2017) 155} [\href{https://arxiv.org/abs/1707.05800}{{\ttfamily
  1707.05800}}].

\bibitem{Eberhardt:2018vho}
L.~Eberhardt and K.~Ferreira, \emph{{Long strings and chiral primaries in the
  hybrid formalism}},  \href{https://arxiv.org/abs/1810.08621}{{\ttfamily
  1810.08621}}.

\end{thebibliography}\endgroup

\end{document}